\documentclass[12pt]{article}

\usepackage{tikz}
\usetikzlibrary{matrix,arrows}

\usepackage{a4wide}

\usepackage{amsmath,amssymb,amsfonts} 
\usepackage[T1]{fontenc}
\usepackage{yfonts}

\usepackage{extarrows}
\usepackage{undertilde}

\usepackage{mathrsfs,latexsym}

\newcommand{\II}{{\boldsymbol{1}}}

\newcommand{\CC}{{\mathbb C}}

\newcommand{\RR}{{\mathbb R}}

\newcommand{\NN}{{\mathbb N}}

\newcommand{\CoinX}[1]{C_0^\infty({#1})}

\newtheorem{Thm}{Theorem}[section]

\newtheorem{Def}[Thm]{Definition}
\newtheorem{Lem}[Thm]{Lemma}
\newtheorem{Prop}[Thm]{Proposition}

\numberwithin{equation}{section}

\newcommand{\sN}{{\mathsf N}}

\newcommand{\HH}{{\mathscr H}}
\newcommand{\KK}{{\mathscr K}}

\newcommand{\Ac}{{\mathcal A}}
\newcommand{\Bc}{{\mathcal B}}

\newcommand{\cI}{{\mathcal I}}
\newcommand{\Ic}{{\mathcal I}}

\newcommand{\OO}{{\mathscr O}}

\newcommand{\kb}{{\boldsymbol{k}}}

\newcommand{\fb}{{\boldsymbol{f}}}
\newcommand{\gb}{{\boldsymbol{g}}}
\newcommand{\hb}{{\boldsymbol{h}}}

\newcommand{\ogth}{{\mathfrak o}}
\newcommand{\tgth}{{\mathfrak t}}
\newcommand{\wgth}{{\mathfrak w}}

\newcommand{\supp}{{\rm supp}\,}

\DeclareMathOperator{\pr}{pr}

\renewcommand{\Im}{{\rm Im}\,}

\newcommand{\Ob}{{\boldsymbol{0}}}

\newcommand{\Bb}{{\boldsymbol{B}}}
\newcommand{\Cb}{{\boldsymbol{C}}}
\newcommand{\Db}{{\boldsymbol{D}}}
\newcommand{\Fb}{{\boldsymbol{F}}}
\newcommand{\Ib}{{\boldsymbol{I}}}
\newcommand{\Lb}{{\boldsymbol{L}}}
\newcommand{\Mb}{{\boldsymbol{M}}}
\newcommand{\Nb}{{\boldsymbol{N}}}
\newcommand{\Pb}{{\boldsymbol{P}}}

\newcommand{\Tb}{{\boldsymbol{T}}}
\newcommand{\Ub}{{\boldsymbol{U}}}

\newcommand{\Lc}{{\mathcal{L}}}
\newcommand{\Mc}{{\mathcal{M}}}
\newcommand{\Nc}{{\mathcal{N}}}

\newcommand{\Ct}{{\sf C}}

\newcommand{\It}{{\sf I}}

\newcommand{\LCTo}{{\sf LCT}_0}
\newcommand{\LCT}{{\sf LCT}}
\newcommand{\Loco}{{\sf Loc}_0}
\newcommand{\Loc}{{\sf Loc}}
\newcommand{\Man}{\Loco}
\newcommand{\Mand}{\Loc}

\newcommand{\Sys}{{\sf Sys}}

\newcommand{\Alg}{{\sf Alg}}
\newcommand{\CAlg}{{\sf C^*\hbox{-}Alg}}
\newcommand{\TAlg}{{\sf TAlg}}

\newcommand{\Phys}{{\sf Phys}}

\newcommand{\Af}{{\mathscr A}}

\newcommand{\Bf}{{\mathscr B}}

\newcommand{\Df}{{\mathscr D}}

\newcommand{\Ff}{{\mathscr F}}

\newcommand{\If}{{\mathscr I}}

\newcommand{\id}{{\rm id}}

\newcommand{\nto}{\stackrel{.}{\to}}

\newcommand{\Aut}{{\rm Aut}}

\newcommand{\Funct}{{\rm Funct}}

\DeclareMathOperator{\cl}{cl}

\newcommand{\rce}{{\rm rce}}

\newcommand{\Cpts}{{\rm Cpts}}
\newcommand{\dyn}{{\rm dyn}}
\newcommand{\kin}{{\rm kin}}

\begin{document}

%

\renewcommand{\thefootnote}{\fnsymbol{footnote}}
\begin{center}
{ \Large \bf Dynamical locality and covariance: What makes a physical theory the same in all spacetimes?}
\\[20pt]
{\large  Christopher J.\ Fewster${}^{1}$\footnote{E-mail: chris.fewster@york.ac.uk}
 {\rm and}   Rainer Verch${}^{2}$\footnote{E-mail: verch@itp.uni-leipzig.de}}
\\[20pt]  
                 ${}^1$ Department of Mathematics,
                 University of York, 
                 Heslington,
                 York YO10 5DD, U.K.
                 \\
                 ${}^2$\,
Institut f\"ur Theoretische Physik,
Universit\"at Leipzig,
04009 Leipzig, Germany \\[18pt]
\today
\end{center}
${}$\\[10pt]
{\small {\bf Abstract. }    
The question of what it means for a theory to describe the same physics on
all spacetimes (SPASs) is discussed. As there may be many answers to this
question, we isolate a necessary condition, the SPASs property, that should
be satisfied by any reasonable notion of SPASs. This requires that if two
theories conform to a common notion of SPASs, with one a subtheory of the
other, and are isomorphic in some particular spacetime, then they should be
isomorphic in all globally hyperbolic spacetimes (of given dimension). The
SPASs property is formulated in a functorial setting broad enough to
describe general physical theories describing processes in spacetime,
subject to very minimal assumptions. By explicit constructions, the full
class of locally covariant theories is shown not to satisfy the SPASs
property, establishing that there is no notion of SPASs encompassing all
such theories. It is also shown that all locally covariant theories obeying
the time-slice property possess two local substructures, one kinematical
(obtained directly from the functorial structure) and the other dynamical
(obtained from a natural form of dynamics, termed relative Cauchy
evolution). The covariance properties of relative Cauchy evolution and the
kinematic and dynamical substructures are analyzed in detail. Calling local
covariant theories dynamically local if their kinematical and dynamical
local substructures coincide, it is shown that the class of dynamically
local theories fulfills the SPASs property. As an application in quantum
field theory, we give a model independent proof of the impossibility of
making a covariant choice of preferred state in all spacetimes, for theories
obeying dynamical locality together with typical assumptions.
}
${}$

\renewcommand{\thefootnote}{\arabic{footnote}}

\section{Introduction}

Terrestrial experiments in particle physics are conducted in a weak gravitational field. 
To interpret their results
in terms of QFT models it is therefore necessary that these models can,
in principle, be formulated in curved spacetimes without altering their
essential physical content, and that one can study and control the limit
of weak gravitational fields. This paper is devoted to the first of
these issues: specifically, to understanding what requirements should
be imposed on a theory formulated on a large class of spacetimes to ensure that the physical
content is the same in all cases. 

Operational concerns dictate a number of restrictions. Experiments are
performed in finite regions of spacetime; local causality~\cite{HawkingEllis} requires
that these experiments should be insensitive to the geometry in the
casual complement of the region concerned. Furthermore, the geometrical
description of the theory should not be based
on preferred systems of reference. 

In the context of quantum field theory in curved spacetime, the
requirements mentioned so far are implemented within a framework of
locally covariant QFT developed by Brunetti, Fredenhagen and Verch
(hereafter abbreviated to BFV) in \cite{BrFrVe03} (see also \cite{Verch01}; 
antecedents of these ideas may be found, e.g., in~\cite{Dimock1980,Kay79,Kay_Flocality:1992}). 
There, a quantum field 
theory defined on all spacetimes is modelled by a functor between a
category of globally hyperbolic manifolds and a category of unital $(C)^*$-algebras. Thus to each spacetime
$\Mb$ the theory assigns a $(C)^*$-algebra $\Af(\Mb)$ which might be an algebra of smeared fields, or of local observables; importantly, 
to each morphism $\psi:\Mb\to\Nb$ between spacetimes\footnote{The morphisms are
isometric embeddings, preserving orientation and time-orientation, with causally convex image.
See Section~\ref{sect:spacetimes}.}
 there is a
corresponding morphism $\Af(\psi): \Af(\Mb)\to\Af(\Nb)$ of $(C)^*$-algebras,
so that $\Af(\psi\circ\varphi)=\Af(\psi)\circ\Af(\varphi)$, and with identity morphisms
of spacetimes mapped to identity morphisms of $(C)^*$-algebras.  

The BFV approach, which we review and develop in Section~\ref{sect:LOCCOV}, has significantly
advanced the programme of extending results of flat spacetime QFT to
curved spacetimes: particular instances include a
spin-statistics theorem \cite{Verch01}, analogues of the Reeh-Schlieder theorem \cite{Sanders_ReehSchlieder},
superselection theory~\cite{Br&Ru05,BrunettiRuzzi_topsect}, and the perturbative construction of interacting theories in curved spacetime \cite{BrFr2000,Ho&Wa01,Ho&Wa02}. 
Applications to {\em a priori} bounds on Casimir energy densities~\cite{Few&Pfen06, Fewster2007} and new viewpoints in cosmology~\cite{DapFrePin2008,DegVer2010,VerchRegensburg} have also resulted from
this circle of ideas. 

Somewhat surprisingly, however, it turns out that one may formulate
theories in the BFV framework that (at least intuitively) do {\em not} represent the same
physical content in all spacetimes. We will give specific examples in
section~\ref{sect:pathologies},  although these should be regarded
as illustrating the range of pathological behaviour, rather than completely describing it.
This raises the questions: (a)  can one make precise the sense in which such
theories fail to have the same content in all spacetimes, and (b) what additional conditions should be imposed to remedy this shortcoming? While we will not completely resolve these issues, we are able to give
a framework in which it may be addressed and at least partly resolved. 

A fundamental problem is that it is unclear how the `physical content'
of a theory is to be defined in an axiomatic framework. Even
a recourse to a Lagrangian setting does not resolve all the issues: see~\cite{FewsterRegensburg} for 
examples of covariantly defined Lagrangian field theories that do not  represent 
the same physics in all spacetimes. This being so, it is even harder to make precise,
by a purely intensional definition, what it means for this content to `be the same' in different spacetimes. 

Given this situation, it seems advisable to allow that there may be many cogent notions 
of what it means for a theory to represent the same physics in all spacetimes (often abbreviated as
SPASs in this paper).\footnote{In principle we
even allow that there might even be {\em no} such notion.} Our first aim is to 
assert principles that should be obeyed by any notion of SPASs and investigate the consequences.
In order to do this, we represent any candidate definition of SPASs by the class of theories
that conform to it (i.e., an extensional viewpoint); our
principles can therefore be expressed as necessary conditions on a class $\frak{T}$ of theories in order
that it can serve as a notion of SPASs. Stated as physical principles, they are:
\begin{enumerate}\addtolength{\itemsep}{-0.5\baselineskip}
\item[S1] Every theory in $\frak{T}$ should be locally covariant.
\item[S2] If $\Af$ and $\Bf$ are  (not necessarily distinct) theories in $\frak{T}$, with $\Af$ a subtheory of $\Bf$, and
$\Af$ and $\Bf$ coincide in one spacetime, then they should coincide in all spacetimes.
\end{enumerate}
We do not by any means claim that this is an exhaustive prescription and emphasise again
that this is not a definition of any particular notion of SPASs but rather a set of principles 
that should be obeyed by all reasonable notions. Moreover, the term `coincide' requires precise definition, which will be given below.  However, the two conditions together will turn out to be surprisingly strong. 

Implementing these principles mathematically, S1 is exactly implemented in the BFV framework and immediately
restricts attention to theories that are covariant functors from the category of globally hyperbolic
spacetimes to a category $\Phys$ of mathematical objects representing `the physics'. Principle S2
is new, and can be implemented in the BFV framework as follows: if $\Af$ and $\Bf$ are functors 
representing locally covariant theories, any natural transformation $\zeta:\Af\nto\Bf$ is interpreted as  embedding $\Af$ as a subtheory of $\Bf$. The collection of locally covariant theories becomes a category on adopting such subtheory embeddings as morphisms. We will regard the theories as coinciding in some spacetime $\Mb$ if this embedding is an isomorphism in $\Mb$,
in which case $\zeta$ is called a {\em partial isomorphism}; the theories coincide in all spacetimes if
this condition holds for all $\Mb$, in which case $\zeta$ is a natural isomorphism. Principle S2 is
then implemented by requiring that all partial isomorphisms between theories in $\frak{T}$ are in fact isomorphisms. In this paper, we will refer to S2, implemented in this way, as {\em the SPASs property}; however, as indicated above, the axioms above are not expected to be exhaustive.  It is conceivable that
S2 should be strengthened, to cover situations in which $\Af$ and $\Bf$
may be regarded as coinciding in one spacetime but without the assumption that one is
a subtheory of the other. At present it is not known how to implement this mathematically.

Before proceeding, we wish to emphasise the nature of the subtheory embedding with an
example. Consider the quantum field theory of the nonminimally coupled scalar field. The field equations 
$(\Box+\xi R+m^2)\phi=0$ are evidently independent of the coupling constant in Ricci-flat spacetimes, and this allows the construction of an obvious isomorphism between the algebras of observables for different values of $\xi$ in such spacetimes. However, this does not extend to give a natural transformation between the theories labelled by distinct $\xi$: the `obvious isomorphism' does not qualify as a coincidence of the theories, in our sense, even in Ricci-flat spacetimes. A proof of this is sketched at the end of Sect.~\ref{sect:rce}.

The main result of Sect.~\ref{sect:pathologies} is that the SPASs property does not hold in the category
of all locally covariant theories unless $\Phys$ has rather trivial content; indeed, one can
give pairs of theories (which can be otherwise well-behaved) that cannot satisfy the SPASs property; 
accordingly there is no common notion of SPASs that can accommodate both theories.\footnote{At the end of Sect.~\ref{sect:pathologies} we even construct {\em single} theories that cannot satisfy
{\em any} notion of SPASs.} This is done by an explicit construction that may provide a useful supply of nonstandard
locally covariant theories for other purposes. To give a simple outline of one version of our construction, suppose that $\Phys$ is the category of $*$-algebras and suppose
that $\Af$ is a well-behaved theory. We will show that it is possible to construct nonconstant functions 
on the category of spacetimes, valued in the natural numbers, that are monotonic in the sense that  $\chi(\Mb)\le \chi(\Nb)$ for all pairs of spacetimes linked by a morphism $\Mb\to\Nb$. 
We then define a new theory $\widetilde{\Af}$ on objects by
$\widetilde{\Af}(\Mb) = \Af(\Mb)^{\otimes\chi(\Mb)}$,
where the tensor product is the algebraic tensor product on $\Alg$.
To any morphism $\psi:\Mb\to\Nb$, we assign a morphism 
$\widetilde{\Af}(\psi):\widetilde{\Af}(\Mb)\to\widetilde{\Af}(\Nb)$
given by
\[
\widetilde{\Af}(\psi)(A) =\Af(\psi)^{\otimes\chi(\Mb)}(A)\otimes
\left(\II_{\Af(\Nb)}\right)^{\otimes(\chi(\Nb)-\chi(\Mb))}.
\]
A simple computation shows that $\widetilde{\Af}$ is a functor from the category of spacetimes to $\Alg$. To check this, note that
\[
\widetilde{\Af}(\id_\Mb)=\Af(\id_\Mb)^{\otimes\chi(\Mb)}=\id_{\Af(\Mb)}^{\otimes\chi(\Mb)}=
\id_{\widetilde{\Af}(\Mb)},
\] 
and that, if
$\Mb_1\stackrel{\psi_1}{\to}\Mb_2\stackrel{\psi_2}{\to}\Mb_3$ then
\begin{eqnarray*}
\widetilde{\Af}(\psi_2\circ\psi_1)(A)&=&\Af(\psi_2\circ\psi_1)^{\otimes\chi(\Mb_1)}(A)\otimes
\left(\II_{\Af(\Mb_3)}\right)^{\otimes(\chi(\Mb_3)-\chi(\Mb_1))}\nonumber\\
&=&\left(\Af(\psi_2)^{\otimes\chi(\Mb_1)}(\Af(\psi_1)^{\otimes\chi(\Mb_1)}(A))\right)\otimes
\II_{\Af(\Mb_3)}^{\otimes(\chi(\Mb_2)-\chi(\Mb_1))}\otimes
\II_{\Af(\Mb_3)}^{\otimes(\chi(\Mb_3)-\chi(\Mb_2))}\nonumber\\
&=&\Af(\psi_2)^{\otimes\chi(\Mb_2)}\left(\Af(\psi_1)^{\otimes\chi(\Mb_1)}(A)\otimes
\II_{\Af(\Mb_2)}^{\otimes(\chi(\Mb_2)-\chi(\Mb_1))}\right)\otimes
\II_{\Af(\Mb_3)}^{\otimes(\chi(\Mb_3)-\chi(\Mb_2))}\nonumber\\
&=& \widetilde{\Af}(\psi_2)(\widetilde{\Af}(\psi_1)(A))\nonumber\\
&=& \left(\widetilde{\Af}(\psi_2)\circ\widetilde{\Af}(\psi_1)\right)(A)
\end{eqnarray*}
for any $A\in\Af(\Mb_1)$, using the unit-preserving property of
$\Alg$-morphisms. Thus the functor $\widetilde{\Af}$ satisfies the
definition of a locally covariant quantum field theory. However one cannot
expect both $\Af$ and $\widetilde{\Af}$ to have the same physical 
content in all spacetimes as the theory consists of $\chi(\Mb)$ copies of the basic theory $\Af(\Mb)$ in each spacetime $\Mb$. Developing the example further, if $\chi$ has $1$ and $\ell$ as its minimum and
maximum values, then there are successive subtheory embeddings $\Af\nto\widetilde{\Af}\nto \Af^{\otimes\ell}$, each of which is a partial isomorphism, but whose composite is not an isomorphism
[this is a mild assumption on $\Af$]; so at least one of the partial isomorphisms cannot be an isomorphism. Thus the three
theories $\{\Af,\widetilde{\Af}, \Af^{\otimes\ell}\}$ cannot conform to a single notion of SPASs, and
indeed the same is true of at least one of the pairs $\{\Af,\widetilde{\Af}\}$ or $\{\Af^{\otimes\ell},\widetilde{\Af}\}$. Of course, if $\Af$ is a familiar theory that one would regard intuitively as representing SPASs, then one would regard $\widetilde{\Af}$ as `obviously' not representing SPASs by
the same intuitive standard. Our aim in formalising these questions is to provide a framework in which 
such judgements can be made without relying on intuition.

This result raises the question as to what conditions might produce a class of theories obeying the SPASs property. Our answer to this involves a closer examination of the specification of the physics associated with local regions in globally hyperbolic spacetimes. One of the attractive features of the BFV framework
is that it gives a definition for the local physics associated to a region $O$ in spacetime $\Mb$, 
essentially by considering the region $O$ (with the geometry restricted from $\Mb$) as a spacetime in its own right. We shall regard this as a kinematical description of local physics.  
In Section~\ref{sect:intrinsic}, we introduce a new description of the local physics in $O$ that is based on dynamics: 
the local physics in $O$ is that portion of the physics on the whole spacetime that is invariant under modifications of the spacetime metric in the causal complement of $O$ in $\Mb$. The effect of a modification to the metric is captured by the {\em relative Cauchy evolution} introduced by BFV, which
is closely related to the dynamics of the theory.\footnote{The functional derivative of the relative Cauchy
evolution with respect to the metric perturbation can be interpreted as the stress-energy tensor, from
which viewpoint the relative Cauchy evolution is a replacement for a classical action in this framework.}
We investigate the basic properties of the resulting `dynamical net'; while it has a number of features in
common with the `kinematic net' it lacks others, notably the local covariance property of the
kinematic net does not hold for the dynamical net in general.

The situation in which the kinematic and dynamical nets coincide is of particular interest, and
those theories for which it holds will be said to be {\em dynamically local}. As we show in
Section~\ref{sect:dynamical_locality}, dynamically local theories have a number of good
properties: they are additive, have good covariance properties for the dynamical net, and (under a mild additional assumption) obey {\em extended locality} in the sense that the local physics for
spacelike separated regions intersect only trivially.\footnote{What `trivial' means here will depend on the
category $\Phys$ employed to describe the physics. In the categories of $(C)^*$-algebras 
employed in QFT, this means that the intersection consists of complex multiples of the algebra unit. 
Extended locality was originally introduced in~\cite{Schoch1968, Landau1969} in the
context of algebraic QFT in Minkowski space.} The scope for constructing pathological theories
of the sort discussed in Section~\ref{sect:pathologies} is significantly reduced and even
eliminated if the theory has no nontrivial automorphisms (as is expected for a 
theory of local observables). Moreover, as is shown in Theorem~\ref{thm:SPASs}, 
the class of dynamically local theories has the SPASs property. Accordingly, the concept of dynamical locality
provides a first answer to the problem of isolating those theories that can be regarded as representing the same physics in all spacetimes, and appears to be a useful addition to the axiomatic framework in curved
spacetimes.  

As an application of these results to QFT, we give the first model independent proof of
the impossibility of selecting a single `natural' state in each spacetime (Section~\ref{sect:dlQFT})
for any nontrivial dynamically local theory with the extended locality property, on the assumption that the
supposed natural state has the Reeh--Schlieder property in some spacetime. (Here, we say that a
theory is trivial if it is equivalent to the theory
whose algebra of observables consists of complex multiples of the unit in every spacetime.) Neither of these
additional assumptions seem unreasonable; in particular, our result applies to any
theory that reduces, in Minkowski space, to a Wightman or Haag--Kastler theory obeying standard
conditions and with the natural state reducing to the Minkowski vacuum state. It is worth noting 
that the SPASs property is used as a technical input to the proof: the given theory is shown
to coincide with the trivial theory in one spacetime, and must therefore do so in all. 

In addition to these results, and as a necessary technical tool in proving them, we make 
a thorough study of the relative Cauchy evolution, deepening the investigation begun by 
BFV. We particularly study the covariance properties of the relative Cauchy evolution, 
and the way in which subtheory embeddings intertwine the relative Cauchy evolutions
of different theories. Our methods, wherever possible, are adapted to the widest possible
categorical setting, to emphasise the applicability of underlying ideas; all the key concepts
are expressed in terms of universal properties, which makes for efficient proofs that 
are portable between different physical settings. On the geometrical side, we also 
adapt and extend the spacetime deformation methods
introduced in~\cite{FullingNarcowichWald}: in spacetime dimension $n\ge 2$, these techniques 
allow us to partition the category of spacetimes into connected components labelled by equivalence classes
of Riemannian manifolds of dimension $n-1$, modulo orientation-preserving
diffeomorphisms. Here, connectedness is understood in terms of the existence of
chains of `Cauchy wedges' from one spacetime to another. One might conjecture that a
more detailed study of the category of spacetimes from this viewpoint would give
a cohomology theory with many ramifications. Indeed, following our suggestion, Sanders
has shown that one may regard various freedoms arising in the construction of the Dirac field in 
curved spacetimes in precisely such a cohomological way~\cite{Sanders_dirac:2010}. 
Appendix~\ref{appx:geom} provides a body of material on spacetime
structure, required in the body of the paper, particularly in relation to different notions
of causal complement. We hope that a number of these developments will be useful for other purposes. 

A separate paper~\cite{FewVer:dynloc2} is devoted to an investigation of
the dynamical locality for various linear theories, both as classical and quantum fields. 
It is shown that  dynamical locality is satisfied by the massive minimally coupled free scalar field. At zero mass, dynamical locality fails; however, this can be understood as an expression of the 
rigid gauge symmetry of the minimally coupled massless field. When the theory is
quantised as a (rather simple) gauge theory, dynamical locality is restored in dimension $n>2$ (and
even in dimension $n=2$ if one restricts to connected spacetimes). What significance can be read into this special case is currently unclear. Dynamical locality is known to hold for the nonminimally coupled scalar field at any value of the mass~\cite{Ferg_in_prep}, and work on other models, including the algebra of Wick products is under way.

\section{Categories of spacetimes}\label{sect:spacetimes}

We begin by defining the categories of spacetimes that will be used
as the arena for locally covariant theories. This serves to fix our notation and
terminology; while much of this material is fairly standard,  
our study of the connectedness properties of the categories with respect to wedges
gives a new viewpoint on classical results on deformations of globally
hyperbolic spacetimes~\cite{FullingNarcowichWald}. Some of the details 
are deferred to Appendix~\ref{appx:geom}, which also contains 
a number of useful results on causal structure. 

\subsection{Globally hyperbolic spacetimes}

A {\em globally hyperbolic spacetime} of dimension $n$ is a quadruple
$(\Mc,\gb,\ogth,\tgth)$  such that
\begin{itemize}\addtolength{\itemsep}{-0.5\baselineskip}
\item $\Mc$ is a smooth paracompact orientable nonempty 
$n$-manifold with finitely many connected components
\item $\gb$ is a smooth time-orientable metric of signature $+-\cdots-$
on $\Mc$
\item $\ogth$ is a choice of orientation, i.e., one of the
connected components of the set of nowhere-zero smooth $n$-form
fields on $\Mc$
\item $\tgth$ is a choice of time-orientation for $\gb$, i.e., one of the
connected components of the set of nowhere-zero smooth
$\gb$-timelike $1$-form fields on $\Mc$
\end{itemize}
and such that the resulting causal structure is globally hyperbolic,
i.e., there are no closed causal curves and the intersection of the 
causal past and future of any pair of points is
compact.\footnote{This appears weaker than the definition
given, e.g., in~\cite{HawkingEllis}, but the two are equivalent by
Thm.~3.2 in~\cite{Bernal:2006xf}.} For global hyperbolicity, it is sufficient that $\Mc$ contains
a Cauchy surface \cite[Cor.~14.39]{ONeill}, that is, a subset met exactly once by every inextendible timelike curve in the spacetime.\footnote{A slightly stronger definition is employed
in~\cite{HawkingEllis}, where a  Cauchy surface is defined to be an edgeless
acausal set intersected (exactly once) by every inextendible causal curve. This
equates to an acausal Cauchy surface in our terminology.} A Cauchy surface is necessarily a closed achronal topological hypersurface met (at least once) by every inextendible causal
curve~\cite[Lem.~14.29]{ONeill}. All Cauchy surfaces of a given globally hyperbolic spacetime $\Mb$ are homeomorphic~\cite[Cor.~14.27]{ONeill}. Further, $\Mb$
admits smooth spacelike Cauchy surfaces~\cite[Thm~1.1]{Bernal:2003jb}; given any such surface $\Sigma$,  it is possible to construct a diffeomorphism $\rho:\RR\times\Sigma\to \Mc$ with the following properties (see \cite[Thm 1.2]{Bernal:2005qf} and \cite[Thm 2.4]{Bernal:2004gm}):
\begin{itemize}\addtolength{\itemsep}{-0.5\baselineskip}
\item $\rho_0(\cdot)=\rho(0,\cdot)$ is the inclusion $\Sigma\hookrightarrow \Mb$
\item for each $t\in\RR$, $\rho(\{t\}\times\Sigma)$ is a smooth spacelike Cauchy surface 
\item $\rho_*\partial/\partial t$ is future-directed
\item the pulled back metric splits in the form $\rho^*\gb = \beta dt\otimes dt - \hb_t$ where $\beta\in C^\infty(\RR\times\Sigma)$ is positive, and $t\mapsto \hb_t$ is a smooth map into the smooth Riemannian metrics on $\Sigma$.
\end{itemize}
The Cauchy surface $\Sigma$ has a unique orientation $\wgth$ such that $\ogth=\tgth\wedge\wgth$
(extending the wedge product to equivalence classes of forms in an obvious way) and we will
regard this as the canonical orientation on $\Sigma$. Equipping $\RR\times\Sigma$ with the orientation
corresponding to $dt\wedge\wgth$, the diffeomorphism $\rho$ is promoted to an orientation-preserving
diffeomorphism (abbreviated as oriented-diffeomorphism); this preserves time-orientations on declaring $\partial/\partial t$ to be future-pointing. We refer to the result of the above construction as the {\em normal form} for globally hyperbolic spacetimes. 

There are, of course, many globally hyperbolic spacetimes. 
\begin{Prop}
Every smooth, paracompact oriented $(n-1)$-manifold that is connected (resp., has
finitely many connected components) is oriented-diffeomorphic to a smooth spacelike Cauchy surface of a spacetime in $\Man$ (resp., $\Mand$).  
\end{Prop}
{\noindent\em Proof:}  In the connected case, suppose an $(n-1)$-manifold $\Sigma$ is given 
with orientation defined by a nonvanishing $(n-1)$-form $\omega$. Equip $\Sigma$ with a complete Riemannian metric $\hb$ \cite{NomizuHideki1961} and endow $\RR\times\Sigma$ with metric
$dt\otimes dt-\hb$, orientation $dt\wedge\omega$ and time-orientation $\partial/\partial t$. 
Then the resulting structure is globally hyperbolic with each $\{t\}\times \Sigma$ as a Cauchy surface \cite[Prop. 5.2]{Kay1978} that is oriented-diffeomorphic to $\Sigma$ with orientation $\omega$. In the disconnected case, we perform this construction on each connected component and form the union. $\square$

\subsection{The categories $\Mand$ and $\Man$}

The globally hyperbolic spacetimes (of dimension $n$) form the
objects of a category $\Mand$. By definition, a morphism $\psi$ in $\Mand$ between
$\Mb=(\Mc,\gb,\ogth,\tgth)$ and $\Mb'=(\Mc',\gb',\ogth',\tgth')$ is a smooth embedding (also denoted
$\psi$) of $\Mc$ in $\Mc'$ whose image is causally convex in $\Mb'$ and
such that $\psi^*\gb'=\gb$, $\psi^*\ogth'=\ogth$ and $\psi^*\tgth'=\tgth$. 
Thus the embedding is isometric and respects orientation and
time-orientation. In particular, any diffeomorphism putting a globally hyperbolic spacetime
into normal form is itself an isomorphism in $\Mand$. 

Causal convexity of the image entails that every smooth causal curve 
with ends contained in the image is contained entirely in it. In particular,  if $O_1$ and $O_2$ are any distinct connected components
of the image of $\Mb$ there can be no causal curve joining a point of $O_1$ to a point of $O_2$:
i.e., $O_1\subset O_2^\perp:= \Nb\setminus J_\Nb(O_2)$ and likewise $O_2\subset O_1^\perp$. 
In fact, as the $O_i$ and hence $J_\Nb(O_i)$ are necessarily open (see, e.g., Lem.~\ref{lem:Jofopenisopen}), we have the slightly stronger condition
$O_1\subset O_2':=\Nb\setminus\cl J_\Nb(O_2)$ and $O_2\subset O_1'$. 
It is possible, nonetheless, that the closures of $O_1$ and $O_2$ can intersect nontrivially. 
Note that we have introduced two distinct notions of causal complement, both of which will be
needed in what follows. Some relations between these two definitions and their various properties are discussed in Appendix~\ref{appx:geom}, in which standard definitions of causal structure 
(such as the set $J_\Nb(O)$ just used) are also recalled -- see Appendix~\ref{sect:cov_hyp}.

We will also study the full subcategory of $\Mand$ with connected spacetimes as objects,
which will be denoted $\Man$. Each connected component of an $\Mand$ object $\Mb$ is 
an $\Man$ object; we denote the set of components of $\Mb\in\Mand$ by $\Cpts(\Mb)$. 
Each $\Mand$ morphism $\psi:\Mb\to\Nb$ comprises one or more $\Man$ morphisms: to
each component $\Bb\in\Cpts(\Mb)$ there is a unique component $\Cb\in\Cpts(\Nb)$ containing the image $\psi(\Bb)$ of $\Bb$, and the restriction of $\psi$ to $\Bb$ yields a $\Man$-morphism $\psi_\Bb^\Cb:\Bb\to\Cb$. Conversely, any collection of $\Man$-morphisms $(\psi_\Bb^{f(\Bb)})_{\Bb\in\Cpts(\Mb)}$ where $f:\Cpts(\Mb)\to\Cpts(\Nb)$ and $\psi_\Bb^{f(\Bb)}:\Bb\to f(\Bb)$ defines a $\Mand$ morphism, provided that their images are all causally disjoint in the above sense. It is not required that
every component of $\Nb$ should contain the image of one or more components of $\Mb$.

Two particular classes of $\Mand$ and $\Man$ morphisms will be used extensively in what follows: {\em canonical inclusions} and {\em Cauchy morphisms}. Inclusions arise as follows. 
For any $\Mb$ in $\Mand$ (and hence $\Man$) let $\OO(\Mb)$ be the set of 
open globally hyperbolic subsets\footnote{See~\cite[Def.~14.20]{ONeill}. Note that the open globally hyperbolic subsets
of a globally hyperbolic spacetime are precisely the open causally convex
subsets.} of $\Mb$ with at most finitely many connected components all of which are mutually causally disjoint, and let $\OO_0(\Mb)$ be the set of connected open globally hyperbolic subsets of $\Mb$. 
For each $\Mb=(\Mc,\gb,\ogth,\tgth)\in\Mand$, any nonempty $O\in \OO(\Mb)$ induces an object 
$\Mb|_O=(O,\gb|_O,\ogth|_O,\tgth|_O)$ of $\Mand$, which
we call the {\em restriction} of $\Mb$ to $O$, and the subset inclusion of $O$ in $\Mb$
induces a $\Mand$-morphism $\iota_{\Mb;O}:\Mb|_O\to\Mb$ that we call a canonical inclusion. 
Any morphism $\Lb\stackrel{\psi}{\to}\Mb$ induces a canonical
isomorphism $\tilde{\psi}:\Lb\stackrel{\cong}{\to}\Mb|_{\psi(\Lb)}$ so that
$\psi=\iota_{\Mb;\psi(\Lb)}\circ \tilde{\psi}$. If $O\in\OO_0(\Mb)$ for $\Mb\in\Man$ then
$\iota_{\Mb;O}$ is also a $\Man$-morphism, provided $O$ is nonempty.

A morphism $\psi:\Mb\to\Nb$ will be described as a {\em Cauchy morphism}, or simply as {\em Cauchy} if its image contains a Cauchy surface for $\Nb$. All identity morphisms in $\Man$ and $\Mand$ are Cauchy and compositions of Cauchy morphisms
are Cauchy (Lem.~\ref{lem:Cauchy_comp} in Appendix~\ref{appx:geom}), so the
globally hyperbolic spacetimes with  Cauchy morphisms define subcategories of $\Man$ and $\Mand$.  
As there are slightly different definitions of Cauchy surface 
in the literature, of which we have adopted the weakest, the following 
observation is worth recording (see Appendix~\ref{appx:geom} for the proof).
\begin{Prop}\label{prop:Cauchy}
If $\psi:\Mb\to\Nb$ is Cauchy (in $\Man$ or $\Mand$) then $\psi(\Mb)$ contains a Cauchy surface of $\Nb$ 
that is smooth, spacelike and acausal. Moreover, the Cauchy surfaces of $\Mb$ and $\Nb$ are homeomorphic and their smooth spacelike Cauchy surfaces are oriented-diffeomorphic.
\end{Prop}

A key fact for our purposes is that morphisms in $\Man$ whose domain has compact Cauchy surfaces are 
always Cauchy. The following is an immediate consequence of Prop.~\ref{prop:embeddings_appx} in Appendix~\ref{appx:geom} together with Prop.~\ref{prop:Cauchy}.
\begin{Prop}\label{prop:embeddings} 
(a) Suppose $\psi:\Mb\to\Nb$ in $\Man$, where $\Mb$ has
compact Cauchy surfaces. Then $\psi$ is Cauchy and the smooth spacelike Cauchy surfaces of $\Nb$ are oriented-diffeomorphic to those of $\Mb$. (b) Suppose $\psi:\Mb\to\Nb$ in $\Mand$
and suppose $\Bb\in\Cpts(\Mb)$ has compact Cauchy surfaces. If $\Cb$ is the component of $\Nb$
containing $\psi(\Bb)$, then $\psi_\Bb^\Cb$ is Cauchy and $\Cb$ has smooth spacelike Cauchy surfaces oriented-diffeomorphic to those of $\Bb$. Moreover, $\Cb$ cannot contain the image of any 
component of $\Mb$ other than $\Bb$ (because $\psi(\Bb)$ has trivial causal complement in $\Cb$). 
\end{Prop}

\subsection{Deformation arguments and ``wedge connectedness''}

Globally hyperbolic spacetimes with oriented-diffeomorphic Cauchy surfaces can be deformed into one another, a result going back to~\cite{FullingNarcowichWald} (although the emphasis on orientation here is new). In the present language, this can be stated as follows: 
\begin{Prop} \label{prop:Cauchy_chain}
Two spacetimes $\Mb$, $\Nb$ in $\Man$ (resp., $\Mand$) have oriented-diffeomorphic Cauchy surfaces
if and only if there exists a chain of Cauchy morphisms in $\Man$ (resp., $\Mand$) forming a diagram
\begin{equation} \label{eq:Cauchy_chain}
\Mb\leftarrow \Fb \rightarrow \Ib \leftarrow \Pb \rightarrow \Nb.
\end{equation}
\end{Prop}
{\noindent\em Proof:} If such a chain of Cauchy morphisms exists, then $\Mb$ and $\Nb$ have oriented-diffeomorphic smooth spacelike Cauchy surfaces by Prop.~\ref{prop:Cauchy}. The converse is an
elaboration of \cite[Appx~C]{FullingNarcowichWald} and is given for completeness in Appendix~\ref{appx:geom}. 
 $\square$. 

The chain of morphisms here is far from unique. We will find it useful to regard this result
in the following manner. In a general category, a {\em wedge} is any pair of
morphisms with common domain, i.e., a diagram of form $B\stackrel{f}{\leftarrow} A\stackrel{g}{\rightarrow} C$. Proposition~\ref{prop:Cauchy_chain} then asserts that spacetimes 
with oriented-diffeomorphic Cauchy surfaces are connected by two {\em Cauchy wedges}, i.e., 
wedges consisting of Cauchy morphisms. This shows that $\Mand$ decomposes into ``Cauchy-wedge-connected'' components labelled by equivalence classes of 
Cauchy surfaces modulo oriented-diffeomorphisms; the same is true for $\Man$ on restriction
to connected $(n-1$)-manifolds. We remark in passing that some oriented $(n-1)$-manifolds belong
to the same equivalence class as their orientation reverse (e.g., $\RR^{n-1}$, $S^{n-1}$) while
others (e.g., the three-dimensional lens space $L_5(1,1)$) are
not, and are sometimes called {\em chiral} (see, e.g., \cite{Muellner2009}). Thus any two spacetimes with Cauchy surfaces diffeomorphic to $\RR^{n-1}$ (with the standard differential structure) are linked by a chain of Cauchy morphisms, but
spacetimes with inequivalently oriented chiral Cauchy surfaces belong to different Cauchy components of $\Mand$. 

We also have another connectedness result, this time for the general class of wedges. 
To this end, we first introduce the particularly useful class of {\em
diamond} subsets of a globally hyperbolic spacetime following Brunetti
and Ruzzi \cite{BrunettiRuzzi_topsect}. We will also consider {\em multi-diamonds}, 
that is, unions of finitely many causally disjoint diamonds. 
\begin{Def}
Let $\Mb$ be a spacetime in $\Mand$. A {\em Cauchy ball} in a Cauchy surface $\Sigma$ of $\Mb$ is a subset $B\subset\Sigma$ for which there is a chart $(U,\phi)$ of $\Sigma$ such that $\phi(B)$ a nonempty open ball in $\RR^{n-1}$ whose closure is contained in $\phi(U)$.
A {\em diamond} in $\Mb$ is any open relatively compact subset of the form $D_\Mb(B)$
where $B$ is a Cauchy ball in some Cauchy surface $\Sigma$. We say that the diamond has base $B$ and that it is based
on the Cauchy surface $\Sigma$.\footnote{Neither the base $B$ nor the Cauchy surface $\Sigma$ 
are uniquely associated with the diamond.}
A {\em multi-diamond} is a union of finitely many causally disjoint diamonds, and
therefore takes the form $D_\Mb(B)$ where $B$ is a {\em Cauchy multi-ball}, i.e., a union of finitely many
causally disjoint Cauchy balls. 
\end{Def}
Using Theorem~4.1 and Remark~4.14 in \cite{Bernal:2005qf}, for any Cauchy multi-ball $B$ there is
a (nonunique) Cauchy surface in which it is contained. This observation allows us to extend
the properties of diamonds established in \cite{BrunettiRuzzi_topsect} to show that, 
in spacetime dimension $n\ge 3$, any (multi)-diamond is (among other properties) open, 
relatively compact, simply connected, and has a nonempty causal complement
$O'=\Mb\setminus \cl(J_\Mb(O))$, whose intersection with any connected component of $\Mb$ is
itself connected. Diamonds are connected. A number of further properties of(multi-)diamonds 
are given in Appendix~\ref{appx:geom}. In particular, in Lemma~\ref{lem:causal_completeness} we
demonstrate for completeness that any (multi-)diamond is causally complete in
the sense that $O=O''$. 

In what follows, we will say that a spacetime $\Db$ is a (multi-)diamond if it is isomorphic to a restriction
$\Mb|_O$, where $O$ is a (multi-)diamond of some $\Mb$ in $\Mand$ or $\Man$. A {\em truncated
(multi-)diamond} will refer to any intersection of a (multi-)diamond with an open globally hyperbolic neighbourhood of a Cauchy surface on which it is based.

\begin{Prop} \label{prop:chain}
If $\Mb$ and $\Nb$ are any globally hyperbolic spacetimes in $\Mand$ (resp., $\Man$)
then there exists a chain of (not necessarily Cauchy) morphisms in $\Mand$  (resp., $\Man$)
creating a diagram of the form \eqref{eq:Cauchy_chain}. 
\end{Prop}
{\noindent\em Proof:} Let $O_1$ and $O_2$ be diamond regions in $\Mb$ and $\Nb$
respectively. The Cauchy surfaces of $\Mb|_{O_1}$ and $\Nb|_{O_2}$ 
are oriented-diffeomorphic (they are homeomorphic to $\RR^{n-1}$), so we may apply Prop.~\ref{prop:Cauchy_chain} to obtain
a chain of Cauchy morphisms $\Mb|_{O_1}\leftarrow \Fb \rightarrow \Ib \leftarrow \Pb \rightarrow \Nb|_{O_2}$
and we compose at the two ends with $\iota_{\Mb;O_1}$ and $\iota_{\Nb;O_2}$ to obtain the required
result.  $\square$

\section{Locally covariant theories}\label{sect:LOCCOV}

\subsection{Categories of physical systems}

The focus of BFV was on quantum field theories, described in terms of algebras of
observables and suitable state spaces. Here, we wish take a more general approach in order to encompass a broader range of physical theories.

Suppose a certain type of physical system is to be 
formulated in a locally covariant way on globally hyperbolic spacetimes.
We suppose that the physical systems
concerned can be represented mathematically by objects of a category
$\Phys$, whose morphisms correspond to embeddings of
one such system in another. 

The general conditions imposed on $\Phys$ will be that all its
morphisms are monic, that it has equalisers, intersections and unions [in the categorical sense, which do
not necessarily coincide with the set-theoretic notions; the relevant
definitions are given in Appendix~\ref{appx:subob}], and
that it possesses an initial object, denoted $\Ic$ and representing the trivial physical system of the given type, which is uniquely embedded in every system $\Ac$
via a morphism denoted $\Ic_\Ac$ (we have $\alpha\circ\cI_\Ac=\Ic_\Bc$
for every $\alpha:\Ac\to\Bc$). As general references on category theory, see~\cite{Maclane,AdamekHerrlichStrecker}; our discussion of subobjects and their intersections and unions follows~\cite{DikranjanTholen}.

Possible candidates for the category $\Phys$ abound. The BFV setting 
corresponds to categories such as:
(a) the category $\Alg$ of unital $*$-algebras with unit preserving 
faithful $*$-homomorphisms; (b) the category $\CAlg$ of unital $C^*$-algebras with unit preserving  faithful $*$-homomorphisms; (c) the category $\TAlg$ of unital topological $*$-algebras with
continuous unit preserving faithful $*$-homomorphisms as morphisms; in each 
case the initial object $\Ic$ is the complex number field $\CC$ with $1$ as the unit, complex conjugation as the $*$-operation and  additional topological structure as appropriate to the category concerned.  Elsewhere, we will discuss a category $\Sys$, whose objects are $*$-algebras or $C^*$-algebras together with a suitable subset of the states thereon. More widely, our discussion could also
be applied to classical mechanical or field systems -- the use of a general category $\Phys$ emphasises these possibilities. As a classical example, $\Phys$  
could be the category of presymplectic vector spaces with injective symplectic linear
maps as morphisms, and the trivial symplectic space as the initial object. 

The categorical notions mentioned above can be illustrated easily in $\CAlg$: the equalizer of
$\alpha,\beta:\Ac\to\Bc$ can be described as the inclusion map in $\Ac$ of the maximal $C^*$-subalgebra of $\Ac$ on which $\alpha$ and $\beta$ agree; given a family $(\alpha_i)_{i\in I}$ of morphisms $\alpha_i:\Ac_i\to\Bc$, their
intersection $\bigwedge_{i\in I}\alpha_i$ is the inclusion map of the set-theoretic intersection $\bigcap_{i\in I} \alpha_i(\Ac_i)$ in $\Bc$, while the union
$\bigvee_{i\in I}\alpha_i$ is the inclusion of the $C^*$-subalgebra of $\Bc$ generated by the $\alpha_i(\Ac_i)$ [i.e., the intersection of all $C^*$-subalgebras containing the set-theoretic union].

\subsection{The category of locally covariant theories}\label{sect:LCT}

Once the category $\Phys$ has been selected, we may follow the line of BFV
and define a locally covariant physical theory of the given type to be any
(covariant) functor $\Af$ from $\Mand$ to $\Phys$ (equally we may use $\Man$ as the
domain category if we wish to restrict to connected spacetimes). Thus,
to each spacetime $\Mb\in\Mand$
there is an object $\Af(\Mb)$ of $\Phys$ and to each
$\Mand$-morphism $\psi:\Mb\to\Nb$ there is an $\Phys$-morphism
$\Af(\psi):\Af(\Mb)\to\Af(\Nb)$ such that $\Af(\psi\circ\psi')
=\Af(\psi)\circ\Af(\psi')$ for arbitrary compositions of morphisms
and $\Af(\id_\Mb)=\id_{\Af(\Mb)}$ for all $\Mb$. For BFV, where
$\Phys$ is a suitable category of $*$-algebras, the
$\Af(\Mb)$ is the algebra of observables or of smeared fields describing the theory
in spacetime $\Mb$. 

There is always at least one theory, namely the trivial theory $\If$
with $\If(\Mb)=\Ic$, $\If(\psi)=\id_\Ic$, where $\Ic$ is the initial
object of $\Phys$. As shown in BFV, the standard example of the
Klein--Gordon field provides another example (with $\Phys$ 
chosen as a category of $*$- or $C^*$-algebras according to the
quantization method); the same is true of the extended algebra of
Wick products~\cite{Ho&Wa01} (refined, to remove the dependence
on choice of Hadamard function, as in \cite[\S 5.5.3]{BruFre_LNP:2009}) 
and (passing to the category of globally hyperbolic spacetimes with spin structure) the Dirac field~\cite{Sanders_dirac:2010}
and its corresponding extended algebra~\cite{DHP_dirac:2009}. (Strictly, 
these examples were discussed in the context of functors from $\Man$ to $\Alg$ or $\CAlg$, but 
they generalise to $\Mand$.)

The functorial nature of a theory $\Af$ ensures that it respects local
general covariance, as we will see in Sect.~\ref{sect:kinematic}. In practice various other properties would normally
be expected of the theory. Here, the most important will be the {\em
time-slice property} which requires that $\Af$ maps Cauchy morphisms of $\Mand$ to
isomorphisms in $\Alg$.\footnote{In BFV, the timeslice property was
phrased in terms of surjectivity of $\Af(\psi)$ -- an equivalent
formulation in the category of $C^*$-algebras. In general, however, what
is needed is the invertibility of $\Af(\psi)$ (in $\Phys$) when $\psi$ is Cauchy.}
The time-slice property essentially asserts the existence of a dynamical law for
the theory and will hold in this form for many different physical theories.

For our purposes, it will be important to regard locally covariant theories as objects within
the functor category $\LCT=\Funct(\Mand,\Phys)$ (or, $\LCTo=\Funct(\Man,\Phys)$) 
in which the morphisms are natural transformations $\zeta:\Af\nto\Bf$.
Thus, to each $\Mb\in\Mand$ there is a morphism
$\zeta_\Mb:\Af(\Mb)\to\Bf(\Mb)$ such that $\Bf(\psi)\circ\zeta_\Mb =
\zeta_\Nb\circ\Af(\psi)$ for all morphisms $\psi:\Mb\to\Nb$. 
The physical interpretation of a morphism $\zeta:\Af\nto\Bf$ is that it
provides a way of embedding the theory $\Af$ as a subtheory of $\Bf$. In
the special case where every component $\zeta_\Mb$ is an isomorphism,
$\zeta$ is said to be a natural isomorphism; we interpret this as
indicating that the theories are equivalent. 

Simple examples of morphisms in $\LCT$ may be constructed as follows.
First, the trivial theory $\If$ is a subtheory of every theory $\Af$, and indeed is 
an initial object for $\LCT$ because there is a unique natural $\If_\Af:\If\nto\Af$, 
whose typical component is $(\If_\Af)_\Mb=\Ic_{\Af(\Mb)}$, the unique morphism $\Ic\to\Af(\Mb)$.

Second, given an endofunctor $\Ff$ of $\Phys$ and a natural
$\eta:\Ff\nto\id_\Phys$, any $\Af\in\LCT$ has a subtheory
$\zeta:\Ff\circ\Af\nto\Af$. In the case $\Phys=\TAlg$, an example is given
as follows: to each object $\Ac$, let $\Ff(\Ac)$ be the same $*$-algebra but
equipped with the discrete topology, and let $\eta_\Ac:\Ff(\Ac)\to\Ac$
have the identity as its underlying $*$-homomorphism; then
$\Ff$ can be defined on morphisms in the obvious way so as to ensure
naturality of $\eta: \Ff\nto\id_\TAlg$.

Third, if $\Phys$ is a monoidal category (see, e.g., \cite{Maclane}), with the initial object as the unit, it induces  a
monoidal structure on $\LCT$: given $\Af,\Bf\in\LCT$, define $(\Af\otimes\Bf)(\Mb)
=\Af(\Mb)\otimes\Bf(\Mb)$ and
$(\Af\otimes\Bf)(\psi)=\Af(\psi)\otimes\Bf(\psi)$; this is easily
checked to define a new functor $\Af\otimes\Bf\in\LCT$. The theory $\If$
is the unit for the tensor product in $\LCT$ and the associators and unitors
all lift immediately. For example, recall that the right unitor $\rho$ of a monoidal category
$\Phys$ is a natural isomorphism with components $\rho_\Ac:\Ac\otimes\Ic\cong \Ac$ (obeying certain properties).  This lifts to a natural $\hat{\rho}$,  with components $\hat{\rho}_\Af:\Af\otimes\If\cong\Af$, where $(\hat{\rho}_\Af)_\Mb=\rho_{\Af(\Mb)}$ and which functions as the right unitor in $\LCT$. 
One may check that all the coherence properties required of a monoidal structure lift in this way. 
Writing $\hat{\lambda}$ for the left-unitor in $\LCT$, we obtain $\LCT$ morphisms $\eta_{\Af,\Bf}:\Af\nto\Af\otimes\Bf$ and
$\zeta_{\Af,\Bf}:\Bf\nto\Af\otimes\Bf$ for any pair of theories
$\Af,\Bf\in\LCT$, given by
\[
\eta_{\Af,\Bf} = (\id_\Af\otimes
\If_\Bf)\circ\hat{\rho}^{-1}_\Af
\qquad
\zeta_{\Af,\Bf}=  (\If_\Af\otimes\id_\Bf)\circ\hat{\lambda}^{-1}_\Bf.
\]
Given these structures we can define arbitrary monoidal powers
of a given theory $\Af\in\LCT$, by setting, for example,
$\Af^{\otimes 1}:=\Af$ and $\Af^{\otimes (k+1)}:= 
\Af^{\otimes k}\otimes\Af$ for each $k\in\NN$.\footnote{
Thus $\Af^{\otimes k} = ((\cdots ((\Af\otimes\Af)\otimes\Af)\otimes\cdots)\otimes \Af)\otimes\Af$.
In a monoidal category in which associators are not necessarily identities, there would
be other possible (isomorphic) definitions of the monoidal powers by different placement of
brackets.} Then $\gamma(k):=\eta_{\Af^{\otimes k},\Af}$ provides a natural transformation
$\gamma(k):\Af^{\otimes
k}\nto \Af^{\otimes (k+1)}$; and if $k<k'$ are any natural numbers we may set
\begin{equation}\label{eq:beta_comp}
\beta(k,k') = \gamma(k'-1)\circ \cdots\circ \gamma(k),
\end{equation}
giving a natural transformation $\beta(k,k'):\Af^{\otimes k}\nto\Af^{\otimes k'}$. Defining,
additionally, $\beta(k,k)=\id_{\Af^{\otimes
k}}$ (the identity morphism of $\Af^{\otimes k}$ in $\LCT$), it is
clear that $\beta(k',k'')\circ\beta(k,k')=\beta(k,k'')$ whenever $k\le k'\le k''$
and that $\beta$ defines a functor $\beta:{\sf N}\to\LCT$. Here, ${\sf N}$ is the
category whose morphisms are ordered pairs $(k,k')$ of natural numbers with
$k\le k'$ and composition $(k',k'')\circ (k,k') = (k,k'')$; that is,
${\sf N}$ is the partially ordered set $(\NN,\le)$ regarded as a category.


In the case $\Phys=\Alg$, using the algebraic tensor product, these constructions
reduce to 
\[
\Af^{\otimes k}(\Mb) = \Af(\Mb)^{\otimes k}, \qquad \beta(k,k')_\Mb A=A\otimes \II_{\Af(\Mb)}^{\otimes (k'-k)}  \quad (A\in \Af(\Mb)^{\otimes k})
\]
for $k<k'$; the duals of the $\beta(k,k')_\Mb$ are of course partial traces.

There are many similar ways of constructing functors from ${\sf N}$ to $\LCT$,
of course, but the above will suffice for our purposes and provide useful building blocks
in the sequel.

\subsection{The kinematic net}\label{sect:kinematic}

One of the aims of the BFV paper was to formulate QFT in curved spacetime in such a way that algebraic quantum field theory in Minkowski space could be recovered as a special case. This requires that every suitable subregion of a spacetime $\Mb$ should be associated with a subalgebra of the algebra $\Af(\Mb)$ assigned to $\Mb$ by the theory $\Af$ (for the moment, we take $\Phys=\Alg$ or $\CAlg$). 

For any $\Mb\in\Mand$, recall that $\OO(\Mb)$ is the set of globally hyperbolic open subsets of $\Mb$
with at most finitely many connected components, all of which are mutually causally disjoint, and $\OO_0(\Mb)$ those which are connected. 
For each nonempty $O\in\OO(\Mb)$ (resp., 
$O\in\OO_0(\Mb)$), we have a canonical inclusion
$\iota_{\Mb;O}:\Mb|_O\to\Mb$, an algebra $\Af(\Mb|_O)$ and a morphism $\Af(\iota_{\Mb;O}):\Af(\Mb|_O)\to\Af(\Mb)$. BFV took the image of $\Af(\iota_{\Mb;O})$ as the subalgebra associated with $O$ and showed that this assignment generalises AQFT. To facilitate the discussion of arbitrary categories $\Phys$ it is better to focus attention on the morphism $\Af(\iota_{\Mb;O})$ than its `image' (which is not defined in general categories). 

Accordingly, let $\Phys$ be any category obeying our minimal assumptions and let 
$\Af\in\LCT$ (resp., $\LCTo$). For $\Mb\in\Mand$ (resp., $\Man$) and nonempty $O\in\OO(\Mb)$ (resp., $O\in\OO_0(\Mb)$), we define
\[
\Af^\kin(\Mb;O) = \Af(\Mb|_O), \qquad \text{and}\qquad
\alpha^\kin_{\Mb;O}=\Af(\iota_{\Mb;O}):\Af^\kin(\Mb;O)\to\Af(\Mb).
\]
We refer to the assignment $O\mapsto \alpha^{\kin}_{\Mb;O}$ as the {\em kinematic net}.
Strictly, BFV only considered local algebras corresponding to relatively compact globally hyperbolic
subsets; however it is useful (and natural, in the functorial setting) to extend the assignment of local algebras to regions with noncompact closure. 
Note, however, that the pathologies discussed below are already visible for local algebras of 
relatively compact regions.

The following result shows that the subobject depends only on $\Mb$ and $O$. In the statement
of this result, the $\cong$ symbol between two morphisms with a common codomain asserts their isomorphism as subobjects of the codomain object; i.e., $\alpha\cong\beta$ holds iff there is a (necessarily unique) isomorphism $\gamma$ such that $\alpha=\beta\circ\gamma$; see Appendix~\ref{appx:subob}.
\begin{Lem} \label{lem:ext_M_O}
If $\psi:\Lb\to\Mb$ then $\Af(\psi)\cong\alpha^\kin_{\Mb;\psi(\Lb)}$.
\end{Lem}
{\noindent\em Proof:} We may factor $\psi=\iota_{\Mb;\psi(\Lb)}\circ\tilde{\psi}$ where
$\tilde{\psi}:\Lb\to\Mb|_{\psi(\Lb)}$ is an isomorphism; as functors preserve isomorphisms
we therefore have $\Af(\psi)=\alpha^\kin_{\Mb;\psi(\Lb)}\circ \Af(\tilde{\psi}) \cong
\alpha^\kin_{\Mb;\psi(\Lb)}$. $\square$

The basic properties of $O\mapsto \Af^\kin(\Mb;O)$ are discussed in Prop.~2.3 of BFV [in the case of 
connected $O$]. In particular,  if $O_1\subset O_2$ then $\iota_{\Mb;O_1}$ factorises via 
$\iota_{\Mb;O_2}$ as $\iota_{\Mb;O_1}=\iota_{\Mb;O_2}\circ \iota_{\Mb|_{O_2};O_1}$ and 
the functorial property of $\Af$ implies
\begin{equation}\label{eq:isotony}
\alpha^{\kin}_{\Mb;O_1}=\alpha^{\kin}_{\Mb;O_2}\circ \Af(\iota_{\Mb|_{O_2};O_1}),
\end{equation}
which can also be written in the form $\alpha^{\kin}_{\Mb;O_1}\le \alpha^{\kin}_{\Mb;O_2}$, 
where $\le$ is the order relation in the subobject lattice of $\Af(\Mb)$ (see, e.g., \cite{DikranjanTholen}). That is, the kinematic net is {\em isotonous}. 

If, additionally, $O_1$ contains a Cauchy surface for $O_2$, then the morphism 
$\iota_{\Mb|_{O_2};O_1}$ is Cauchy and is mapped to an isomorphism if $\Af$ obeys the timeslice property. Then the factorisation \eqref{eq:isotony} asserts that $\alpha^{\kin}_{\Mb;O_1}$ and $\alpha^{\kin}_{\Mb;O_2}$ determine isomorphic subobjects of $\Af(\Mb)$: we write
$\alpha^{\kin}_{\Mb;O_1}\cong \alpha^{\kin}_{\Mb;O_2}$. [This is an improved formulation of Prop.~2.3(d) in BFV. Compare also Thm.~\ref{thm:A_int_MO} below.]

Now suppose that $\psi:\Mb\to\Nb$. If $O\in\OO(\Mb)$ (resp., $\OO_0(\Mb)$) is nonempty then $\psi(O)\in\OO(\Nb)$ (resp., $\OO_0(\Nb)$) and there is an isomorphism $\tilde{\psi}:\Mb|_O\to\Nb|_{\psi(O)}$ such that $\psi\circ \iota_{\Mb;O} =
\iota_{\Nb;\psi(O)}\circ \tilde{\psi}$. Applying the functor $\Af$, this gives
a commuting diagram
\[
\begin{tikzpicture}[baseline=0 em,description/.style={fill=white,inner sep=2pt}]
\matrix (m) [ampersand replacement=\&,matrix of math nodes, row sep=3em,
column sep=2.5em, text height=1.5ex, text depth=0.25ex]
{ \Af^\kin(\Mb;O) \&  \Af^\kin(\Nb;\psi(O)) \\
 \Af(\Mb) \&  \Af(\Nb)\\ };
\path[->,font=\scriptsize]
(m-1-1) edge node[left] {$\alpha^\kin_{\Mb;O}$} (m-2-1)
(m-1-2) edge node[auto] {$\alpha^\kin_{\Nb;\psi(O)}$} (m-2-2)
(m-2-1) edge node[auto] {$\Af(\psi)$} (m-2-2);
\path[->,font=\scriptsize]
(m-1-1) edge node[above] {$\Af(\tilde{\psi})$} node[below]  {$\cong$} (m-1-2);
\end{tikzpicture}\label{eq:kinematic_covariance_diagram}
\]
and thus the equivalence of subobjects
\begin{equation}\label{eq:kinematic_covariance}
\alpha^\kin_{\Nb;\psi(O)} \cong \Af(\psi)\circ\alpha^\kin_{\Mb;O},
\end{equation}
which expresses the covariance of the kinematic net. In particular, this gives the
action of automorphisms of $\Mb$ (i.e., a (time-)orientation preserving isometric diffeomorphism) on the kinematic net: the functor $\Af$ provides a representation of the automorphism group $\Aut(\Mb)$ in the
automorphism group of $\Af(\Mb)$ by $\kappa\mapsto\Af(\kappa)$, and the formula
\[
\alpha^\kin_{\Mb;\kappa(O)} \cong \Af(\kappa)\circ\alpha^\kin_{\Mb;O}
\]
shows that this has the expected geometrical action on the kinematic net.

\subsection{Relative Cauchy evolution}\label{sect:rce}

Let $\Mb=(\Mc,\gb,\ogth,\tgth)\in\Mand$ be a globally hyperbolic spacetime. Given any symmetric
$\hb\in\CoinX{T^0_2\Mb}$ such that $\gb+\hb$ is a time-orientable
Lorentz metric on $\Mc$, there is a
unique choice of time-orientation $\tgth_\hb$ for $\gb+\hb$ that agrees with $\tgth$ outside
$K$. If $\Mb[\hb]=(\Mc,\gb+\hb,\ogth,\tgth_\hb)$ is a globally hyperbolic
spacetime, we say that $\hb$ is a {\em globally hyperbolic perturbation}
of $\Mb$ and write $\hb\in H(\Mb)$. The subset of $\hb\in H(\Mb)$ with
support in $K\subset\Mc$ is denoted $H(\Mb;K)$. Clearly, $\Mb=\Mb[\Ob]$, 
where $\Ob$ is identically zero, so $H(\Mb)$ is nonempty; in fact it contains an open neighbourhood of
$\Ob$ in the usual test-function topology on symmetric smooth
compactly supported sections of $T^0_2\Mb$ (see \S 7.1
of~\cite{BeemEhrlichEasley}). 
We endow $H(\Mb)$ with the subspace topology induced from $\Df(T^0_2M)$.

If a theory $\Af\in\LCT$  has the timeslice property then, as shown by BFV, we may compare
the dynamics on $\Mb$ and its perturbations via a {\em relative Cauchy evolution}. We now
describe the construction in more depth than BFV, paying attention to
the covariance properties of the relative Cauchy evolution and the
relation between the evolutions of theories related by morphisms in
$\LCT$. A number of geometrical lemmas, including the following, will be proved in
Appendix~\ref{appx:geom}.

\begin{Lem} \label{lem:geom1}
Let $\Mb\in\Mand$ and $\hb\in H(\Mb)$, and set
$\Mc^\pm=\Mc\setminus J^\mp_\Mb(\supp\hb)$. Then (a) $\Mc^\pm$ are
globally hyperbolic subsets of both $\Mb$ and $\Mb[\hb]$, and
$\Mb^\pm[\hb]\stackrel{{\rm def}}{=}\Mb|_{\Mc^\pm} =\Mb[\hb]|_{\Mc^\pm}$; (b) the canonical inclusions $\imath_\Mb^\pm[\hb]\stackrel{{\rm def}}{=}\iota_{\Mb;\Mc^\pm}$ and 
$\jmath_\Mb^\pm[\hb]\stackrel{{\rm def}}{=}\iota_{\Mb[\hb];\Mc^\pm}$ are Cauchy morphisms.
If $\Mb\in\Man$, then $\Mb[\hb]$, $\Mb^\pm[\hb]$ are also in $\Man$, and the
morphisms $\iota_\Mb[\hb]^\pm$, $\jmath_\Mb^\pm[\hb]$ are $\Man$-morphisms. 
\end{Lem}

Among other things, this result shows that we can work consistently in either $\Mand$ or $\Man$.
For the rest of this section, we will not distinguish between $\Mand$ or $\Man$ in the statement
of our results (with the exception of Prop.~\ref{prop:rce_locality}, where there is a slight difference) but it should be understood that all spacetimes and morphisms should be
taken consistently from one or other of $\Mand$ or $\Man$, and that the locally covariant
theories mentioned are taken consistently from $\LCT$ or $\LCTo$ respectively. (In some cases, the proofs of these statements differ slightly depending on which category is being used.)

Proceeding in this way, if $\Mb$ is a spacetime, each $\hb\in H(\Mb)$ induces a {\em past Cauchy wedge}, i.e., the diagram 
\[
\Mb\stackrel{\imath_\Mb^-[\hb]}{\longleftarrow} \Mb^-[\hb]\stackrel{\jmath_\Mb^-[\hb]}{\longrightarrow} \Mb[\hb]
\]
and a {\em future Cauchy wedge}, namely,
\[
\Mb\stackrel{\imath_\Mb^+[\hb]}{\longleftarrow} \Mb^+[\hb]\stackrel{\jmath_\Mb^+[\hb]}{\longrightarrow} \Mb[\hb].
\]
Any locally covariant theory $\Af$ obeying the timeslice axiom will map each morphism in the past and future Cauchy
wedges to an isomorphism. In particular there are isomorphisms
\[
\tau^\pm_\Mb[\hb]=\Af(\jmath_\Mb^\pm[\hb])
\circ(\Af(\imath_\Mb^\pm[\hb]))^{-1}:\Af(\Mb)\to\Af(\Mb[\hb])
\]
and an automorphism $\rce_\Mb[\hb]$ of $\Af(\Mb)$ given by
\[
\rce_\Mb[\hb] = (\tau^-_\Mb[\hb])^{-1} \circ \tau^+_\Mb[\hb],
\]
which is called the {\em relative Cauchy evolution} induced by $\hb$. 
Not all metric perturbations are physically significant: for example, if
$\Mb=(\Mc,\gb,\ogth,\tgth)$ and $\psi$ is a diffeomorphism of $\Mc$ acting as the identity outside a
compact set $K$, then $\psi$ induces a morphism (also denoted $\psi$) from $\Mb$ to
$\Mb'=(\Mc,\psi_*\gb,\psi_*\ogth,\psi_*\tgth)$ which can be regarded as a globally
hyperbolic perturbation $\Mb'=\Mb[\hb]$ for $\hb=\psi_*\gb-\gb$. It is
easily seen that 
\[
\psi\circ \imath^\pm_\Mb[\hb] = \jmath^\pm_\Mb[\hb]
\]
for both choices of sign; accordingly, we have
$\tau_\Mb^+[\hb]=\Af(\psi)=\tau_\Mb^-[\hb]$ and hence
$\rce_\Mb[\hb]=\id_{\Af(\Mb)}$, which reflects the fact that $\Mb$ and
$\Mb'$ are physically equivalent and have equivalent dynamics. 
 
The definition of relative Cauchy evolution given here differs slightly from that given in
BFV, where the Cauchy morphisms used were not fixed by the
perturbation; our approach avoids the necessity of demonstrating that
the definition does not depend on the choices made by introducing the
preferred past and future Cauchy wedges. In order to make contact with
the original definition, however, we give the following result, which is
also useful for computations (and, in passing, establishes the
independence mentioned above).
\begin{Prop} \label{prop:BFV_contact}
Let $K$ be a compact subset of $\Mc$ and suppose 
$\psi^\pm:\Lb^\pm\to\Mb$ are Cauchy morphisms with image contained
in $\Mc\setminus J^\mp_\Mb(K)$. For each $\hb\in H(\Mb;K)$ there are
morphisms $\psi^\pm[\hb]:\Lb^\pm\to\Mb[\hb]$ with the same underlying
embedding as $\psi^\pm$ such that 
\[
\tau_\Mb^\pm[\hb] = \Af(\psi^\pm[\hb])\circ \Af(\psi^\pm)^{-1}
\]
and hence
\[
\rce_\Mb[\hb] = \Af(\psi^-)\circ \Af(\psi[\hb]^-)^{-1} \circ \Af(\psi^+[\hb])\circ
\Af(\psi^+)^{-1}.
\]
\end{Prop}
{\noindent\em Proof:} The following lemma will be proved in Appendix~\ref{appx:geom}.
\begin{Lem} \label{lem:geom2}
Let $K$ be a compact subset of $\Mc$ and suppose $\psi:\Lb\to\Mb$ has
its range contained in one or both of $\Mc\setminus J^\mp_\Mb(K)$. Then the underlying
embedding of $\psi$ induces $\psi[\hb]:\Lb\to\Mb[\hb]$ for any $\hb\in
H(\Mb;K)$ (independent of the choice $\pm$ in the hypothesis). Moreover,
there is $\varphi^\pm:\Lb\to\Mb^\pm[\hb]$ such that
\[
\psi=\imath_\Mb^\pm[\hb]\circ\varphi^\pm, \qquad \psi[\hb]=\jmath_\Mb^\pm[\hb]\circ\varphi^\pm.
\]
If $\psi$ is Cauchy then so are $\psi[\hb]$ and $\varphi^\pm$. 
\end{Lem}
The immediate consequence is that 
\[
\tau_\Mb^\pm[\hb]\circ\Af(\psi) =
\tau_\Mb^\pm[\hb]\circ\Af(\imath_\Mb^\pm[\hb])\circ\Af(\varphi^\pm)
=\Af(\jmath_\Mb^\pm[\hb])\circ\Af(\varphi^\pm)
=\Af(\psi[\hb]).
\]
Applying the $+$ (resp., $-$) case to the $\psi^+$ (resp., $\psi^-$) in the
hypothesis, Prop.~\ref{prop:BFV_contact} follows. $\square$

Much of the present paper depends crucially on locality and
covariance properties of the relative Cauchy evolution that were not
addressed in BFV. Locality can be obtained from Lem.~\ref{lem:geom2}.
\begin{Prop} \label{prop:rce_locality}
Let $K$ be a compact subset of $\Mc$ and suppose $\psi:\Lb\to\Mb$ has
its range contained in the causal complement $K^\perp=\Mc\setminus
J_\Mb(K)$ of $K$ (hence, in particular, if $K\subset \psi(\Lb)'$). Then
\[
\rce_\Mb[\hb]\circ\Af(\psi) = \Af(\psi)
\]
for all $\hb\in H(\Mb;K)$. In particular, this implies that the kinematical net obeys
\[
\rce_\Mb[\hb]\circ\alpha^\kin_{\Mb;O}=\alpha^\kin_{\Mb;O}
\]
for all nonempty $O\in\OO(\Mb)$ (or $\OO_0(\Mb)$ for theories in $\LCTo$) with $O\subset(\supp \hb)^\perp$.
\end{Prop}
{\noindent\em Proof:} The morphism $\psi$ obeys the hypothesis of Lem.~\ref{lem:geom2}
in both the $+$ and $-$ cases. Accordingly
\[
\tau_\Mb^+[\hb]\circ\Af(\psi) = \Af(\psi[\hb]) = \tau_\Mb^-[\hb]\circ\Af(\psi)
\]
and the result follows on composing with $\tau_\Mb^-[\hb]^{-1}$. $\square$

We remark that the hypotheses of this result allow for nontrivial intersection of 
$\cl(\psi(\Lb))$ and $J_\Mb(K)$.

Our covariance result depends on the following geometrical lemma.
\begin{Lem}\label{lem:geom3}
For each morphism $\psi:\Mb\to\Nb$, we have $\psi_* H(\Mb)\subset
H(\Nb)$. Moreover, for each $\hb\in H(\Mb)$ and $\psi:\Mb\to\Nb$ there
are morphisms $\psi^\pm[\hb]:\Mb^\pm[\hb]\to \Nb^\pm[\psi_*\hb]$ and 
$\psi[\hb]:\Mb[\hb]\to \Nb[\psi_*\hb]$ so that the following diagram commutes:
\begin{equation}\label{eq:wedge_morphism}
\begin{tikzpicture}[baseline=0 em, description/.style={fill=white,inner sep=2pt}]
\matrix (m) [ampersand replacement=\&,matrix of math nodes, row sep=3em,
column sep=4em, text height=1.5ex, text depth=0.25ex]
{ \Mb \& \Mb^\pm[\hb] \& \Mb[\hb] \\
\Nb \& \Nb^\pm[\psi_*\hb] \& \Nb[\psi_*\hb]\\ };
\path[->]
(m-1-2) edge node[above] {$ \imath^\pm_\Mb[\hb] $} (m-1-1)
        edge node[above] {$ \jmath^\pm_\Mb[\hb] $} (m-1-3)
(m-2-2) edge node[below] {$ \imath^\pm_\Nb[\psi_*\hb] $} (m-2-1)
        edge node[below] {$ \jmath^\pm_\Nb[\psi_*\hb] $} (m-2-3)
(m-1-1) edge node[auto] {$ \psi $} (m-2-1)
(m-1-2) edge node[auto] {$ \psi^\pm[\hb] $} (m-2-2)
(m-1-3) edge node[auto] {$ \psi[\hb] $} (m-2-3);
\end{tikzpicture}.
\end{equation}
\end{Lem}
{\noindent\em Proof:} The most involved
aspect is to prove that $\psi_*\hb\in H(\Nb)$. This is accomplished by
Lem.~\ref{lem:H_functorial} below. As the horizontal morphisms in
diagram~\eqref{eq:wedge_morphism} are inclusions it is now sufficient to
show that there are morphisms
$\psi^\pm[\hb]:\Mb^\pm[\hb]\to\Nb^\pm[\psi_*\hb]$ and
$\psi[\hb]:\Mb[\hb]\to\Nb[\psi_*\hb]$ with the same underlying embedding
as $\psi$; the diagram will then automatically commute. The existence of
$\psi[\hb]$ is obvious. As the image of $\psi$ is causally convex in
$\Nb$,
$J_\Nb^\mp(\supp\psi_*\hb)\cap\psi(\Mc)=\psi(J_\Mb^\mp(\supp\hb))$ and
hence
$\psi(\Mc\setminus J_\Mb^\mp(\supp\hb))\subset \Nc\setminus 
J_\Nb^\mp(\supp\psi_*\hb))$. Hence the underlying embedding induces
$\psi^\pm[\hb]:\Mb^\pm[\hb]\to \Nb^\pm[\hb]$ as required. 
$\square$ 

This result shows that the sets of hyperbolic perturbations are functorially assigned to
spacetimes of $\Mand$ and $\Man$, and the push-forward induces a mapping between
Cauchy wedges, which could also be interpreted as a morphism in a
suitable category of wedges. We do not pursue this here. 
The main use of the above lemma is to establish covariance of the
relative Cauchy evolution.

\begin{Prop} \label{prop:rce_covariance}
If $\psi:\Mb\to\Nb$ and $\hb\in H(\Mb)$, then 
\begin{equation}\label{eq:tau_covariance}
\tau^\pm_\Nb[\psi_*\hb]\circ \Af(\psi) = \Af(\psi[\hb])\circ
\tau^\pm_\Mb[\hb],
\end{equation}
and consequently
\begin{equation}\label{eq:rce_covariance}
\rce_\Nb[\psi_*\hb]\circ \Af(\psi) = \Af(\psi)\circ
\rce_\Mb[\hb].
\end{equation}
\end{Prop}
{\noindent\em Proof:} Lemma~\ref{lem:geom3} demonstrates that
the $\tau_\Nb^\pm[\psi_*\hb]$ and $\rce_\Nb[\psi_*\hb]$ exist. Taking
the image under $\Af$ of diagram~\eqref{eq:wedge_morphism} and using the definitions of
$\tau_\Mb^\pm[\hb]$ and $\tau_\Nb^\pm[\psi_*\hb]$, 
we obtain the commutative diagrams
\[
\begin{tikzpicture}[baseline=0 em, description/.style={fill=white,inner sep=2pt}]
\matrix (m) [ampersand replacement=\&,matrix of math nodes, row sep=2em,
column sep=3em, text height=1.5ex, text depth=0.25ex]
{ \Af(\Mb) \&                 \& \Af(\Mb[\hb])  \\
           \& \Af(\Mb^\pm[\hb]) \& \\
           \& \Af(\Nb^\pm[\hb]) \& \\
 \Af(\Nb) \&                 \& \Af(\Nb[\hb])\\ };
\path[->]
(m-1-1) edge node[above] {$ \tau^\pm_\Mb[\hb] $} (m-1-3)
        edge node[auto] {$ \Af(\psi) $} (m-4-1)
(m-1-3) edge node[auto] {$ \Af(\psi[\hb]) $} (m-4-3)
(m-4-1) edge node[below] {$ \tau^\pm_\Nb[\psi_*\hb] $} (m-4-3)
(m-2-2) edge node[auto] {$ \Af(\psi^\pm[\hb]) $} (m-3-2)
        edge node[below,sloped] {} (m-1-1)
        edge node[below,sloped] {} (m-1-3)
(m-3-2) edge node[above,sloped] {} (m-4-1)
        edge node[above,sloped] {} (m-4-3);
\end{tikzpicture}
\]
(we suppress labels on the slanted arrows) from which
Eqs.~\eqref{eq:tau_covariance} and \eqref{eq:rce_covariance} follow
immediately. $\square$

So far we have defined relative Cauchy evolution for a single theory
$\Af$ obeying the timeslice property. Where a number of theories are
considered, we will distinguish the relative Cauchy evolution and
related structures by a superscript to indicate the theory concerned.
The relative Cauchy evolution interacts in an elegant way with the
morphisms of $\LCT$ and $\LCTo$:

\begin{Prop}\label{prop:rce_intertwine}
Suppose locally covariant theories $\Af$ and $\Bf$ both satisfy the timeslice property
and let $\zeta:\Af\nto\Bf$. For any spacetime $\Mb$ and metric perturbation $\hb\in H(\Mb)$ we have
\[
\zeta_{\Mb[\hb]}\circ \tau_\Mb^{(\Af)\pm}[\hb] =
\tau_\Mb^{(\Bf)\pm}[\hb]\circ\zeta_{\Mb}
\]
and therefore 
\[
\rce_\Mb^{(\Bf)}[\hb]\circ\zeta_\Mb =
\zeta_\Mb\circ\rce_\Mb^{(\Af)}[\hb].
\]
\end{Prop}
{\noindent\em Proof:} Introducing the past and future Cauchy wedges as
before, we have
\begin{align*}
\tau_\Mb^{(\Bf)\pm}[\hb]\circ\zeta_{\Mb}\circ\Af(\imath_\Mb^\pm[\hb]) &= 
\tau_\Mb^{(\Bf)\pm}[\hb]\circ\Bf(\imath_\Mb^\pm[\hb])\circ\zeta_{\Mb^\pm[\hb]} =
\Bf(\jmath_\Mb^\pm[\hb])\circ\zeta_{\Mb^\pm[\hb]} \\
&= \zeta_{\Mb[\hb]}\circ\Af(\jmath_\Mb^\pm[\hb]) 
= \zeta_{\Mb[\hb]}\circ\tau_\Mb^{(\Af)\pm}[\hb]\circ\Af(\imath_\Mb^\pm[\hb])
\end{align*}
and since $\Af(\imath_\Mb^\pm[\hb])$ is epic, the first result holds. Hence
\begin{align*}
\tau_\Mb^{(\Bf)+}[\hb]\circ\rce^{(\Bf)}_\Mb[\hb] \circ\zeta_{\Mb}
&= \tau_\Mb^{(\Bf)-}[\hb]\circ\zeta_{\Mb} = \zeta_{\Mb[\hb]}\circ
\tau_\Mb^{(\Af)-}[\hb] = \zeta_{\Mb[\hb]}\circ\tau_\Mb^{(\Af)+}[\hb]
\circ\rce_\Mb^{(\Af)}[\hb] \\
&= \tau_\Mb^{(\Bf)+}[\hb]\circ\zeta_{\Mb}\circ\rce_\Mb^{(\Af)}[\hb]
\end{align*}
and as $\tau_\Mb^{(\Bf)+}[\hb]$ is monic the second part follows.
$\square$

An important observation in BFV is that the functional derivative of the relative Cauchy evolution with respect to the metric can be interpreted as a stress-energy tensor of the theory, so that (in the case $\Phys=\Alg$ or $\CAlg$)
\[
[\Tb_\Mb[\fb],A] = 2i \left.\frac{d}{ds}\rce_\Mb[s\fb] A\right|_{s=0},
\]
where $\Tb_\Mb$ is the stress-energy tensor in $\Mb$; the left-hand side should be
regarded as the definition of a (not necessarily inner) derivation, and we suppress
all technicalities regarding the sense in which differentiation is intended. 
Prop.~\ref{prop:rce_intertwine} then has an immediate consequence that
\begin{equation}\label{eq:T_intertwine}
[\Tb^{(\Bf)}_\Mb[\fb],\zeta_\Mb A] =\zeta_\Mb [\Tb^{(\Af)}_\Mb[\fb], A] 
\end{equation}
i.e., a subtheory embedding necessarily intertwines the stress-energy tensors of the two theories. 

As an immediate application, consider the quantum field theory of the nonminimally coupled
scalar field, with field equation $(\Box_\Mb+\xi R_\Mb+m^2)\varphi=0$,
where $R_\Mb$ is the scalar curvature. For each 
value of the coupling $\xi$ and the mass $m$, there is a locally covariant 
theory $\Af^{(m,\xi)}$ so that  each $\Af^{(m,\xi)}(\Mb)$ has
generators $\Phi_\Mb^{(m,\xi)}(f)$ labelled by $f\in\CoinX{\Mb}$ subject
to relations depending only on the field equation and its Green functions (together
with basic structures of linearity and complex conjugation). In  Ricci-flat spacetimes, 
therefore, the map $\Phi^{(m,\xi)}_\Mb(f)\mapsto \Phi^{(m,\xi')}_\Mb(f)$ extends
to an isomorphism $\Af^{(m,\xi)}(\Mb)\to \Af^{(m,\xi')}(\Mb)$ for any $\xi,\xi'$. 
We shall call this the `obvious isomorphism'. Similarly, if $\Mb$ has constant scalar curvature, there is an obvious isomorphism 
$\Af^{(m,\xi)}(\Mb)\cong \Af^{(m',\xi')}(\Mb)$ whenever $m^2+\xi R_\Mb=m'{}^2+\xi'R_\Mb$. 
However, none of these isomorphisms (for distinct values of the labels) can be the components  
of natural transformations between these theories for the simple reason that the
commutators of the stress-energy tensor with the smeared fields  
(which yield further smeared fields) depend nontrivially on the parameters $m$ and $\xi$ {\em even in spacetimes
that have constant or vanishing scalar curvature}. Thus Eq.~\eqref{eq:T_intertwine} 
cannot hold if $\zeta_\Mb$ is one of these `obvious' isomorphisms.

\section{Failure of SPASs in $\LCT$}
\label{sect:pathologies}

The BFV definition of a locally covariant QFT is that it is a functor
$\Af:\Mand\to\Alg$. However, in the absence of further assumptions this
does not fully answer the question of what it means for the theory to
have the same physical content, i.e., to be `the same theory' in
different spacetimes of the same dimension.\footnote{The question of
whether there is a sensible notion of `the same theory' in
spacetimes of {\em different} dimensions is an interesting one, to which we
hope to return elsewhere.} 

A definition of what it means for a single theory to represent the same physics
in all spacetimes (abbreviated SPASs) is not easy to give, and risks the introduction of possibly over-restrictive assumptions on the nature of the theory in question. However it seems reasonable
that if we are given two theories, each of which represents the same physics in all spacetimes 
(by some reasonable definition) and these theories coincide in {\em some} spacetime, then
they should coincide in {\em all} spacetimes. This motivates the following definition, 
in which we refer to a natural transformation between functors as a {\em partial isomorphism} if 
at least one of its components is an isomorphism. 
\begin{Def} 
A class of theories $\frak{T}$ in $\LCT$ (or $\LCTo$) is said to have the {\em SPASs property} if
all partial isomorphisms (in $\LCT$ or $\LCTo$) between theories in $\frak{T}$  are isomorphisms.
\end{Def}
As explained in the introduction, any candidate definition of SPASs can be represented by the class of theories that obey it; the SPASs property can then be used as a necessary criterion on `good' 
notions of SPASs. In this section, we will show by examples that neither $\LCTo$ nor $\LCT$ has the SPASs property except
where the category $\Phys$ is rather trivial; we will use this to demonstrate the existence of individual theories that cannot be regarded as representing the same physics in all spacetimes by any reasonable definition. This will be remedied in Section~\ref{sect:SPASs}, where we will exhibit particular subclasses that do enjoy the SPASs property.

\subsection{Diagonal functors}

In the introduction we gave a simple example of a pathological locally covariant theory with target category $\Alg$. 
As we now show, this example may be placed within a more general setting,
which provides a broader class of pathological theories and enables the
consideration of more general categories for $\Phys$. 

We begin with
a simple categorical construction. Given any two categories 
$\Ct$ and $\Ct'$, the functors between $\Ct$ and $\Ct'$ form the objects
of a category $\Funct(\Ct,\Ct')$ (also written ${\Ct'}^\Ct$ in the
literature) in which morphisms are natural transformations between pairs of
functors. In particular, this applies to the
locally covariant theories, which (in the BFV definition) are precisely the objects
of $\LCTo = \Funct(\Man,\Phys)$. Iterating this construction, we may also
consider functors from $\Ct$ to $\Funct(\Ct,\Ct')$; any such functor then
induces a functor in $\Funct(\Ct,\Ct')$ by the following `diagonal construction'.

\begin{Prop} \label{prop:diagonal_construction}
Given $\varphi\in\Funct(\Ct,\Funct(\Ct,\Ct'))$, define maps
of objects $A\in\Ct$ and morphisms $f\in\Ct(A,B)$ of $\Ct$ to objects
and morphisms of $\Ct'$ by
\begin{align*}
\varphi_\Delta(A) &=  \varphi(A)(A) \\
\varphi_\Delta(f) &=  \varphi(f)_B\circ\varphi(A)(f).
\end{align*}
Then $\varphi_\Delta$ is a functor from $\Ct$ to $\Ct'$; we refer to
$\varphi_\Delta$ as the {\em diagonal} of $\varphi$. Moreover, if $\varphi$, $\varphi'$ are elements of
$\Funct(\Ct,\Funct(\Ct,\Ct'))$ and $\xi:\varphi\nto\varphi'$ is natural, there
is a natural transformation
$\xi_\Delta:\varphi_\Delta\nto\varphi'_\Delta$ with components
$(\xi_\Delta)_A = (\xi_A)_A$. The map $\xi\mapsto \xi_\Delta$ is in
fact a functor from $\Funct(\Ct,\Funct(\Ct,\Ct'))$ to $\Funct(\Ct,\Ct')$.
\end{Prop}
{\noindent\em Remarks:} (1) The expressions above are
well-defined because $\varphi(f):\varphi(A)\nto\varphi(B)$; diagrammatically,
$\varphi_\Delta(f)$ is the diagonal of the naturality square
\[
\begin{tikzpicture}[description/.style={fill=white,inner sep=2pt}]
\matrix (m) [ampersand replacement=\&,matrix of math nodes, row sep=3em,
column sep=4em, text height=1.5ex, text depth=0.25ex]
{ A \& \varphi(A)(A) \&  \varphi(B)(A) \\
B \&   \varphi(A)(B) \&  \varphi(B)(B)\\ };
\path[->]
(m-1-1) edge node[left] {$ f $} (m-2-1)
(m-1-2) edge node[auto] {$ \varphi(f)_A $} (m-1-3)
        edge node[left] {$ \varphi(A)(f) $} (m-2-2)
(m-1-3) edge node[auto] {$ \varphi(B)(f) $} (m-2-3)
(m-2-2) edge node[below] {$ \varphi(f)_B $} (m-2-3)
(m-1-2) edge node[above,sloped] {$ \varphi_\Delta(f) $} (m-2-3);
\end{tikzpicture}
\]
%
and we also have $\varphi_\Delta(f) = \varphi(B)(f)\circ\varphi(f)_A$.\\
(2) Given any functor $F:\Ct\to\Ct'$, let $\varphi$ be the constant
functor $\Ct\to\Funct(\Ct,\Ct')$ taking the value $F$ on all objects.
Then $F=\varphi_\Delta$.\\
{\noindent\em Proof:} As $\varphi$ and $\varphi(A)$ are functors, we have
\[
\varphi_\Delta(\id_A) = \varphi(\id_A)_A\circ\varphi(A)(\id_A)
= (\id_{\varphi(A)})_A\circ\id_{\varphi(A)(A)} = \id_{\varphi(A)(A)} =
\id_{\varphi_\Delta(A)}
\]
for any $A\in\Ct$. Moreover, if $g:B\to C$, we have
\[
\varphi_\Delta(g\circ f) = \varphi(C)(g\circ f)\circ\varphi(g\circ f)_A
=\varphi(C)(g)\circ\underbrace{\varphi(C)(f)\circ\varphi(g)_A}_{=\varphi(g)_B\circ\varphi(B)(f)}
\circ\varphi(f)_A = \varphi_\Delta(g)\circ\varphi_\Delta(f),
\]
in which we have used the naturality of
$\varphi(g):\varphi(B)\nto\varphi(C)$. 

Now suppose that $\xi:\varphi\nto\varphi'$. Noting that $\xi_A:\varphi(A)\to\varphi'(A)$ is
itself a natural transformation, each $(\xi_A)_A$ is a morphism from
$(\xi_A)_A:\varphi(A)(A)\to \varphi'(A)(A)$. Given $f:A\to B$ we compute
\begin{align*}
\varphi'_\Delta(f)\circ (\xi_\Delta)_A &= 
\varphi'(B)(f)\circ\varphi'(f)_A\circ (\xi_A)_A =
\varphi'(B)(f)\circ(\varphi'(f)\circ\xi_A)_A \\
&= 
\varphi'(B)(f)\circ(\xi_B\circ\varphi(f))_A = 
\varphi'(B)(f)\circ(\xi_B)_A\circ \varphi(f)_A \\ 
&= 
(\xi_B)_B\circ\varphi(B)(f)\circ \varphi(f)_A = 
(\xi_\Delta)_B \circ \varphi_\Delta(f),
\end{align*}
thus establishing naturality. (The above computation may be displayed diagrammatically
using a commuting cube). It is simple to check the
functor property and we skip the proof.
$\square$

In our examples, it will be convenient to construct functors from 
the category of spacetimes to the category of locally covariant theories
using a construction of the following type.
\begin{Lem} \label{lem:factor_construction}
Let $\Ct$ and $\Ct'$ be categories and $\It$ be a partially ordered set, which we may
regard as a category [with a single arrow $\iota\to\iota'$ if and only
if $\iota\preccurlyeq\iota'$], and suppose a functor
$\beta:\It\to\Funct(\Ct,\Ct')$ is given. Then every functor
$\lambda:\Ct\to\It$ determines a functor
$\varphi=\beta\circ\lambda:\Ct\to\Funct(\Ct,\Ct')$ and hence a
diagonal functor $\varphi_\Delta\in\Funct(\Ct,\Ct')$. Moreover, any natural transformation
$\zeta:\lambda\nto\lambda'$ between
$\lambda,\lambda'\in\Funct(\Ct,\It)$ induces a natural transformation
between the corresponding diagonal functors.
\end{Lem}
{\noindent\em Remark:} A functor $\lambda:\Ct\to\It$ is equivalent to
labelling each object $A$ of $\Ct$ with an element $\lambda(A)\in\It$, subject to the
requirement that $\lambda(A)\preccurlyeq\lambda(B)$ if
there is a $\Ct$-morphism from $A$ to $B$. The existence of a natural transformation
between $\lambda$ and $\lambda'$ amounts to the condition that
$\lambda(A)\preccurlyeq \lambda'(A)$ for all $A$.  The use of partially
ordered sets is simply for convenience and familiarity.

{\noindent\em Proof:} Given functors $\beta$ and $\lambda$ as described,
it is obvious that $\varphi=\beta\circ\lambda\in\Funct(\Ct,\Funct(\Ct,\Ct'))$. Given
$\zeta:\lambda\nto\lambda'$, the maps $\xi_A=\beta(\zeta_A)$ form the
components of a natural $\xi:\varphi\nto\varphi'$ by functoriality of
$\beta$. Hence $\xi_\Delta:\varphi_\Delta\nto\varphi'_\Delta$ has
components $(\xi_\Delta)_A = \beta(\zeta_A)_A$.  
$\square$

\subsection{Diagonal theories in $\LCTo$ and $\LCT$}

Any functor $\varphi:\Mand\to\LCT$ assigns to each spacetime $\Mb\in\Mand$ a locally covariant theory
defined on {\em all} spacetimes, i.e., a functor
$\varphi(\Mb):\Mand\to\Phys$, and assigns to each embedding $\psi:\Mb\to\Nb$ a
natural transformation $\varphi(\psi):\varphi(\Mb)\nto\varphi(\Nb)$
between the theories assigned to $\Mb$ and $\Nb$ respectively. The
diagonal functor $\varphi_\Delta$ 
is then again an object of $\LCT$ and hence a theory 
in its own right; we will refer to it as a {\em diagonal theory}.

By Remark (2) following Prop.~\ref{prop:diagonal_construction}, 
every theory $\Af\in\LCTo$ is a diagonal theory
in which $\varphi:\Man\to\Phys$ is a constant functor taking the value $\Af$
in all spacetimes. So diagonal theories certainly exist. Our aim in this subsection
is to investigate some of the general properties of diagonal theories and to 
develop criteria that would give various types of desirable or pathological properties;
in particular, that violations of SPASs can, in principle, be achieved with theories that are otherwise well-behaved.  In the following subsection we will show that such diagonal theories exist under fairly mild restrictions on the category $\Phys$.

Our discussion is expressed for diagonal theories in $\LCT$; all our remarks in this subsection apply
equally to diagonal theories in $\LCTo$ on replacing $\Mand$ by $\Man$, $\LCT$ by $\LCTo$, and $\OO(\Mb)$ by $\OO_0(\Mb)$.

\paragraph{The kinematic net}  If $\Mb\in\Mand$ and $O\in\OO(\Mb)$, the 
kinematic local algebra is
\begin{align*}
(\varphi_\Delta)^\kin_{\Mb;O} &=\varphi_\Delta(\iota_{\Mb;O}) = 
\varphi(\Mb)(\iota_{\Mb;O})\circ \varphi(\iota_{\Mb;O})_{\Mb|_O}= \varphi(\Mb)^\kin_{\Mb;O}
\circ \varphi(\iota_{\Mb;O})_{\Mb|_O} \\
&\le \varphi(\Mb)^\kin_{\Mb;O}.
\end{align*}
If there exists any morphism $\psi:\Lb\to\Mb$ such that $\varphi(\psi)_\Mb$ is not an isomorphism then
$(\varphi_\Delta)^\kin_{\Mb;\psi(\Lb)}$ is a proper subobject of $\varphi(\Mb)^\kin_{\Mb;\psi(\Lb)}$. 

\paragraph{The timeslice property} Suppose $\varphi:\Mand\to\LCT$. For any morphism $\psi:\Mb\to\Nb$ we have $\varphi_\Delta(\psi) = \varphi(\psi)_\Nb\circ\varphi(\Mb)(\psi)$. Accordingly, a 
{\em sufficient} condition for $\varphi_\Delta$ to satisfy the timeslice property is that both the
following hold: (i) for every
$\Mb\in\Mand$, $\varphi(\Mb)$ satisfies the timeslice property and (ii) $\varphi$ obeys the timeslice
property in that $\varphi(\psi)$ is a natural isomorphism whenever $\psi$ is Cauchy. 

In particular, suppose that $\varphi=\beta\circ\lambda$, where $\It$ is a poset (regarded as a category) and
$\beta$ and $\lambda$ are functors. Then the sufficient condition just mentioned becomes 
(i) for each $\ell\in\Im\lambda$, $\beta(\ell)$ obeys the timeslice axiom, and (ii) $\lambda$ is constant
on Cauchy-wedge-connected components of $\Mand$. To see this, note that (ii) implies that if
$\psi:\Mb\to\Nb$ is Cauchy, then $\lambda(\psi) = \id_{\lambda(\Mb)}$ and hence 
$\varphi(\psi) = \id_{\varphi(\Mb)}$. 

\paragraph{The relative Cauchy evolution} 

Suppose $\varphi:\Mand\to\LCT$ is such that every
$\varphi(\Mb)$ obeys the timeslice axiom and so does $\varphi_\Delta$. 

\begin{Lem} \label{lem:diagonal_Cauchy_morphisms}
If $\Mb\stackrel{\psi}{\to}\Nb$ is a Cauchy morphism then
$\varphi(\psi)_\Mb$ and $\varphi(\psi)_\Nb$ are isomorphisms.
\end{Lem}
{\noindent\em Proof:} As $\varphi(\Mb)$ and $\varphi(\Nb)$ obey the
timeslice axiom, $\varphi(\Mb)(\psi)$ and $\varphi(\Nb)(\psi)$ are
isomorphisms. As $\varphi_\Delta(\psi)$ is also an isomorphism, the
result follows because
$\varphi(\psi)_\Mb=(\varphi(\Nb)(\psi))^{-1}\circ \varphi_\Delta(\psi)$
and $\varphi(\psi)_\Nb=\varphi_\Delta(\psi)\circ
(\varphi(\Mb)(\psi))^{-1}$. $\square$
 
\begin{Prop}\label{prop:rce_diagonal}
For any $\hb\in H(\Mb)$ we have
\begin{equation}\label{eq:tau_diagonal}
\tau_\Mb^{(\varphi_\Delta)\pm}[\hb] =
\tau_\Mb^{(\varphi(\Mb[\hb]))\pm}[\hb] \circ
\varphi(\jmath_\Mb^\pm[\hb])_\Mb\circ \left(\varphi(\imath_\Mb^\pm[\hb])_\Mb\right)^{-1},
\end{equation}
where
$\Mb\stackrel{\imath_\Mb^\pm[\hb]}{\longleftarrow}\Mb^\pm[\hb] 
\stackrel{\jmath_\Mb^\pm[\hb]}{\longrightarrow}\Mb[\hb]$
are the future and past Cauchy wedges induced by $\hb$. Hence
\[
\rce_\Mb^{(\varphi_\Delta)}[\hb] = \varphi(\imath_\Mb^-[\hb])_\Mb\circ  
\varphi(\jmath_\Mb^-[\hb])_\Mb^{-1}\circ
\varphi(\jmath_\Mb^+[\hb])_\Mb\circ \left(\varphi(\imath_\Mb^+[\hb])_\Mb\right)^{-1}\circ
\rce_\Mb^{(\varphi(\Mb))}[\hb].
\]
If $\varphi$ also obeys timeslice [i.e., $\varphi(\psi)$ is a natural
isomorphism for each Cauchy morphism $\psi$] then these results may be
written more compactly as
\[
\tau_\Mb^{(\varphi_\Delta)\pm}[\hb] =
\tau_\Mb^{(\varphi(\Mb[\hb]))\pm}[\hb] \circ (\tau_\Mb^{(\varphi)\pm}[\hb])_\Mb
\]
and
\[
\rce_\Mb^{(\varphi_\Delta)}[\hb] = (\rce_\Mb^{\varphi}[\hb])_\Mb\circ\rce_\Mb^{(\varphi(\Mb))}[\hb].
\]
\end{Prop}
{\noindent\em Proof:} As usual, $\tau_\Mb^{(\varphi_\Delta)\pm}[\hb]$
is the unique morphism such that
$\tau_\Mb^{(\varphi_\Delta)\,\pm}[\hb]\circ\varphi_\Delta(\imath_\Mb^\pm[\hb])=
\varphi_\Delta(\jmath_\Mb^\pm[\hb])$, i.e.,
\begin{align*}
\tau_\Mb^{(\varphi_\Delta)\pm}[\hb]\circ\varphi(\imath_\Mb^\pm[\hb])_\Mb
\circ\varphi(\Mb^\pm)(\imath_\Mb^\pm[\hb])
&=
\varphi(\jmath_\Mb^\pm[\hb])_{\Mb[\hb]}\circ\varphi(\Mb^\pm)(\jmath_\Mb^\pm[\hb])
\\
&=
\varphi(\jmath_\Mb^\pm[\hb])_{\Mb[\hb]}\circ 
\tau_\Mb^{(\varphi(\Mb^\pm))\pm}[\hb]\circ\varphi(\Mb^\pm)(\imath_\Mb^\pm[\hb]).
\end{align*}
As $\varphi(\Mb^\pm)(\imath_\Mb^\pm[\hb])$ and (by
Lemma~\ref{lem:diagonal_Cauchy_morphisms}) $\varphi(\imath_\Mb^\pm[\hb])_\Mb$ are
isomorphisms, Eq.~\eqref{eq:tau_diagonal} holds. Accordingly, 
\begin{align*}
\rce_\Mb^{(\varphi_\Delta)}[\hb] &= \left(\tau_\Mb^{(\varphi_\Delta)-}[\hb]\right)^{-1}\circ
\tau_\Mb^{(\varphi_\Delta)+}[\hb]  \\
&= \varphi(\imath_\Mb^-[\hb])_\Mb\circ \left(\varphi(\jmath_\Mb^-[\hb])_\Mb\right)^{-1}
\circ\rce_\Mb^{(\varphi(\Mb[\hb]))}[\hb]\circ
\varphi(\jmath_\Mb^+[\hb])_\Mb\circ
\left(\varphi(\imath_\Mb^+[\hb])_\Mb\right)^{-1}\\
&= \varphi(\imath_\Mb^-[\hb])_\Mb\circ \left(\varphi(\jmath_\Mb^-[\hb])_\Mb\right)^{-1}
\circ\varphi(\jmath_\Mb^+[\hb])_\Mb\circ\rce_\Mb^{(\varphi(\Mb^+))}[\hb]\circ
\left(\varphi(\imath_\Mb^+[\hb])_\Mb\right)^{-1} \\
&= \varphi(\imath_\Mb^-[\hb])_\Mb\circ \left(\varphi(\jmath_\Mb^-[\hb])_\Mb\right)^{-1}
\circ\varphi(\jmath_\Mb^+[\hb])_\Mb\circ
\left(\varphi(\imath_\Mb^+[\hb])_\Mb\right)^{-1}\circ\rce_\Mb^{(\varphi(\Mb))}[\hb]
\end{align*}
as required, where we have used Prop.~\ref{prop:rce_intertwine} in the last two
steps. The remaining statements are straightforward.
$\square$ 

It is clear from the above result that the diagonal theory
$\varphi_\Delta$ does not
necessarily have the same relative Cauchy evolution in spacetime $\Mb$
as $\varphi(\Mb)$. In principle, this allows the stress--energy tensor to have a component that reflects the dynamics of the functor $\varphi$ as well as the
dynamics of the theory in spacetime $\Mb$. We do not know whether this
can be realised in actual examples, however. Certainly, if $\varphi$
factors through a poset, then (as we have already seen) $\varphi$ maps
any Cauchy morphism to an identity and so we have the simpler formulae
\begin{align}
\tau_\Mb^{(\varphi_\Delta)\pm}[\hb] &= \tau_\Mb^{(\varphi(\Mb[\hb]))\pm}[\hb]
\\
\rce_\Mb^{(\varphi_\Delta)}[\hb] &= \rce_\Mb^{(\varphi(\Mb))}[\hb].
\end{align}
Any diagonal theory in which these relations hold will be described as
{\em ordinary}.

\paragraph{Comparison of theories and failure of SPASs in $\LCT$} Suppose $\varphi=\beta\circ\lambda$,
where $\lambda:\Mand\to\It$ and $\beta:\It\to\LCT$ with $\It$ a poset. 
Suppose that there are $\ell,\ell'\in\It$ such that $\ell\preccurlyeq\lambda(\Mb) \preccurlyeq\ell'$
for all $\Mb\in\Mand$, with both $\ell$ and $\ell'$ being attained on certain spacetimes,
and assume that $\beta(\ell,\ell')$ is not an isomorphism.
By the remark following Lem.~\ref{lem:factor_construction}, this 
gives natural transformations $\kappa_\ell\nto\lambda\nto\kappa_{\ell'}$, where $\kappa_p$
is the constant functor taking the value $p$ on all objects; hence, by  Lem.~\ref{lem:factor_construction}, there are natural transformations
\begin{equation}\label{eq:diagonal_comp}
\beta(\ell) = (\beta\circ\kappa_\ell)_\Delta \nto (\beta\circ\lambda)_\Delta\nto 
(\beta\circ\kappa_{\ell'})_\Delta = \beta(\ell')
\end{equation}
whose components in an arbitrary spacetime $\Mb$ are
\[
\beta(\ell)(\Mb)\xlongrightarrow{\beta(\ell,\lambda(\Mb))_\Mb}
 (\beta\circ\lambda)_\Delta(\Mb) \xlongrightarrow{\beta(\lambda(\Mb),\ell')_\Mb} \beta(\ell')(\Mb),
\]
composing to $\beta(\ell,\ell')_\Mb$. Accordingly the two naturals in Eq.~\eqref{eq:diagonal_comp}
compose to $\beta(\ell,\ell')$. 

Now let $\Lb$ and $\Lb'$ be spacetimes with $\lambda(\Lb)=\ell$, $\lambda(\Lb')=\ell'$. 
Then the first natural is an identity in spacetime $\Lb$, while the second is an identity in 
spacetime $\Lb'$. Thus both are partial isomorphisms. If the SPASs property were to hold on (any class of theories including) $\beta(\ell)$, $(\beta\circ\lambda)_\Delta$ and $\beta(\ell')$, then both
naturals would have to be isomorphisms, which contradicts the fact that their composite, $\beta(\ell,\ell')$,
is not an isomorphism. 

In particular, if one or both of the theories $\beta(\ell)$ and $\beta(\ell')$ are regarded as individually
representing the same physics in all spacetimes (by some reasonable definition) then it is clearly
impossible for $(\beta\circ\lambda)_\Delta$ to represent the same physics in all spacetimes (by the same definition). 

This discussion shows that the failure of SPASs can be exhibited quite straightforwardly, given 
suitable functors $\beta$ and $\lambda$. In the next subsection, we will give some concrete
constructions which achieve this goal. We have presented the discussion so far in fairly abstract terms, 
partly to facilitate discussion of general categories $\Phys$ and partly because a wide range of 
constructions can be given and we wish to emphasise that the issue runs more deeply than 
a few isolated counterexamples (each of which, perhaps, could be removed by some {\em ad hoc} 
additional assumptions). In addition, it may be that diagonal theories may
provide useful examples in other contexts, e.g., locally covariant theories
that do not obey the timeslice axiom. 

\subsection{Specific Examples}

To start, let us consider the problem of constructing a functor from $\Man$ to a poset. 
There are many ways of doing this, and the reader should regard the examples
presented here as indicative rather than exhaustive.

For a first example, fix a constant $R_0>0$ with dimensions of $\text{length}^{-2}$ and define
\begin{equation}\label{eq:poisoned_spacetimes}
\lambda(\Mb) = \begin{cases} 2 &\sup R_\Mb >R_0 \\ 1 &\sup R_\Mb \le R_0,
\end{cases}
\end{equation}
where $R_\Mb$ is the scalar curvature on $\Mb\in\Man$ and the supremum is taken over all of $\Mb$. It is clear that if $\psi:\Mb\to\Nb$ then
$\lambda(\Mb)\le \lambda(\Nb)$, so $\lambda$ is indeed a functor from $\Man$ to 
$\sN$, i.e., the natural numbers with their usual ordering. 

This particular functor is not constant on Cauchy-wedge-connected components
of $\Man$, however. To see this, consider a spacetime containing Cauchy surfaces $\Sigma_1$
and $\Sigma_2$ so that the scalar curvature exceeds $R_0$ near $\Sigma_1$, but is
everywhere less than $R_0$ in a globally hyperbolic neighbourhood of $\Sigma_2$. 
This induces a Cauchy wedge connecting a spacetime with $\lambda=1$ to a spacetime
where $\lambda=2$. Thus diagonal
theories based on such functors would not be expected to have the timeslice property. 
However, we will find a use for this example below.

A different type of example is constructed by choosing any function $\mu:\Man\to \NN$ 
such that (i) $\mu(\Mb)$ depends only on the oriented-diffeomorphism class of the
smooth spacelike Cauchy surfaces of $\Mb$; (ii) $\mu$ takes its minimum value
on all spacetimes with noncompact Cauchy surfaces. This is obviously constant
on Cauchy-wedge-connected components by Prop.~\ref{prop:Cauchy_chain}.
To see that it is a functor from $\Man$ to $\sN$, we take any morphism $\psi:\Mb\to\Nb$
in $\Man$. If $\Mb$ has noncompact Cauchy surface, then $\mu(\Mb)\le \mu(\Nb)$
by condition (ii). If, on the other hand, $\Mb$ has compact Cauchy surfaces, then 
Prop.~\ref{prop:embeddings}(a) entails that $\Mb$ and $\Nb$ have oriented-diffeomorphic 
Cauchy surfaces and hence $\mu(\Mb)=\mu(\Nb)$. Thus  $\mu\in\Funct(\Man,\sN)$.
[Equally, this construction gives a functor to $\Im\mu$, equipped with the 
partial ordering in which $p\preccurlyeq q$ iff $p=q$ or $p=\min \Im\mu$]. 

In view of the comments in the previous subsection, diagonal theories $(\beta\circ\mu)_\Delta$ will obey
the timeslice property provided that $\beta(\ell)$ obeys timeslice for each $\ell\in\Im \mu$.  

Turning to the case of possibly disconnected spacetimes,  one way of
constructing a functor from $\Mand$ to $\sN$ is to take any functor $\lambda_0:\Man\to\sN$ and to define
\[
\lambda(\Mb) = \max_{\Cb\in\Cpts(\Mb)} \lambda_0(\Cb)
\]
for $\Mb\in\Mand$. Consider any $\Mand$-morphism $\psi:\Mb\to\Nb$ and
let $\Bb$ be a component of $\Mb$ such that $\lambda(\Mb)=\lambda_0(\Bb)$.
Then there is a component $\Cb$ of $\Nb$ so that $\psi(\Bb)\subset \Cb$ and
a $\Man$-morphism $\psi^\Bb_\Cb:\Bb\to\Cb$. Then
\[
\lambda(\Mb) =\lambda_0(\Bb) \le \lambda_0(\Cb) \le \lambda(\Nb),
\]
which suffices to show that $\lambda\in\Funct(\Mand,\sN)$. Moreover,
$\lambda$ will be constant on Cauchy-wedge-connected components of $\Mand$
if $\lambda_0$ is constant on Cauchy-wedge-connected components of $\Man$. 

There are many other possibilities. Let ${\rm Surf}$ be the set of smooth connected {\em compact} orientable $(n-1)$-manifolds modulo oriented-diffeomorphisms ($n$ being the spacetime dimension). 
To every $\Mb\in\Mand$ there is a function $\nu_\Mb:\text{Surf}\to\NN_0$ such that $\nu_\Mb(\Sigma)$ is the number of
connected components of $\Mb$ whose Cauchy surfaces are oriented-diffeomorphic to $\Sigma$. 
Evidently $\nu_\Mb(\Sigma)$ is nonzero for at most finitely many $\Sigma\in\text{Surf}$; 
using Prop.~\ref{prop:embeddings}(b) it is easily seen that the existence of a morphism $\psi:\Mb\to\Nb$ entails that $\nu_\Mb(\Sigma)\le \nu_\Nb(\Sigma)$ for all $\Sigma\in \text{Surf}$ (we are only counting compact connected components). 
A wide variety of functors $\lambda:\Mand\to\sN$ may now be constructed, such as 
\[
\lambda(\Mb) =  a + \sum_\Sigma m(\Sigma) \nu_\Mb(\Sigma)^{p(\Sigma)}
\]
for $a\in\NN$ and any functions $m,p:\text{Surf}\to \NN_0$. All such functors
are constant on Cauchy-wedge-connected components of $\Mand$, because
Cauchy-wedge-connected spacetimes $\Mb$ and $\Nb$ have oriented-diffeomorphic Cauchy surfaces, so the functions $\nu_\Mb$ and $\nu_\Nb$ coincide.

We have shown that it is possible to construct functors from $\Man$ and $\Mand$
to various posets in various ways. There are also various ways of obtaining
functors from a poset to $\LCT$ as shown by the following examples (all of which
adapt straightforwardly to $\LCTo$):
\begin{enumerate}\addtolength{\itemsep}{-0.5\baselineskip}
\item If $\It$ is the poset $\NN$ with the ordering $p\preccurlyeq q$ iff 
$p=1$ or $p=q$, we may proceed by setting $\beta(1)=\If$, the initial theory, and choose $\beta(p)\in\LCT$ arbitrarily for $p\ge 2$. To the arrow $1\to p$ assign the natural $\If_{\beta(p)}:\If\nto \beta(p)$ that arises because $\If$ is initial. 
All other arrows in $\It$ are identities, and we assign to each $\id_p$ the
morphism $\id_{\beta(p)}$ [evidently this is compatible with the previous
assignment for $p=1$]. Then $\beta\in\Funct(\It,\LCT)$.

\item Suppose $\Phys$ admits an endofunctor $\Ff$ and a natural $\eta:\Ff\nto\id_{\Phys}$.\footnote{See Sect.~\ref{sect:LCT} for an example
in $\TAlg$.}
 Given any $\Af\in\LCT$ there is a functor $\beta: (\{1,2\},\le)\to \LCT$ with
\[
\beta(1) = \Ff\circ\Af, \quad \beta(2) = \Af, \qquad 
\beta(\id_1) = \id_{\Ff\circ\Af},\quad \beta(1\to 2) = \eta, \quad \beta(\id_2)=
\id_{\Af}.
\]
\item If $\Phys$ has a monoidal structure then, as discussed in Sect.~\ref{sect:LCT}, we obtain a functor $\beta:\sN\to\LCT$ with $\beta(k)=\Af^{\otimes k}$ and naturals $\beta(k,k'):\beta(k)\nto\beta(k')$ for
any $k\le k'$.
\end{enumerate}

Pursuing the third of these examples, let us suppose that $\mu_0:\Man\to\NN$ 
is constant on Cauchy-wedge-connected components of $\Man$, with $\mu_0(\Mb)=1$ if $\Mb$ has noncompact Cauchy surfaces and $\mu_0(\Mb)\neq 1$ for some spacetimes. Let us suppose that the basic
theory $\Af$ has the timeslice property and is not idempotent, meaning that
there is no $k \ge 2$ for which $\beta(1,k)$ is an isomorphism. Setting
$\varphi=\beta\circ\lambda$, 
the $\varphi_\Delta$ is an ordinary diagonal theory in $\LCTo$, that
will be denoted $\Af^{[\mu_0]}$; it obeys the timeslice axiom because each
$\Af^{\otimes k}$ does. 

In any spacetime $\Mb$, we have
$\Af^{[\mu_0]}(\Mb) = \Af^{\otimes \mu_0(\Mb)}(\Mb)$; if 
$\psi:\Mb\to\Nb$ and $\mu_0(\Mb)\le \mu_0(\Nb)$ then 
\[
\Af^{[\mu_0]}(\psi) = \beta(\mu_0(\Mb),\mu_0(\Nb))_\Nb\circ \Af^{\otimes\mu_0(\Mb)}(\psi) = 
\beta(\mu_0(\Mb),\mu_0(\Nb))_\Nb\circ \Af(\psi)^{\otimes\mu_0(\Mb)}
\]
(if the category $\Phys$ is $\Alg$, with the algebraic tensor product as
the monoidal structure, then this has the action 
\[
\Af^{[\mu_0]}(\psi) X = (\Af^{\otimes\mu_0(\Mb)}(\psi) X)\otimes 
\II_{\Af(\Nb)}^{\otimes(\mu_0(\Nb)-\mu_0(\Mb))}
\]
on $X\in\Af^{\otimes\mu_0(\Mb)}(\Mb)$). 
The kinematic net for $\varphi_\Delta$ produces subobjects 
$(\varphi_\Delta)^\kin_{\Mb;O}$ that are proper subobjects of $\varphi(\Mb)^\kin_{\Mb;O}$ whenever $\mu_0(\Mb)>1$ and $O\in\OO_0(\Mb)$
has noncompact Cauchy surface.

If, additionally, $\mu_0$ is bounded with maximum value $\ell'$, then
we may argue as in the previous subsection that the SPASs property cannot
hold on any class of theories including $\Af$, $\Af^{[\mu_0]}$ and $\Af^{\otimes\ell'}$; if either $\Af$ or $\Af^{\otimes\ell'}$ is regarded as
representing the same physics in all spacetimes (by some definition), it follows that $\Af^{[\mu_0]}$ cannot have this property (by the same definition). 

This example is enough to show that $\LCTo$ will generally fail to have the SPASs property, except in the case that all its theories are idempotent. Similarly, in $\LCT$, if we define $\mu(\Mb)=\max\{ \mu_0(\Cb):\Cb\in\Cpts(\Mb)\}$, then the theory
$\Af^{[\mu]}:=(\beta\circ\mu)_\Delta$ [with $\beta$ now giving monoidal powers
in $\LCT$] has analogous properties and demonstrates the failure of SPASs
in $\LCT$.
%
%

We conclude this section by sketching two other examples to illustrate the range
of bad behaviour that can occur. For the first, we return to the functor $\lambda:\Man\to\sN$ of Eq.~\eqref{eq:poisoned_spacetimes} and compose with the functor $\beta(k)=\Af^{\otimes k}$,
where $\Af$ is nontrivial and has the timeslice property and is additive, in the sense that $\Af(\Mb)$
is generated by the $\Af^{\kin}(\Mb;O_i)$ whenever the $O_i$ form a cover of $\Mb$ by open
globally hyperbolic spacetimes. The upshot is a theory $\Bf = (\beta\circ\lambda)_\Delta$
that coincides with $\Af^{\otimes 2}$ in spacetimes whose scalar curvature somewhere exceeds $R_0$,
and otherwise coincides with $\Af$. (The theory $\Bf$ does not have the timeslice property.) 
Now consider a spacetime $\Mb$ that has a Cauchy surface $\Sigma$ on which the scalar 
curvature is everywhere greater than $R_0$, but which also has an open globally hyperbolic region $U$ on which the scalar curvature is everywhere less than $R_0$. Consider any cover 
$\Mb=\bigcup_i O_i$ by nonempty open globally hyperbolic spacetimes $O_i$. 
Then the $O_i$ also cover $\Sigma$, and every $O_i$ that intersects
$\Sigma$ nontrivially must have $\lambda(\Mb|_{O_i}) = 2$, so $\Bf^{\kin}(\Mb;O_i) = \Af^{\otimes 2\kin}(\Mb;O_i)$ for these particular regions. As $\Af^{\otimes 2}$ has the timeslice property, this proves that $\Bf(\Mb)$ is generated by the $\Bf^{\kin}(\Mb;O_i)$ with $O_i\cap\Sigma\neq 0$, and hence 
{\em a fortiori} by the full collection of $\Bf^{\kin}(\Mb;O_i)$. Thus
the theory $\Bf$ is additive on $\Mb$ and has $\Bf(\Mb)=\Af^{\otimes 2}(\Mb)$; 
but at the same time, $\Mb$ contains a region $U$ for which the local kinematic subobject $\beta^{\kin}_{\Mb;U} = \beta(1,2)\circ \alpha^{\kin}_{\Mb;U}$ corresponds to only {\em one} copy of the theory $\Af$. This example stands as a counterpoint to the previous examples, where 
additivity would not be expected to hold in spacetimes with $\lambda=2$.  

Finally, as an extreme example, suppose $\Phys$ admits
infinite monoidal products indexed over the naturals (i.e., a colimit of the functor giving finite monoidal powers). Then we may also form infinite powers $\Bf^{\otimes\infty}$ of any theory $\Bf$ in $\LCT$. There is a right-shift
endomorphism $\sigma:\Bf^{\otimes\infty}\nto\Bf^{\otimes\infty}$ which
is given (for $\Phys=\Alg$, say) by 
\[
\sigma_\Mb X = \II_{\Bf(\Mb)}\otimes X,
\]
which realises any such $\Bf^{\otimes\infty}$ as a proper subtheory of itself [i.e., $\sigma$ is a non-automorphic endomorphism] except if $\Bf$ is the trivial theory. 
Now suppose $\Af$ is any nontrivial locally covariant 
theory and let $\Bf$ be a diagonal theory that coincides with $\If$ in some spacetimes and $\Af$ in others. Then the right-shift $\sigma$ on $\Bf^{\otimes\infty}$ is a partial isomorphism, as $\sigma_\Mb$ is an isomorphism in every spacetime where $\Bf(\Mb)$ is trivial. Of course, the theory $\Bf^{\otimes\infty}(\Mb)$ is also trivial in such spacetimes, but by passing to the theory
$\Af\otimes\Bf^{\otimes\infty}$, we obtain a theory that is nontrivial in all spacetimes
and admits an endomorphism $\id_{\Af}\otimes\sigma$ that is a partial isomorphism but not an automorphism. Theories of this type cannot be regarded 
as obeying the same physics in all spacetimes by any reasonable notion: 
even the singleton $\{\Af\otimes\Bf^{\otimes\infty}\}$ fails to have the SPASs property.
One might suspect that theories admitting proper endomorphisms are always
unphysical; elsewhere it will be shown that they conflict with natural requirements of nuclearity/energy compactness, which supports the idea that they must have 
infinitely many degrees of freedom available in bounded regions at finite energies~\cite{Fewster:gauge}.

We have described these examples in some detail to illustrate that a wide variety of bad behaviour can be exhibited by locally covariant theories. It seems likely that yet worse behaviour could be found.

\section{Dynamical determination of local observables}\label{sect:intrinsic}

\subsection{The dynamical net}\label{sect:intrinsic_net}

In Sect.~\ref{sect:kinematic}, we saw how BFV used the functorial structure
of a locally covariant theory to reconstruct a net structure of local observables.
The idea was to regard the theory in a subregion of a spacetime as the theory
assigned to that subregion when considered as a spacetime in its own right. 
We regard this as a kinematic description of the local physics. In this section
we use the dynamics of the relative Cauchy evolution to give another description
of local physics; the theory will be said to be {\em dynamically local} when these
two descriptions of the local physics coincide. The diagonal theories, as we will see,
include examples of theories that are not dynamically local; in~\cite{FewVer:dynloc2}
we will show that the Klein--Gordon theory is dynamically local both as a classical and a quantum theory (at nonzero mass; the massless case involves further subtleties).

To illustrate the general idea, suppose that $\Alg$ has been taken as the category $\Phys$,
and that $\Af$ is a locally covariant theory in this setting. Fix a spacetime $\Mb$ and
a compact set $K$ therein. Any hyperbolic perturbation $\hb\in H(\Mb;K^\perp)$
represents a modification in the spacetime in regions causally inaccessible from $K$; one
would expect that observables localised within $K$ should be insensitive to such 
changes. Taking this as a {\em definition} of what it means to be localised in $K$, we
are led to study the subalgebra
\[
\Af^{\bullet}(\Mb;K) = \{A\in\Af(\Mb): \rce_\Mb[\hb]A = A ~\textrm{for all
$\hb\in H(\Mb;K^\perp)$}\}
\]
as the candidate for the description of the local physics. Given an open globally hyperbolic subset with
finitely many components (though not necessarily nonempty) $O\in\OO(\Mb)$
we may define the subalgebra $\Af^\dyn(\Mb;O)$ of $\Af(\Mb)$ generated by the
$\Af^\bullet(\Mb;K)$ for a suitable class of compact subsets of $O$. (The simpler possibility
of defining the $\Af^\bullet(\Mb;\cl(O))$ as the local algebra of a relatively compact open
globally hyperbolic set $O$ would not generally give a match with the kinematic algebra $\Af^\kin(\Mb;O)$ 
as can be seen in the example of the Klein--Gordon field~\cite{FewVer:dynloc2}).
To this end, for each nonempty $O\in \OO(\Mb)$ we define $\KK(\Mb;O)$ to be
the set of compact subsets contained in $O$ and having a multi-diamond neighbourhood whose base is contained in $O$. In particular, this condition is obeyed by the empty set, 
so $\emptyset\in \KK(\Mb;O)$ for all nonempty $O\in\OO(\Mb)$. By convention we also set
$\KK(\Mb;\emptyset)= \{\emptyset\}$. We use $\KK(\Mb)$ as a
shorthand for $\KK(\Mb;\Mc)$. 

This class is chosen for various reasons. The requirement to have a (multi)-diamond
neighbourhood ensures, for example, that if $K\in\KK(\Mb)$ then $K^{\perp\perp}$ is again compact (see  Lemma~\ref{lem:perp_of_compacts}; the proof relies on the relative compactness of multi-diamonds). 
We use multi-diamonds, rather than diamonds, to facilitate the treatment of sets $O$ with more than 
one connected component; in some (but not all) theories one could insist on diamond neighbourhoods
without loss. These issues will be discussed elsewhere. 

We then define the {\em dynamical net} as the assignment to
each $O\in\OO(\Mb)$ of the subalgebra
\begin{equation}\label{eq:dyndef1}
\Af^\dyn(\Mb;O) = \bigvee_{K\in\KK(\Mb;O)}
\Af^{\bullet}(\Mb;K)
\end{equation}
in which the right-hand side denotes the $\Alg$-subobject of $\Af(\Mb)$ generated by the
$\Af^{\bullet}(\Mb;K)$ for $\KK(\Mb)\owns K\subset O$. As $\emptyset\in\KK(\Mb;O)$, 
we always have $\Af^\bullet(\Mb;\emptyset)\subset \Af^\dyn(\Mb;O)$ for every $O$; 
in particular, $\Af^\dyn(\Mb;\emptyset) = \Af^\bullet(\Mb;\emptyset)$. As we will show in \cite{FewVer:dynloc2}, Eq.~\eqref{eq:dyndef1} gives the correct local algebras for the simple model
of the massive Klein--Gordon field.

More generally, the above ideas can be implemented in any category $\Phys$ satisfying
our standing assumptions. As in the case of the kinematic net it is convenient to focus 
on the subobject morphisms; we will also find it useful to give `universal' definitions
for the various subobjects of interest. 

\begin{Lem} 
For any compact subset $K$ of $\Mb$ there exists a unique (up to isomorphism)
subobject $\alpha^\bullet_{\Mb;K}$ of $\Af(\Mb)$ such that (i)
\begin{equation} \label{eq:alpha_bullet}
\rce_\Mb[\hb]\circ\alpha^\bullet_{\Mb;K} = \alpha^\bullet_{\Mb;K} \qquad
\forall \hb\in H(\Mb;K^\perp);
\end{equation}
and (ii) if any other morphism $\alpha$ satisfies
Eq.~\eqref{eq:alpha_bullet} in place of $\alpha^\bullet_{\Mb;K}$, then
$\alpha \le \alpha^\bullet_{\Mb;K}$ in the subobject lattice of $\Af(\Mb)$.\footnote{Recall
that this means there is a unique $\beta$ such that
$\alpha = \alpha^\bullet_{\Mb;K}\circ\beta$.} 
\end{Lem}
{\noindent\em Proof:} For each $\hb\in H(\Mb;K^\perp)$, let $\alpha_\hb$ be the 
equaliser of $\rce_\Mb[\hb]$ and $\id_{\Af(\Mb)}$ [which exists by assumption on $\Phys$], 
i.e., a morphism such that $\rce_\Mb[\hb]\circ\alpha_\hb = \alpha_\hb$, and so that
any other morphism $\beta_\hb$ obeying this equation in place of $\alpha_\hb$ obeys
$\beta_\hb\le \alpha_\hb$. Then any intersection
\[
\alpha^\bullet_{\Mb;K} \cong \bigwedge_{\hb\in H(\Mb;K^\perp)} \alpha_\hb
\]
(which exists by assumption on $\Phys$) obeys Eq.~\eqref{eq:alpha_bullet}:
see, e.g., Lem.~\ref{lem:intersections}. 
Any $\beta$ also obeying this equation must in particular obey $\beta\le \alpha_\hb$
for all $\hb\in H(\Mb;K^\perp)$ by the definition of the equaliser; accordingly,
$\beta\le \alpha^\bullet_{\Mb;K}$ by the definition of an intersection. $\square$

In the case $\Phys=\Alg$, $\alpha^\bullet_{\Mb;K}$ is of course the inclusion morphism
of $\Af^\bullet(\Mb;K)$ in $\Af(\Mb)$. Returning to the general case, $\Phys$
also has arbitrary categorical unions; accordingly, to each $O\in\OO(\Mb)$
there is a (unique up to isomorphism) subobject
\begin{equation}\label{eq:intrinsic}
\alpha^\dyn_{\Mb;O} \cong 
\bigvee_{K\in\KK(\Mb; O)}\alpha^{\bullet}_{\Mb;K}
\end{equation}
(generalising the inclusion morphism of $\Af^\dyn(\Mb;O)$ in $\Af(\Mb)$ in the category $\Alg$)
that we take as the definition of the dynamical net. 
Denoting the domain of $\alpha^\dyn_{\Mb;O}$ as $\Af^\dyn(\Mb;O)$,
Eq.~\eqref{eq:intrinsic} means that (i) every $\alpha^{\bullet}_{\Mb;K}$
(with $K\in \KK(\Mb;O)$) factorises (uniquely) via $\alpha^\dyn_{\Mb;O}$
as $\alpha^{\bullet}_{\Mb;K}= \alpha^\dyn_{\Mb;O}\circ\alpha_{\Mb;O;K}$; (ii) whenever there are morphisms
$\beta$ and $\gamma$ and $\beta_{K}$ such that
$\beta\circ\beta_K=\gamma\circ \alpha^{\bullet}_{\Mb;K}$ for every $K\in \KK(\Mb;O)$, there exists a unique $\xi:\Af^\dyn(\Mb;O)\to B$ such that
\[
\beta_K = \xi\circ \alpha_{\Mb;O;K} \qquad{\rm and}\qquad \beta\circ\xi
= \gamma\circ\alpha^{\rm
int}_{\Mb;O}
\]
for all $K\in \KK(\Mb;O)$. Diagrammatically, fixing $\beta$ and $\gamma$, if 
the outer portion of every diagram of the following form commutes as $K$ varies in $\KK(\Mb;O)$  then there is a unique $\xi$ to make all the diagrams commute in full:
\[
\begin{tikzpicture}[baseline=0 em, description/.style={fill=white,inner sep=2pt}]
\matrix (m) [ampersand replacement=\&,matrix of math nodes, row sep=3em,
column sep=4em, text height=1.5ex, text depth=0.25ex]
{  \& \Af^\dyn(\Mb;O) \& \Af(\Mb) \\
\Af^\bullet(\Mb;K) \& \& \\
\& B \& C \\ };
\path[->]
(m-2-1) edge node[above,sloped] {$ \alpha_{\Mb;O;K} $} (m-1-2)
        edge node[above,sloped] {$ \beta_K $} (m-3-2)
(m-1-2) edge node[above] {$ \alpha^\dyn_{\Mb;O} $} (m-1-3)
        edge[dotted] node[auto] {$ \xi $} (m-3-2)
(m-3-2) edge node[above] {$ \beta $} (m-3-3)
(m-1-3) edge node[auto] {$ \gamma $} (m-3-3);
\end{tikzpicture}
\]
(see Appendix~\ref{appx:subob} and~\cite{DikranjanTholen} for more details on the union in general categories). 

Although we have given notation for the domains of the morphisms $\alpha^\bullet_{\Mb;K}$, $\alpha^\dyn_{\Mb;O}$, one should bear in mind that it is the morphisms that are the significant entities. For the sake of familiarity we will write
expressions such as $\Af^\bullet(\Mb;K_1)\subset
\Af^\bullet(\Mb;K_2)$, but this must be understood as asserting 
that $\alpha^\bullet_{\Mb;K_1}$ factorizes via $\alpha^\bullet_{\Mb;K_2}$, i.e.,
$\alpha^\bullet_{\Mb;K_1}=\alpha^\bullet_{\Mb;K_2}\circ\beta$ for some $\beta:
\Af^\bullet(\Mb;K_1)\to \Af^\bullet(\Mb;K_2)$. This is the order relation in the subobject lattice of $\Af(\Mb)$ (see, e.g., \cite{DikranjanTholen}). Similarly, $\Af^\bullet(\Mb;K_1)\cong
\Af^\bullet(\Mb;K_2)$ asserts that $\alpha^\bullet_{\Mb;K_1}=\alpha^\bullet_{\Mb;K_2}\circ\beta$
with $\beta$ an isomorphism, i.e.,  $\alpha^\bullet_{\Mb;K_1}\cong \alpha^\bullet_{\Mb;K_2}$ as subobjects. In the case of $\Alg$ or other category in which 
$\Af^\bullet(\Mb;K)$ and $\Af^\dyn(\Mb;O)$ are realised concretely as subsets of $\Af(\Mb)$, 
and the $\alpha^\bullet_{\Mb;K}$, $\alpha^\dyn_{\Mb;O}$ morphisms are set inclusions then
the $\subset$ notation may be taken to indicate a subset and isomorphism can be upgraded to equality.

\subsection{Properties of the dynamical net}

The assignments $K\mapsto \Af^\bullet(\Mb;K)$ and $O\mapsto\Af^\dyn(\Mb;O)$ possess a number of properties that would be expected of a net of local algebras: namely, isotony, causal
dynamics, and covariance with respect to isomorphisms. 

\begin{Thm} \label{thm:A_bullet_MK}
(a) Suppose $K_1,K_2$ are compact and $J_\Mb(K_1)\subset J_\Mb(K_2)$
(in particular, if $K_1\subset K_2$). Then $\Af^\bullet(\Mb;K_1)\subset
\Af^\bullet(\Mb;K_2)$.\\
(b) In consequence, we have
\[
\Af^\bullet(\Mb;K) \cong \Af^\bullet(\Mb;K^{\perp\perp}) 
\]
provided $K^{\perp\perp}$ is also compact (in particular, if $K\in \KK(\Mb)$)
and, for any compact sets $K_1,K_2$,
\begin{align*}
\Af^\bullet(\Mb;K_1)\vee \Af^\bullet(\Mb;K_2) &\subset \Af^\bullet(\Mb; K_1\cup
K_2) \\
\Af^\bullet(\Mb; K_1\cap K_2) &\subset \Af^\bullet(\Mb;K_1)\wedge \Af^\bullet(\Mb;K_2)
\end{align*}
and $\Af^\bullet(\Mb;\emptyset) \subset \Af^\bullet(\Mb;K)$ for all compact $K$. 
(c) If $\psi:\Mb\to\Nb$ is an isomorphism then $\Af(\psi)$ restricts to
an isomorphism $\Af^{\bullet}(\Mb;K)\to\Af^\bullet(\Nb;\psi(K))$
(this applies in particular to the (time-)orientation preserving isometric isomorphisms of $\Mb$).
\end{Thm}
{\noindent\em Proof:} (a) Immediate from the definition. (b) These results
follow from (a) because $J_\Mb(K)=J_\Mb(K^{\perp\perp})$ for
compact $K$ (see Lem.~\ref{lem:perp_of_compacts}(ii)) and the obvious
inclusions $K_1\cap K_2\subset K_i\subset K_1\cup K_2$. (c) As $\psi$
is an isomorphism, $\psi(K)^\perp=\psi(K^\perp)$. Thus the pushforward $\psi_*$
restricts to an isomorphism between $H(\Mb;K^\perp)$ and
$H(\Nb;\psi(K)^\perp)$, with inverse given by the pullback $\psi^*$.
Hence for all $\hb\in H(\Nb;\psi(K))$, 
\[
\rce_\Nb[\hb]\circ\Af(\psi)\circ\alpha^\bullet_{\Mb;K} = 
\Af(\psi)\circ\rce_\Mb[\psi^*\hb]\circ\alpha^\bullet_{\Mb;K} =
\Af(\psi)\circ \alpha^\bullet_{\Mb;K}
\]
by the defining property of $\alpha^\bullet_{\Mb;K}$; it follows that $
\Af(\psi)\circ\alpha^\bullet_{\Mb;K} = \alpha^\bullet_{\Nb;\psi(K)}
\circ \beta$ for some $\beta$ (depending on $\psi$ and $K$). Applying the same argument to
$\psi^{-1}$, it follows easily that $\beta$ is an isomorphism. 
$\square$

These results immediately induce a number of analogous properties of the $\Af^\dyn(\Mb;O)$, in Theorem~\ref{thm:A_int_MO} below. First, we
give a useful simplifying observation.
\begin{Lem}\label{lem:refined_dyn}
Given any $O\in\OO(\Mb)$, we have
\[
\Af^\dyn(\Mb;O) \cong \bigvee_{K\in \KK_b(\Mb;O)} \Af^\bullet(\Mb;K),
\]
where $\KK_b(\Mb;O)$ is the set of those $K\in \KK(\Mb;O)$ obtained as
the closure of a base of a multi-diamond, with $\KK_b(\Mb;\emptyset)=\{\emptyset\}$ by convention. 
If, in fact, $O$ is a multi-diamond, 
then 
\[
\Af^\dyn(\Mb;O) \cong \bigvee_{K\subset\subset B} \Af^\bullet(\Mb;K) ,
\]
where $B$ is any base of $O$ and the union is taken over all compact subsets of $B$. 
\end{Lem}
{\noindent\em Proof:} 
If $O$ is empty, the first statement holds trivially because
$\KK_b(\Mb;\emptyset)=\KK(\Mb;\emptyset)=\{\emptyset\}$; as 
$\emptyset$ is not a multi-diamond the second statement is irrelevant.
Accordingly, now assume that $O$ is nonempty and 
let $K\in\KK(\Mb;O)$. Then there is a multi-diamond with 
base $B\subset O$ such that $K\subset D_\Mb(B)$. By Lemma~\ref{lem:exhaustion}, 
there exists a compact set $\tilde{K}\subset B$ with $K\subset \tilde{K}^{\perp\perp}$; it is clear that $\tilde{K}\subset \KK(\Mb;O)$. 
Hence $\Af^\bullet(\Mb;K)\subset \Af^\bullet(\Mb;\tilde{K}^{\perp\perp}) \cong  \Af^\bullet(\Mb;\tilde{K})$
by parts (a) and (c) of Thm.~\ref{thm:A_bullet_MK}.
In the case of a general nonempty $O\in\OO(\Mb)$ we deduce that the defining union
of $\Af^\dyn(\Mb;O)$ may be taken over $K\in \KK_b(\Mb;O)$
[see Lemma~\ref{lem:union_refine} for a proof in the abstract setting];
in the case where $O$ is a multi-diamond with $B$ as a base, we may evidently
require that each $\tilde{K}$ be a subset of $B$, obtaining the second
refinement (every compact subset of $B$ is clearly a member of $\KK(\Mb;O)$). $\square$

We expect that stronger causality results than (c) below can be obtained along similar lines.
\begin{Thm}\label{thm:A_int_MO}
 (a) If $O_1,O_2\in\OO(\Mb)$ and $O_1\subset O_2$ then $\Af^\dyn(\Mb;O_1)\subset \Af^\dyn(\Mb;O_2)$. In consequence, we also have, for arbitrary $O_1,O_2\in\OO(\Mb)$,
\begin{align*}
\Af^\dyn(\Mb;O_1)\vee \Af^\dyn(\Mb;O_2) &\subset \Af^\dyn(\Mb; O_1\cup
O_2) \\
\Af^\dyn(\Mb; O_1\cap O_2) &\subset \Af^\dyn(\Mb;O_1)\wedge \Af^\dyn(\Mb;O_2)
\end{align*}
and $\Af^\bullet(\Mb;\emptyset)\cong \Af^\dyn(\Mb;\emptyset) \subset \Af^\dyn(\Mb;O)$ for all $O\in\OO(\Mb)$. \\
(b) If $\psi:\Mb\to\Nb$ is an isomorphism then
$\Af(\psi)$ restricts to an isomorphism
$\Af^\dyn(\Mb;O)\to\Af^\dyn(\Nb;\psi(O))$ for each $O\in\OO(\Mb)$.
(In particular,  this applies to automorphisms $\psi\in\Aut(\Mb)$.)\\ 
(c) If $O\in\OO(\Mb)$ and $O''$ is a multi-diamond with a base contained in $O$, then 
\[
\Af^\dyn(\Mb;O'') \cong \Af^\dyn(\Mb;O).
\]
\end{Thm}
{\noindent\em Proof:} (a) is obvious because $\KK(\Mb;O_1)\subset \KK(\Mb;O_2)$. For (b), we
use Thm.~\ref{thm:A_bullet_MK}(c) and the obvious fact that the unions of 
isomorphic subobjects of isomorphic objects are isomorphic. Turning to (c),
we may suppose that $O''=D_\Mb(B)$, where $B\subset O$ is a base of $O''$. 
By Lemma~\ref{lem:refined_dyn} we then have
\[
\Af^\dyn(\Mb;O'') \cong \bigvee_{K\subset\subset B} \Af^\bullet(\Mb;K) 
\subset \Af^\dyn(\Mb;O) \subset \Af^\dyn(\Mb;O''),
\]
where we have also used part (a).  
$\square$

Further light on the relationship between the two species of dynamical
net is shed by the next result. We will need the following definition. 
\begin{Def} 
A compact set $K\subset \Mb$ will be called {\em outer regular} if
there exist relatively compact nonempty $O_n\in\OO(\Mb)$ ($n\in\NN$) with
$\cl(O_{n+1})\subset O_n$ and $K\in\KK(\Mb;O_n)$ for all $n$, such
that $K=\bigcap_n O_n$. [Note that this excludes the empty set from being outer regular.] 
Any such sequence $O_n$ will be called an 
{\em outer approximation} to $K$. The set of outer regular compact
subsets of any nonempty $O\in\OO(\Mb)$ will be denoted $\KK^{o.r.}(\Mb;O)$.
If $K\in\KK^{o.r.}(\Mb;O)$ has an outer approximating sequence
$O_n\in\OO_0(\Mb)$, we write $K\in\KK^{o.r.}_0(\Mb;O)$. 
\end{Def}
Note that $\KK_b(\Mb;O)\subset \KK^{o.r.}(\Mb;O)$. We write $\KK^{o.r.}(\Mb)$
for $\KK^{o.r.}(\Mb;\Mc)$.

\begin{Thm} \label{thm:A_bullet_and_A_int}
(a) For all $O\in\OO(\Mb)$ and $\hb\in H(\Mb;O')$ we have 
$\rce_\Mb[\hb]\circ \alpha^\dyn_{\Mb;O} =  \alpha^\dyn_{\Mb;O}$. \\
(b) If $O\in\OO(\Mb)$ is relatively compact, then
\[
\Af^\dyn(\Mb;O) \subset \Af^\bullet(\Mb;\cl(O)).
\]
(c) If $K\in\KK^{o.r.}(\Mb)$ has outer approximating sequence $O_n$, then
\begin{equation}
\Af^\bullet(\Mb;K)\cong \bigwedge_{n\in\NN} \Af^\dyn(\Mb;O_n)  .
\label{eq:intersections}
\end{equation}
\end{Thm}
{\noindent\em Proof:} (a) If $K\in \KK(\Mb;O)$, then $K^\perp\supset O'$ and hence
$\hb\in H(\Mb;K^\perp)$. Thus $\rce_\Mb[\hb]\circ \alpha^\bullet_{\Mb;K}=\alpha^\bullet_{\Mb;K}$ for all such $K$.
The same then holds for $\alpha^\dyn_{\Mb;O}$ due to Eq.~\eqref{eq:intrinsic}. (For completeness, 
a proof is given in Lem.~\ref{lem:union_invariance} of Appendix~\ref{appx:subob}.)
\\
(b) Lemma~\ref{lem:perp+prime} entails that $O'=(\cl(O))^\perp$. Using (a), we deduce that $\rce_\Mb[\hb]\circ \alpha^\dyn_{\Mb;O} = \alpha^\dyn_{\Mb;O}$ for all 
$\hb\in H(\Mb;\cl(O)^\perp)$ and hence $\alpha^\dyn_{\Mb;O} \le \alpha^{\bullet}_{\Mb;\cl(O)}$, 
establishing the required inclusion. \\
(c) As $K\in\KK(\Mb;O_n)$ for each $n\in\NN$ the right-hand side of Eq.~\eqref{eq:intersections} clearly contains the left-hand side.
On the other hand, Lemma~\ref{lem:perp_of_reg_int}(ii) entails that
\[
K^\perp = \bigcup_{n\in\NN} O_n'
\]
so for any $\hb\in H(\Mb;K^\perp)$, the compact set $\supp\hb\subset K^\perp$ is covered by finitely many of the open sets 
$O_n'$ and hence (as $O_{n}'\subset O_{n+1}'$ for each $n$) is contained in some $O_{n_0}'$. It follows that
$\hb\in H(\Mb;O_{n_0}')$, so $\rce_\Mb[\hb]$ acts trivially on $\Af^\dyn(\Mb;O_{n_0})$ and therefore on the 
intersection in Eq.~\eqref{eq:intersections}. Accordingly, the right-hand side is  contained in $\Af^\bullet(\Mb;K)$.
$\square$

We remark that this result also gives $\Af^\bullet(\Mb;K)$ for 
sets $K$ such that $K^{\perp\perp}$ is outer regular, by virtue of Thm~\ref{thm:A_bullet_MK}(b).

As an example of the various relationships developed above, we note that if $p,q$ 
are distinct timelike separated points with $q$ to the future of $p$, then
\[
\Af^\bullet(\Mb;\{p,q\}) \cong \Af^\bullet(\Mb;\{p,q\}^{\perp\perp}) \cong \Af^\bullet(\Mb;J_\Mb^+(p)\cap J_\Mb^-(q)) \supset 
\Af^\dyn(\Mb;I^+_\Mb(p)\cap I^-_\Mb(q)).
\]
Moreover, if $p_n\to p$ in $I_\Mb^-(p)$, and $q_n\to q$ in $I_\Mb^+(q)$ then
\[
\Af^\bullet(\Mb;\{p,q\}) \cong \bigwedge_n \Af^\dyn(\Mb;I^+_\Mb(p_n)\cap I^-_\Mb(q_n)).
\]
In addition if $K\in\KK(\Mb)$ is the closure of a Cauchy multi-ball, then we may choose
a sequence of Cauchy multi-balls $B_k$ such that $\cl(B_{k+1})\subset B_k$ and
$\bigcap_k B_k = K$. Choose a strictly decreasing sequence $(\epsilon_k)$ with $\epsilon_k\to 0$ such that $O_k= D_\Mb(B_k)\cap {\cal T}^{-1}(-\epsilon_k,\epsilon_k)$ belongs to $\OO(\Mb)$ for each $k$,
where ${\cal T}$ is a Cauchy temporal function~\cite{Bernal:2004gm} such that ${\cal T}^{-1}(0)$ contains all the $B_k$. 
Then $K=\bigcap_k O_k$ and so
\[
\Af^\bullet(\Mb;K) \cong \bigwedge_{k} \Af^\dyn(\Mb;O_k).
\]

Finally, let us compute the dynamical nets of ordinary diagonal models $\varphi_\Delta$.
\begin{Thm} \label{thm:diagonal_int_net}
For any ordinary diagonal theory $\varphi_\Delta$, we have
\begin{align}
\varphi_\Delta^\bullet(\Mb;K) &= \varphi(\Mb)^\bullet(\Mb;K) \\
\intertext{for all compact $K\subset\Mb$, and}
\varphi_\Delta^\dyn(\Mb;O) &= \varphi(\Mb)^\dyn(\Mb;O)
\end{align}
for all $O\in\OO(\Mb)$.
\end{Thm}
{\noindent\em Proof:} The first statement is an immediate consequence from
Prop.~\ref{prop:rce_diagonal} as
$\rce_\Mb^{(\varphi_\Delta)}[\hb]=\rce_\Mb^{(\varphi(\Mb))}[\hb]$; the
second follows immediately. 
$\square$

Thus the ordinary diagonal theories provide examples in which it is the dynamical net, 
rather than the kinematic net, that appears to have the `right' notion of
the local observables on any given spacetime. 
(As we have no examples of extraordinary diagonal theories, it is less clear
what should be expected in that case.)

\section{Dynamical locality}\label{sect:dynamical_locality}

\subsection{Definition and main properties}

The kinematical and dynamical nets give two isotonous nets on each spacetime;
the diagonal theories show that they are not always equal. In general, their
relationship is given as follows. 
\begin{Prop} \label{prop:int_ext}
Let $\Af\in \LCT$ (resp., $\LCTo$). Suppose $O\in\OO(\Mb)$ (resp., $\OO_0(\Mb)$) is nonempty, and that $O\subset K\in\KK(\Mb;\tilde{O})$ for some $\tilde{O}\in\OO(\Mb)$. Then
\[
\Af^\kin(\Mb;O)\subset \Af^{\bullet}(\Mb;K) \subset \Af^\dyn(\Mb;\tilde{O}).
\]
\end{Prop}
{\noindent\em Proof:} By Prop.~\ref{prop:rce_locality}, we have
$\rce_\Mb[\hb]\circ\alpha^\kin_{\Mb;O} = \alpha^\kin_{\Mb;O}$
for all $\hb\in H(\Mb;K^\perp)$, so $\alpha^\kin_{\Mb;O}=
\alpha^{\bullet}_{\Mb;K}\circ \beta$ for some $\beta$ and the first inclusion is proved.
The second follows immediately as $\Af^{\bullet}(\Mb;K)$ is one of the
generating algebras for $\Af^\dyn(\Mb;\tilde{O})$. $\square$

A clear case of interest is that in which these two nets actually coincide; 
in view of Prop.~\ref{prop:int_ext} this is a maximality condition
on the kinematic net. It requires, roughly, that every observable invariant under
changes of metric in the causal complement of $O$ is localised in $O$.

\begin{Def} 
A theory $\Af\in \LCT$ (resp., $\LCTo$) obeys {\em dynamical locality} if it obeys
the timeslice property and, additionally, for each
$\Mb\in\Mand$ (resp., $\Man$) and all nonempty $O\in\OO(\Mb)$ (resp., $\OO_0(\Mb)$) we have $\Af^\kin(\Mb;O)\cong \Af^{\dyn}(\Mb;O)$, i.e., more abstractly,
\[
\alpha^\kin_{\Mb;O} \cong \alpha^\dyn_{\Mb;O}.
\]
\end{Def} 
In view of Lem.~\ref{lem:ext_M_O}, the dynamical locality condition may also be written in the form
\[
\Af(\psi)\cong \alpha^\dyn_{\Nb;\psi(\Mb)} 
\cong \bigvee_{K\in\KK(\Nb;\psi(\Mb))}\alpha^{\bullet}_{\Nb;K}
\]
for all $\psi:\Mb\to\Nb$.

An immediate example is furnished by the initial theory $\If$, because all subobjects
of an initial object are isomorphic. More physically interesting theories will be
considered in \cite{FewVer:dynloc2}. In the remainder of this section we explore various general features of dynamically local theories without restricting $\Phys$; later, in section~\ref{sect:dlQFT}, we will consider applications to quantum field theory by specifying that $\Phys$ should be $\Alg$ or $\CAlg$. 

\paragraph{Additivity}
Dynamical locality imposes a form of additivity on the theory. 
\begin{Thm} \label{thm:additivity}
Suppose $\Af\in\LCT$ (resp., $\LCTo$) is dynamically local in $\LCT$ (resp., $\LCTo$). \\
(a) For any $\Mb\in \Mand$ (resp., $\Man$), 
the maps
$\bigvee_{K\in\KK(\Mb)}\alpha^\bullet_{\Mb;K}$ and
$\bigvee_{K\in\KK_b(\Mb)}\alpha^\bullet_{\Mb;K}$
are isomorphisms, i.e.,
\[
\Af(\Mb) \cong  \bigvee_{K\in\KK(\Mb)}\Af^{\bullet}(\Mb;K) \cong \bigvee_{K\in\KK_b(\Mb)}\Af^{\bullet}(\Mb;K) .
\]
(b) Suppose $\widetilde{\OO}$ 
is a subset of $\OO(\Mb)$ such that
every $K\in\KK_b(\Mb)$ is contained in some $O\in\widetilde{\OO}$. Then
$\bigvee_{O\in\widetilde{\OO}}\alpha^\dyn_{\Mb;O}$ is an
isomorphism, i.e.,
\[
\Af(\Mb) \cong  \bigvee_{O\in\widetilde{\OO}} \Af^\dyn(\Mb;O) \cong
\bigvee_{O\in\widetilde{\OO}} \Af^\kin(\Mb;O).
\] 
\end{Thm}
{\noindent\em Remark:}  In particular, by definition of $\KK_b(\Mb)$,
part~(b) applies when the $\widetilde{\OO}$
consists of the truncated multi-diamonds of $\Mb$. \\
{\noindent\em Proof:} (a) First observe that $\alpha^\dyn_{\Mb;\Mc}\cong\alpha^{\kin}_{\Mb;\Mc} \cong \id_{\Af(\Mb)}$ (by Lemma~\ref{lem:ext_M_O}). Thus $\alpha^{\dyn}_{\Mb;\Mc}$ is an isomorphism. 
The statement follows from the definition of $\alpha^{\dyn}_{\Mb;\Mc}$ and Lemma~\ref{lem:refined_dyn}.

%

(b) For each $K\in\KK(\Mb)$ choose a $O_K\in\widetilde{\OO}$ with $K\subset O_K$,
whereupon there is a factorization $\alpha^\bullet_{\Mb;K} = \alpha^\dyn_{\Mb;O_K}\circ \alpha_{\Mb;O_K;O}$ for each such $K$. By Lemma~\ref{lem:union_refine},
\[
\bigvee_{K\in\KK_b(\Mb)}\alpha^\bullet_{\Mb;K} \le \bigvee_{O\in\widetilde{\OO}}
\alpha^\dyn_{\Mb;O}.
\]
As the left-hand side is an isomorphism, the monic property of $\bigvee_{O\in\widetilde{\OO}}
\alpha^\dyn_{\Mb;O}$ implies that it is an isomorphism. The remaining
statements are immediate. $\square$

\paragraph{Covariance} Theorem~\ref{thm:A_bullet_MK}(b) and Theorem~\ref{thm:A_int_MO}(b) 
provide rather weaker forms of covariance than the relation Eq.~\eqref{eq:kinematic_covariance} that holds  for the kinematic net. Dynamical locality provides the missing ingredient, provided the class of compact indexing regions is restricted slightly. 
\begin{Thm} \label{thm:dl_covariance}
Suppose $\Af\in\LCT$ (resp., $\LCTo$) is dynamically local in $\LCT$ (resp., $\LCTo$)
and let $\psi:\Mb\to\Nb$ in $\Mand$ (resp., $\LCTo$).
Then for all nonempty $\OO\in\OO(\Mb)$  and $K\in\KK^{o.r.}(\Mb)$ (resp., $\OO_0(\Mb)$, $K\in\KK^{o.r.}_0(\Mb)$), we have
\[
\alpha^\dyn_{\Nb;\psi(O)} \cong \Af(\psi)\circ\alpha^\dyn_{\Mb;O}
\qquad\text{and}\qquad
\alpha^\bullet_{\Nb;\psi(K)} \cong \Af(\psi)\circ\alpha^\bullet_{\Mb;K}.
\]
The second formula holds also for compact $K$ such that $K^{\perp\perp}$ is outer regular. 
\end{Thm}
{\noindent\em Proof:} The first statement follows immediately from dynamical locality and
the covariance of the kinematic net of Eq.~\eqref{eq:kinematic_covariance}, by the calculation
\[
\alpha^\dyn_{\Nb;\psi(O)} \cong 
\alpha^\kin_{\Nb;\psi(O)} = \Af(\psi)\circ  \alpha^\kin_{\Mb;O}
\cong  \Af(\psi)\circ  \alpha^\dyn_{\Mb;O}.
\]
For the second, we claim that if $O_n$ is outer approximating to $K$ in $\Mb$, then 
$\psi(O_n)$ is outer approximating to $\psi(K)$ in $\Nb$. We use the fact that $\psi$ maps
diamonds and their bases in $\Mb$ to diamonds and their bases in $\Nb$; this is otherwise straightforward. 
Using this observation and the first part of the result, we calculate
\[
\alpha^\bullet_{\Nb;\psi(K)} \cong \bigwedge_n 
\alpha^\dyn_{\Nb;\psi(O_n)}\cong 
\bigwedge_n  \Af(\psi)\circ \alpha^\dyn_{\Mb;O_n}\cong
\Af(\psi)\circ  \bigwedge_n   \alpha^\dyn_{\Mb;O_n} \cong \Af(\psi)\circ\alpha^\bullet_{\Mb;K},
\]
in  conjunction with Thm.~\ref{thm:A_bullet_MK}(b) and Lem.~\ref{lem:intersections}. 
Finally, if $K^{\perp\perp}\in \KK^{o.r.}(\Mb)$ (resp., $\KK^{o.r.}_0(\Mb)$), we calculate
\[
\alpha^\bullet_{\Nb;\psi(K)}\cong \alpha^\bullet_{\Nb;\psi(K)^{\perp\perp}} = 
 \alpha^\bullet_{\Nb;\psi(K^{\perp\perp})} \cong 
\Af(\psi)\circ\alpha^\bullet_{\Mb;K^{\perp\perp}}\cong 
\Af(\psi)\circ\alpha^\bullet_{\Mb;K}
\]
using the previous result, Thm.~\ref{thm:A_bullet_MK}(b), and the identity
$\psi(K^{\perp\perp})=\psi(K)^{\perp\perp}$ proved in  Lemma~\ref{lem:psi_perpperp}.
$\square$

\paragraph{Extended locality} In Minkowski space algebraic QFT, extended locality~\cite{Schoch1968, Landau1969} is the condition that local algebras of spacelike separated regions should intersect only 
on multiples of the identity. Here, we will give a necessary and sufficient condition for 
a version of extended locality in general locally covariant physical theories subject to dynamical locality. 
\begin{Thm} \label{thm:extended_locality}
Suppose that $\Af\in\LCT$ (resp., $\LCTo$) is dynamically local. Then the following are equivalent:
\begin{enumerate}\addtolength{\itemsep}{-0.5\baselineskip}
\item $\Af$ obeys extended locality, in the sense that 
$\alpha^\kin_{\Mb;O_1}\wedge\alpha^\kin_{\Mb;O_2}$ is trivial for all
causally disjoint nonempty $O_i\in\OO(\Mb)$\footnote{That is, $O_1\subset O_2^\perp$ and $O_2\subset O_1^\perp$, from which it follows that $O_1\subset O_2'$ and $O_2\subset O_1'$.} (resp., $\OO_0(\Mb)$) for arbitrary $\Mb\in\Mand$ (resp., $\Man$);
\item $\alpha^\bullet_{\Mb;\emptyset}$ (or equivalently $\alpha^\dyn_{\Mb;\emptyset}$) is trivial, i.e., equivalent to $\Ic_{\Af(\Mb)}$
for every $\Mb\in\Mand$ (resp., $\Man$). 
\end{enumerate}
\end{Thm}
{\noindent\em Proof:} (1)$\implies$(2): take any two nonempty causally disjoint $O_i\in\OO_0(\Mb)$. We then have, using Thm.~\ref{thm:A_int_MO}(a),
\[
\alpha^\bullet_{\Mb;\emptyset}\cong  \alpha^\dyn_{\Mb;\emptyset} 
= \alpha^\dyn_{\Mb;O_1\cap O_2}
\le \alpha^\dyn_{\Mb;O_1}\wedge \alpha^\dyn_{\Mb;O_2}
\cong \alpha^\kin_{\Mb;O_1}\wedge \alpha^\kin_{\Mb;O_2}\cong \Ic_{\Af(\Mb)}.
\]
(2)$\implies$(1): On the other hand, let $\alpha\cong \alpha^\kin_{\Mb;O_1}\wedge\alpha^\kin_{\Mb;O_2}$ for causally disjoint nonempty $O_i\in\OO_0(\Mb)$. Then
$\alpha= \alpha^\kin_{\Mb;O_i}\circ \alpha_i$ for some $\alpha_i$; we will show that the
$\alpha_i$ are trivial, which implies triviality of $\alpha$. To this end, let  $\hb\in H(\Mb|_{O_1})$ 
be arbitrary and observe that 
\begin{align*}
\Af(\iota_{\Mb;O_1})\circ \rce_{\Mb|_{O_1}}[\hb] \circ\alpha_1 
&=  \rce_\Mb[\iota_{\Mb;O_1*}\hb]\circ \Af(\iota_{\Mb;O_1})\circ\alpha_1 = 
\rce_\Mb[\iota_{\Mb;O_1*}\hb]\circ \Af(\iota_{\Mb;O_2})\circ \alpha_2 \\
&= \Af(\iota_{\Mb;O_2})\circ \alpha_2 = \Af(\iota_{\Mb;O_1})\circ\alpha_1 ,
\end{align*}
where we have used the causal separation of the $O_i$ and Prop.~\ref{prop:rce_locality}. 
Cancelling the monic $ \Af(\iota_{\Mb;O_1})$, we have 
$\rce_{\Mb|_{O_1}}[\hb] \circ\alpha_1=\alpha_1$ for all $\hb\in H(\Mb|_{O_1})$. 
Hence $\alpha_1\le \alpha^\bullet_{\Mb|_O;\emptyset} \cong \Ic_{\Af(\Mb|_{O_1})}$
and is therefore trivial.
$\square$

The subobject $\alpha^\bullet_{\Mb;\emptyset}$ represents those elements of the theory
that are invariant with respect to arbitrary perturbations of the metric, and therefore do not
couple to gravity. Under many circumstances one would want this to be trivial, i.e., 
that $\alpha^\bullet_{\Mb;\emptyset}\cong \Ic_\Mb$ for all spacetimes $\Mb$.
As we will see, this requirement is not always satisfied -- indeed, it is not satisfied
for the theory of the free massless minimally coupled scalar field in spacetimes of
compact spatial section. However, it can be derived from other reasonable
conditions on theories in $\LCT$ as will be discussed elsewhere.

\subsection{The SPASs property}\label{sect:SPASs}

The pathological theories constructed in Sect.~\ref{sect:pathologies}
had the property that there are natural transformations between them
such that some, but not all, of their components are isomorphisms. In
this section we prove that this cannot occur if we restrict to dynamically
local theories. Throughout this section, $\Af$ and $\Bf$ are 
fixed theories  in either $\LCT$ or $\LCTo$ obeying the timeslice property. 

The following preparatory lemmas are elementary, but crucial; we give
proofs for completeness.

\begin{Lem} \label{lem:zeta_restrictions}
Let $\Mb$ be an arbitrary spacetime. Suppose there is a morphism $\zeta_\Mb:\Bf(\Mb)\to\Af(\Mb)$
[not necessarily a component of a natural transformation] such that
\begin{equation}\label{eq:intertwine_hypothesis}
\rce_\Mb^{(\Af)}[\hb]\circ\zeta_\Mb =
\zeta_\Mb\circ\rce_\Mb^{(\Bf)}[\hb]
\end{equation}
for all $\hb\in H(\Mb;K^\perp)$. Then there are unique
morphisms 
\begin{align*}
\zeta^\bullet_{\Mb;K}:\Bf^\bullet(\Mb;K)&\to\Af^\bullet(\Mb;K)
\\
\zeta^\dyn_{\Mb;O}:\Bf^\dyn(\Mb;O)&\to\Af^\dyn(\Mb;O)
\end{align*}
such that 
\begin{align}
\alpha^\bullet_{\Mb;K}\circ \zeta^\bullet_{\Mb;K}& = 
\zeta_\Mb\circ\beta^\bullet_{\Mb;K}\label{eq:zeta_bullet}\\
\alpha^\dyn_{\Mb;O}\circ \zeta^\dyn_{\Mb;O}& = 
\zeta_\Mb\circ\beta^\dyn_{\Mb;O}\label{eq:zeta_int},
\end{align}
where we use $\beta^\bullet_{\Mb;K}$ and $\beta^{\rm
int}_{\Mb;O}$ for the inclusion morphisms of $\Bf^\bullet(\Mb;K)$ and
$\Bf^\dyn(\Mb;O)$ in $\Bf(\Mb)$. Thus $\zeta^\bullet_{\Mb;K}$ and $\zeta^\dyn_{\Mb;O}$ are
restrictions of $\zeta_\Mb$. Moreover, if $\zeta_\Mb$ is an isomorphism,
so are $\zeta^\bullet_{\Mb;K}$ and $\zeta^\dyn_{\Mb;O}$.
In particular, these conclusions hold if $\zeta_\Mb$ is a component of a
natural transformation $\zeta:\Bf\nto\Af$.
\end{Lem}
{\noindent\em Proof:} As $\alpha^\bullet_{\Mb;K}$ and $\alpha^\dyn_{\Mb;O}$ are monic, uniqueness is automatic and one need only
demonstrate existence. First, by Eq.~\eqref{eq:intertwine_hypothesis} and the defining
property of $\beta^\bullet_{\Mb;K}$,
\[
\rce_\Mb^{(\Af)}[\hb]\circ\zeta_\Mb\circ\beta^\bullet_{\Mb;K} = 
\zeta_\Mb\circ\rce_\Mb^{(\Bf)}[\hb]\circ\beta^\bullet_{\Mb;K} = 
\zeta_\Mb\circ\beta^\bullet_{\Mb;K}
\]
for all $\hb\in H(\Mb;K^\perp)$. Hence $\zeta_\Mb\circ\beta^\bullet_{\Mb;K}$
shares the defining property of $\alpha^\bullet_{\Mb;K}$ and we deduce
the existence of unique
$\zeta^\bullet_{\Mb;K}:\Bf^\bullet(\Mb;K)\to\Af^\bullet(\Mb;K)$ such
that Eq.~\eqref{eq:zeta_bullet} holds. 

Second, for each $\KK(\Mb)\owns K\subset O$, the outer portion of
the diagram 
\[
\begin{tikzpicture}[description/.style={fill=white,inner sep=2pt}]
\matrix (m) [ampersand replacement=\&,matrix of math nodes, row sep=4em,
column sep=6em, text height=1.5ex, text depth=0.25ex]
{ \Bf^\bullet(\Mb;K) \&  \Bf^\dyn(\Mb;O)  \& \Bf(\Mb)   \\
\Af^\bullet(\Mb;K) \&   \Af^\dyn(\Mb;O) \&  \Af(\Mb) \\ };
\path[->]
(m-1-1)  edge node[above] {$ \beta_{\Mb;O;K} $} (m-1-2)
             edge node[left] {$ \zeta^{\bullet}_{\Mb;K} $} (m-2-1)
(m-1-2)  edge node[above] {$ \beta^\dyn_{\Mb;O} $} (m-1-3)
             edge[dotted] node[left] {$ \zeta^\dyn_{\Mb;O} $} (m-2-2)
(m-1-3)  edge node[left] {$ \zeta_\Mb $} (m-2-3)
(m-2-1)  edge node[above] {$ \alpha_{\Mb;O;K} $} (m-2-2)
(m-2-2)  edge node[above] {$ \alpha^\dyn_{\Mb;O} $} (m-2-3);
\end{tikzpicture}
\]
%
now commutes, thus inducing a unique $\zeta^\dyn_{\Mb;O}:\Bf^\dyn(\Mb;O)\to\Af^{\rm
int}(\Mb;O)$ such that all the diagrams commute in full; in particular,
we have the required property Eq.~\eqref{eq:zeta_int}. 

Thirdly, if $\zeta_\Mb$ is an isomorphism,
Eq.~\eqref{eq:intertwine_hypothesis} holds with $\zeta_\Mb$ replaced by
$\zeta_\Mb^{-1}$ and $\Af$ and $\Bf$ interchanged. Thus there are unique
morphisms $(\zeta_\Mb^{-1})^\bullet_{;K}$ and $(\zeta_\Mb^{-1})^{\rm
int}_{;O}$ such that
\begin{align*}
\beta^\bullet_{\Mb;K}\circ (\zeta_\Mb^{-1})^\bullet_{;K}& = 
\zeta_\Mb^{-1}\circ\alpha^\bullet_{\Mb;K} \\
\beta^\dyn_{\Mb;O}\circ (\zeta_\Mb^{-1})^\dyn_{;O}& = 
\zeta_\Mb^{-1}\circ\alpha^\dyn_{\Mb;O}.
\end{align*}
Combining with Eqs.~\eqref{eq:zeta_bullet} and~\eqref{eq:zeta_int} and using the
facts that $\alpha^\bullet_{\Mb;K},\beta^\bullet_{\Mb;K}$ are monic, it
is easily seen that $(\zeta_\Mb^{-1})^\bullet_{;K}$ and $(\zeta_\Mb^{-1})^{\rm
int}_{;O}$ are inverses to $\zeta^\bullet_{\Mb;K}$ and $\zeta^{\rm
int}_{\Mb;O}$, which are therefore isomorphisms.

Finally, in the case that $\zeta_\Mb$ is a component of a natural
transformation $\zeta:\Bf\nto\Af$, Eq.~\eqref{eq:intertwine_hypothesis}
holds by Prop.~\ref{prop:rce_intertwine}.
$\square$

\begin{Lem} \label{lem:bigvee_iso}
Suppose $\zeta:\Bf\nto\Af$ and that there exist subobjects
$\psi_i:\Mb_i\to\Mb$ ($i\in I$) such that $\bigvee_{i\in I}\Af(\psi_i)$ and
all the $\zeta_{\Mb_i}$ are isomorphisms. Then $\zeta_{\Mb}$ and $\bigvee_{i\in I}\Bf(\psi_i)$
are isomorphisms.
\end{Lem}
{\noindent\em Proof:} Consider, for each $i\in I$, the diagram 
\[
\begin{tikzpicture}[description/.style={fill=white,inner sep=2pt}]
\matrix (m) [ampersand replacement=\&,matrix of math nodes, row sep=4em,
column sep=6em, text height=1.5ex, text depth=0.25ex]
{ \Af(\Mb_i) \& {\displaystyle\bigvee_{i\in I} \Af(\Mb_i)} \&   \& \Af(\Mb)   \\
\Bf(\Mb_i) \&  {\displaystyle\bigvee_{i\in I} \Bf(\Mb_i)} \&  \Bf(\Mb)\& \Af(\Mb)  \\ };
\path[->]
(m-1-1)  edge  (m-1-2)
             edge node[left] {$ \zeta^{-1}_{\Mb_i} $} (m-2-1)
(m-1-2)  edge node[above] {$\bigvee_{i\in I} \Af(\psi_i)  $} (m-1-4)
             edge[dotted] node[left] {$ \xi $} (m-2-2)
(m-1-4)  edge node[left] {$ \id_{\Af(\Mb)} $} (m-2-4)
(m-2-1)  edge  (m-2-2)
(m-2-2)  edge node[above] {$ \bigvee_{i\in I} \Bf(\psi_i) $} (m-2-3)
(m-2-3)  edge node[above] {$ \zeta_\Mb $} (m-2-4);
\end{tikzpicture}
\]
%
%
in which the unlabelled morphisms are the canonical inclusions
associated with the join. Thus the two horizontal morphisms on the top
line compose to give $\Af(\psi_i)$, and the left two horizontal
morphisms on the bottom line compose to give $\Bf(\psi_i)$. 
The outer portion of the diagram therefore commutes because $\zeta$ is
natural and the universal property of the union induces a unique morphism $\xi$
such that every such diagram commutes in full. Considering the right-hand rectangle,
it is
evident that $\zeta_\Mb$ and $\bigvee_{i\in I}\Bf(\psi_i)$ have inverses
\begin{align*}
\zeta_\Mb^{-1} & = 
\left(\bigvee_{i\in I} \Bf(\psi_i)\right)\circ\xi\circ \left(\bigvee_{i\in I} \Af(\psi_i)\right)^{-1}
\\
\left(\bigvee_{i\in I} \Bf(\psi_i)\right)^{-1} &= \xi\circ \left(\bigvee_{i\in I}
\Af(\psi_i)\right)^{-1} \circ \zeta_\Mb;
\end{align*}
hence they are isomorphisms. 
$\square$

Both the previous results hold regardless of whether $\Af$ and $\Bf$ are
dynamically local (indeed, Lemma~\ref{lem:bigvee_iso} does not even use the timeslice property). 
Given the additional assumption we can use Lem.~\ref{lem:zeta_restrictions} to prove:
\begin{Prop}\label{prop:iso_inheritance} 
Suppose $\Af$ and $\Bf$ are dynamically local and
$\zeta:\Bf\nto\Af$. Suppose in addition that $\zeta_\Nb$ is an isomorphism
for some $\Nb$. Then $\zeta_\Mb$ is
an isomorphism for all $\Mb$ for which there is a morphism $\Mb\to\Nb$.
\end{Prop}
{\noindent\em Proof:} 
We have a diagram
\[
\begin{tikzpicture}[description/.style={fill=white,inner sep=2pt}]
\matrix (m) [ampersand replacement=\&,matrix of math nodes, row sep=4em,
column sep=6em, text height=1.5ex, text depth=0.25ex]
{ \Bf(\Mb) \& \&  \& \Af(\Mb)   \\
\& \Bf(\Nb) \&  \Af(\Nb)  \&    \\
\Bf^\dyn(\psi(\Mb);\Nb)  \&  \&  \& \Af^\dyn(\psi(\Mb);\Nb)
\\};
\path[->]
(m-1-1)  edge node[above] {$ \zeta_\Mb $} (m-1-4)
             edge node[above,sloped] {$ \Bf(\psi) $} (m-2-2)
             edge node[left] {$\cong$} (m-3-1)
(m-2-2)  edge node[above] {$ \zeta_\Nb $} (m-2-3)
(m-3-1)  edge node[above,sloped] {$ \beta^\dyn_{\Nb;\psi(\Mc)} $} (m-2-2)
             edge node[above] {$ \zeta_{\Nb;\psi(\Mb)} $} (m-3-4)
(m-1-4)  edge node[above,sloped] {$\Af(\psi) $} (m-2-3)
             edge node[left] {$ \cong $} (m-3-4)
(m-3-4)  edge node[above,sloped] {$ \alpha^\dyn_{\Nb;\psi(\Mc)} $} (m-2-3);
\end{tikzpicture}
\]
%
in which the two vertical isomorphisms arise because $\Bf(\psi)\cong \beta^\dyn_{\Nb;\psi(\Mc)}$ and $\Af(\psi)\cong \alpha^\dyn_{\Nb;\psi(\Mc)}$ by dynamical locality, whereupon the side triangles commute. The upper trapezium commutes
by naturality of $\zeta$ and the lower trapezium by Lemma~\ref{lem:zeta_restrictions}, which also
entails that $\zeta_{\Nb;\psi(\Mb)}$ is an
isomorphism. Thus the diagram commutes in full, implying that $\zeta_\Mb$ is an isomorphism
by commutativity of the outer rectangle. $\square$

In addition, we will use the following simple result (here dynamical locality is not
assumed):
\begin{Prop}  \label{prop:Cauchy_iso_transfer}
Suppose  $\zeta:\Bf\nto\Af$.
If $\psi:\Mb\to\Nb$ is a Cauchy morphism, then 
$\zeta_\Mb$ is an isomorphism if and only if $\zeta_\Nb$ is an isomorphism.
\end{Prop}
{\noindent\em Proof:} We have $\Af(\psi)\circ\zeta_\Mb =
\zeta_\Nb\circ\Bf(\psi)$, with $\Af(\psi)$ and $\Bf(\psi)$ isomorphisms.
If $\zeta_\Nb$ is an isomorphism then
$\Bf(\psi)^{-1}\circ\zeta_\Nb^{-1}\circ\Af(\psi)$ is inverse for
$\zeta_\Mb$, and hence $\zeta_\Mb$ is an isomorphism. Similarly 
$\Bf(\psi)\circ\zeta_\Mb^{-1}\circ\Af(\psi)^{-1}$ is inverse to
$\zeta_\Nb$ if $\zeta_\Mb$ is an isomorphism. $\square$

Given the above preparation, we may now state and prove our main result
of this section: namely that the dynamically local theories have the SPASs property.
\begin{Thm} \label{thm:SPASs}
Suppose $\Af$ and $\Bf$ dynamically local theories and $\zeta:\Bf\nto\Af$.
If $\zeta_\Mb$ is an isomorphism for some spacetime $\Mb$ then $\zeta$
is a natural isomorphism. 
\end{Thm}
{\noindent\em Proof:} 
Given that $\zeta_\Mb$ is an isomorphism,
Prop.~\ref{prop:iso_inheritance} entails that $\zeta_\Db$ is an
isomorphism for any multi-diamond spacetime $\Db\to\Mb$. Now let $\Db'$ be any
other multi-diamond spacetime with the same number of components as $\Db$; as 
$\Db$ and $\Db'$ have oriented-diffeomorphic Cauchy surfaces, they are linked by
a chain of Cauchy morphisms by  as shown in Prop.~\ref{prop:Cauchy_chain}.
Using Props.~\ref{prop:Cauchy_iso_transfer} and~\ref{prop:iso_inheritance}, we
may conclude that $\zeta_{\Db'}$ is also an isomorphism. As $\Mb$ contains multi-diamonds 
with any finite number of components, it follows that $\zeta_{\Db'}$ is an isomorphism for
every multi-diamond spacetime $\Db'$. 

Now let $\Mb'$ be an arbitrary spacetime; as $\Af$ is
dynamically local, we may deduce that $\zeta_{\Mb'}$ is an isomorphism 
using Theorem~\ref{thm:additivity}(b) and the remark thereafter,
in conjunction with Lemma~\ref{lem:bigvee_iso}. 
$\square$

Thus for any dynamically local theory $\Af$, there is no simpler dynamically local
theory that could account for the physics in any particular spacetime.
In this sense, dynamical locality therefore ensures that $\Af$ has
the same physical content in all spacetimes. Examples of the type presented
in Section~\ref{sect:pathologies} include cases where $\Af$ (resp., $\Bf$) 
is dynamically local, but $\Bf$ (resp., $\Af$) is not and where there is
a partial isomorphism $\Bf\to\Af$ that is not an isomorphism. Let us note
that much of the argument depends largely on the additivity property
(that is a consequence of dynamical locality). The exception is 
Prop.~\ref{prop:iso_inheritance}, where additivity seems to be insufficient,
and one requires the stronger dynamical locality assumption.

To conclude this section, we consider the consequences of dynamical locality for ordinary diagonal theories.
\begin{Thm} 
Suppose $\varphi_\Delta$ is an ordinary diagonal theory such
that $\varphi_\Delta$ and every $\varphi(\Mb)$ are dynamically local. Then
(a) for every morphism $\psi:\Mb\to\Nb$, $\varphi(\psi)$ is a natural 
isomorphism; (b) $\varphi_\Delta$ is gauge-equivalent to any
$\varphi(\Mb)$. In particular, if $\Aut(\varphi(\Mb))$ is trivial, then
$\varphi_\Delta$ is equivalent to each $\varphi(\Mb)$.
\end{Thm}
{\noindent\em Proof:} (a) We have
\[
\varphi_\Delta(\psi)\cong \alpha^{(\varphi_\Delta)\kin}_{\Nb;\psi(\Mb)}
\cong \alpha^{(\varphi_\Delta)\dyn}_{\Nb;\psi(\Mb)}
\cong \alpha^{(\varphi(\Nb))\dyn}_{\Nb;\psi(\Mb)}
\cong \alpha^{(\varphi(\Nb))\kin}_{\Nb;\psi(\Mb)}
\cong \varphi(\Nb)(\psi)
\]
using dynamical locality of $\varphi_\Delta$ and $\varphi(\Nb)$ and Thm.~\ref{thm:diagonal_int_net}
(expressed in subobject language). Hence $\varphi(\Nb)(\psi)\circ\varphi(\psi)_\Mb\cong \varphi(\Nb)(\psi)$
and as $\varphi(\Nb)(\psi)$ is monic, $\varphi(\psi)_\Mb$ is an
isomorphism. As both $\varphi(\Mb)$ and $\varphi(\Nb)$ are dynamically
local, Theorem~\ref{thm:SPASs} entails that $\varphi(\psi)$ is a natural 
isomorphism. 

(b) Writing $\Mb_0$ for Minkowski space, for each $\Mb$ we may choose a chain of morphisms
as in Prop.~\ref{prop:chain}
\[
\Mb_0\leftarrow \Mb_1 \rightarrow \Mb_2 \leftarrow \Mb_3 \rightarrow \Mb
\]
and use part (a) four times, composing the corresponding natural isomorphisms or their inverses, to obtain a natural isomorphism $\zeta_\Mb:\varphi(\Mb_0)\nto\varphi(\Mb)$. Then for each
$\psi:\Mb\to\Nb$, $\eta(\psi):=\zeta_\Nb^{-1}\circ\varphi(\psi)\circ \zeta_\Mb$ is
an automorphism of $\varphi(\Mb_0)$. It is obvious that
$\eta(\psi\circ\psi')=\eta(\psi)\circ\eta(\psi')$ and
$\eta(\id_{\Mb})=\id_{\varphi(\Mb_0)}$. Thus
$\eta\in\Funct(\Man,\Aut(\varphi(\Mb_0)))$ [with the automorphism group
regarded as a category] and we have
\begin{align*}
\varphi_\Delta(\psi)\circ(\zeta_\Mb)_\Mb &=
\varphi(\Nb)(\psi)\circ\varphi(\psi)\circ (\zeta_\Mb)_\Mb =
\varphi(\Nb)(\psi)\circ(\zeta_\Mb)_\Nb\circ \eta(\psi)_\Mb \\
&=(\zeta_\Nb)_\Nb\circ \varphi(\Mb_0)(\psi)\circ \eta(\psi)_\Mb.
\end{align*}
Thus the morphisms $(\zeta_\Mb)_\Mb$ form the components of a
natural transformation up to the twisting $\eta$.

Finally, if the automorphism group is trivial, $\eta(\psi)$ is an
identity for all $\psi$ and the $(\zeta_\Mb)_\Mb$ become components of a natural
isomorphism
$\hat{\zeta}:\varphi(\Mb_0)\nto\varphi_\Delta$.  $\square$

This result raises the interesting issue of how much freedom is available through choice of $\eta$, which can be regarded as a cohomological issue. If $\Aut(\varphi(\Mb_0))$ is
nontrivial, we can see that inequivalent diagonal theories can be constructed in the following
way. Label every homeomorphism equivalence class $[\Sigma]$ of compact connected Riemannian 
manifold by an element $g_{[\Sigma]}$ of $\Aut(\varphi(\Mb_0))$, and for each
morphism $\psi:\Mb\to\Nb$ in $\Man$ define $\eta(\psi)$ to be trivial except
in the case that $\Mb$ has noncompact Cauchy surfaces and $\Nb$ has compact Cauchy 
surface, in which case we set $\eta(\psi)=g_{[\Sigma(\Nb)]}$. It is clear that
this defines a functor into $\Aut(\varphi(\Mb_0))$.

\subsection{A no-go theorem for natural states}\label{sect:dlQFT}

To illustrate the significance of the dynamical locality assumption, 
we prove a model-independent no-go theorem for assignments of a natural choice of
preferred state of a QFT in all spacetimes.  This brings to sharper form an argument
sketched in BFV and \cite{Ho&Wa01} for the free scalar field; essentially
it shows that a preferred state is essentially incompatible with quantum field theory.

Unlike the results above, this result is specific to situations in which $\Phys$ is a category of $*$-algebras [including $\CAlg$]. We realise the subobjects $\Af^{\bullet/\rm dyn/kin}(\Mb;O)$ as subalgebras
of $\Af(\Mb)$ throughout. In this context, a state of the theory in spacetime $\Mb$ is a normalised
positive linear functional on the algebra $\Af(\Mb)$; the space of all states is denoted $\Af(\Mb)^*_{+,1}$. 
The result is stated in $\LCT$ but has an obvious analogue in $\LCTo$. 

\begin{Def} 
A {\em natural state} of a theory $\Af$ in $\LCT$ is an assignment $\Mand\owns\Mb\mapsto\omega_\Mb
\in \Af(\Mb)^*_{+,1}$ such that $\Af(\psi)^*\omega_\Nb=\omega_\Mb$ for all morphisms $\psi:\Mb\to\Nb$. 
\end{Def}

\begin{Thm}
Suppose $\Af$ is a dynamically local theory in $\LCT$, and has a natural state $(\omega_\Mb)_{\Mb\in\Mand}$. If there is a spacetime $\Mb$ with noncompact Cauchy surfaces such that $\omega_\Mb$ induces a 
faithful GNS representation with the Reeh--Schlieder property 
[i.e., the GNS vector corresponding to $\omega_\Mb$ is cyclic for the induced representation of 
$\Af(\Mb|_O)$ for all relatively compact $O\in\OO_0(\Mb)$],  then the relative Cauchy evolution is trivial in $\Mb$. If, additionally, $\Af$ obeys extended locality, then $\Af$ is equivalent to the trivial theory $\If$. 
\end{Thm}
{\noindent\em Proof:} Let $\Mb$ be as in the statement of the theorem. 
As the relative Cauchy evolution is a composition of (inverses of) morphisms $\Af(\psi)$, 
we have $\omega_\Mb\circ\rce_{\Mb}[\hb] = \omega_\Mb$ for each $\Mb$ and all $\hb\in H(\Mb)$. 
Consequently, in the GNS representation $\pi_\Mb$ induced by $\omega_\Mb$, the relative Cauchy evolution
may be unitarily implemented as 
\[
\pi_\Mb(\rce_\Mb[\hb] A) = U_\Mb[\hb]\pi_\Mb(A) U_\Mb[\hb]^{-1}
\]
for unitaries $U_\Mb[\hb]$ defined by
$U_\Mb[\hb]\pi_\Mb(A)\Omega_\Mb = \pi_\Mb(\rce_\Mb[\hb]A)\Omega_\Mb$, leaving the  GNS vector $\Omega_\Mb$ invariant.

Now let $\hb\in H(\Mb)$ and choose a nonempty relatively compact $O\in \OO_0(\Mb)$ such that $O\subset (\supp \hb)^\perp$ (here we use the noncompactness of the Cauchy surfaces). Then by 
Prop.~\ref{prop:rce_locality}  and  Lem.~\ref{lem:ext_M_O}, 
we have
\[
\rce_\Mb[\hb]\circ\alpha^\kin_{\Mb;O} = \alpha^\kin_{\Mb;O}
\]
and hence that
\[
U_\Mb[\hb]\pi_\Mb(\Af(\iota_{\Mb;O})A)\Omega_\Mb =\pi_\Mb(\Af(\iota_{\Mb;O})A)\Omega_\Mb 
\]
for all $A\in\Af(\Mb|_O)$. Using the Reeh--Schlieder assumption on $\omega_\Mb$ we may deduce that
$U_\Mb[\hb]$ agrees with the identity operator on a dense set and hence $U_\Mb[\hb]=\II_{\HH_\Mb}$ for all $\hb\in H(\Mb)$. 

As the representation $\pi_\Mb$ is assumed faithful, the relative Cauchy evolution is trivial on $\Af(\Mb)$
as claimed. Consequently, $\Af^{\bullet}(\Mb;K)=\Af(\Mb)$ for all compact sets $K$ and hence by dynamical locality $\Af^\kin(\Mb;O) = \Af^\dyn(\Mb;O)=\Af(\Mb)$ for each nonempty $O\in\OO(\Mb)$. 

Now consider two causally disjoint nonempty $O_1,O_2\in\OO(\Mb)$ (it suffices that they are each connected). It is clear that $\Af$ can obey extended locality only if $\Af(\Mb)=\CC\II_{\Af(\Mb)}$. (The same
would also be true if the $\Af^\kin(\Mb;O_i)$ are required to be algebraically independent:
otherwise we can find a linearly independent set $\{\II_{\Af(\Mb)},A\}$ common to the two algebras, whose list of products are of course linearly dependent.)

Thus the subtheory embedding $\If_\Af:\If\nto\Af$ is an isomorphism in spacetime $\Mb$. As both $\If$ and $\Af$ are assumed dynamically local, it follows from 
Theorem~\ref{thm:SPASs} that $\If_\Af$ is a natural isomorphism.
$\square$

We remark that the assumption of commutation at spacelike separation (in place of
extended locality) results in 
$\Af(\Mb)$ being abelian, from which we can deduce that $\Af(\Nb)$ is abelian if $\Nb$ is any truncated multi-diamond spacetime, or any spacetime in which the truncated multi-diamonds form a directed net.

\section{Conclusion}

We conclude with a brief discussion of further work and related approaches. 
First, now that the basic framework has been established, it is necessary to 
show that familiar models satisfy dynamical locality. As already mentioned, 
we show in~\cite{FewVer:dynloc2} that the minimally coupled
free scalar field is dynamically local for nonzero mass, and that the failure of dynamical locality
at zero mass is understood as an expression of the gauge symmetry. Once this is taken into account
the massless theory is again dynamically local, with the single exception of the two-dimensional
theory on $\Mand$. Work is under way on other models, including the algebra of Wick products.

Second, we again emphasise that we do not expect that 
the two principles S1 and S2 described in the 
introduction completely characterise what a notion of SPASs should be. 
For example, it is conceivable that there are (as yet unknown) dynamically local theories
that one might not wish to regard as representing the same physics
in all spacetimes; in that case, it would be clear that S1 and S2
are insufficient and that further conditions should be imposed. Furthermore our
discussion is conducted for the most part at the level of local
observables. Even in the algebraic approach to quantum
field theory in curved spacetimes there are several levels of
description and the present work addresses only those aspects that are
independent of choices of state spaces which can bring in properties
deriving from the global structure of spacetime. In addition, it is possible
that the formulation of dynamical locality can be refined further. For example, 
one might base the theory on the requirement that $\alpha^\bullet_{\Mb;K}$
should be isomorphic to any intersection $\bigwedge_n \alpha^\kin_{\Mb;O_n}$ where $O_n$ is
an outer approximating sequence to $K$.

Finally, we conclude with some remarks that may help to clarify
the relation of the present work to other approaches 
studying the interplay of covariance, locality and dynamics in
abstract (operator-algebraic) quantum field theory. If our
setting is specialized to the case that $\Phys$ is $\CAlg$,
the category of unital $C^*$ -algebras, then our discussion remains
purely at the $C^*$-algebraic level, in that we do not discuss special
classes of states or their GNS representations, from which, in a next step, the
$C^*$-algebraic setting would be taken to the von Neumann-algebraic level.
This step, together with the analysis of distinguished states and their induced 
representations, is one of the central issues in the model-independent approach
to quantum field theory, as is laid out in \cite{Haag}, and other work devoted to
the relations between covariance, locality and dynamics is mostly tied to
distinguished states, often the vacuum state in Minkowski spacetime.
Some authors have attempted to derive a concept of dynamical localization
of observables for quantum field theory in Minkowski spacetime, making use
of the properties of the vacuum representation \cite{Landau:1974,Kuckert:2000}; however this
concept of dynamical localization is different from ours. Another major theme in
operator-algebraic quantum field theory is the concept of ``geometric modular action''
\cite{BDFS:2000} which has at its roots the famous Bisognano--Wichmann theorem
(see \cite{Haag}, and references cited there). This theorem says that the Tomita--Takesaki
modular objects corresponding to von Neumann algebras of observables localized in
special regions, and to the vacuum vector, carry geometrical significance. In fact,
in some situations one can gain the full local net structure and covariance group
from such modular objects \cite{BDFS:2000,Wiesbrock:1998}. This is of interest as the modular
objects also encode dynamical information \cite{Haag}, and in some works,
this dynamical information has been related to concepts of locality and covariance
\cite{CoRo:1994,BruGuiLo:2002,BuMuSu:2002,BuLe:2004}. While these cited works are not directly related
to the approach taken in the present article, they also focus on the relation between
covariance, locality and dynamics. Closer connections between the cited works
and the present article may possibly be revealed once our setup can suitably be
extended at the von Neumann algebraic level, incorporating distinguished classes of
states.

\vspace{0.5cm}

{\small\noindent CJF thanks the Research Academy Leipzig for financial support
and the Institute for Theoretical Physics, University of Leipzig, for
hospitality at various stages of this work. In addition, CJF thanks the Insitute for Theoretical Physics, University of G\"ottingen for hospitality and the participants and organisers of the meetings
AQFT50 (G\"ottingen, July--August 2009) and Quantum Field Theory and Gravity (Regensburg, September--October 2010) at which preliminary versions of this work were presented. 
CJF also thanks Henning Bostelmann, Matthew Ferguson, Klaus Fredenhagen and Ko Sanders for useful conversations at various stages during this work. RV thanks the Department of Mathematics,  University of York, for hospitality and financial support.}

\appendix

\section{Geometrical lemmas} \label{appx:geom}

\subsection{Cauchy morphisms}

\begin{Prop}\label{prop:embeddings_appx} 
Suppose $\Mb\in\Man$ admits a compact Cauchy surface $\Sigma$.
If $\Mb\stackrel{\psi}{\to}\Nb$ in $\Man$ then $\psi$ is Cauchy.
\end{Prop}
{\noindent\em Proof.} Using homeomorphism equivalence of Cauchy surfaces in $\Mb$~\cite[Cor.~14.32]{ONeill} and \cite[Thm.~1.1]{Bernal:2003jb}, we may assume without loss of generality that $\Sigma$ is a smooth spacelike Cauchy surface, which is connected~\cite[Prop.~14.31]{ONeill}, compact and embedded in $\Mb$. As $\psi$ is an isometric embedding, $\psi(\Sigma)$ is (in particular) a smoothly immersed spacelike submanifold of $\Nb$ that is also compact and connected as a result of the properties of $\Sigma$. Theorem 1 of \cite{BILY} then entails that $\psi(\Sigma)$ is an acausal Cauchy surface\footnote{Note that `acausal' is included
in the definition of Cauchy surface in \cite{BILY}.} of $\Nb$, so $\psi$ is Cauchy.
$\square$

\begin{Lem}\label{lem:Cauchy_image}
Suppose $\psi:\Mb\to\Nb$ is Cauchy. Then if $\Sigma$
is any Cauchy surface of $\Mb$, $\psi(\Sigma)$ is a Cauchy surface of $\Nb$.
\end{Lem}
{\noindent\em Proof:} 
Any inextendible timelike curve $\gamma:\RR\to\Nb$ in $\Nb$
enters $\psi(\Mb)$, and $I = \gamma^{-1}(\psi(\Mb))$ is open and 
connected by causal convexity of the embedding. 
We therefore obtain a timelike curve
$\hat{\gamma}:I\to\Mb$ so that $\psi\circ\hat{\gamma} = \gamma|_I$.
Now $\hat{\gamma}$ has no endpoint in $\Mb$ and is therefore inextendible; accordingly it intersects $\Sigma$ exactly once. Hence $\gamma|_I$ intersects 
$\psi(\Sigma)$ exactly once and so the same is true of $\gamma$.
$\square$

\begin{Lem} \label{lem:Cauchy_comp}
The composite of Cauchy morphisms is Cauchy.
\end{Lem}
{\noindent\em Proof:} If $\varphi:\Lb\to\Mb$ and $\psi:\Mb\to\Nb$ are
Cauchy and $\Sigma$ is a Cauchy surface of $\Lb$, then we apply Lem.~\ref{lem:Cauchy_image} 
successively, to show that
$\varphi(\Sigma)$ is a Cauchy surface of $\Mb$ and hence $(\psi\circ\varphi)(\Sigma)$ is a Cauchy surface of $\Nb$. Hence $\psi\circ\varphi$ is Cauchy.
$\square$

\medskip
We now give two proofs deferred from section~\ref{sect:spacetimes}.

{\noindent\em Proof of Prop.~\ref{prop:Cauchy}:} We are given a Cauchy morphism $\psi:\Mb\to\Nb$
in $\Man$ or $\Mand$, and must prove that $\psi(\Mb)$ contains a smooth, spacelike and acausal Cauchy surface for $\Nb$ and that the smooth spacelike Cauchy surfaces of $\Mb$ and $\Nb$ are oriented-diffeomorphic. 

By virtue of \cite[Thm 1.1]{Bernal:2003jb} $\Mb$ has a smooth spacelike Cauchy surface $\Sigma$;  Lem.~\ref{lem:Cauchy_image} shows that the smooth spacelike surface $\Sigma'=\psi(\Sigma)$ is a Cauchy surface for $\Nb$, and is therefore
acausal by \cite[Lem. 14.42]{ONeill}. Putting $\Mb$ and $\Nb$ into normal form, we may construct oriented-diffeomorphisms 
$\rho:\RR\times\Sigma\to\Mb$, $\rho':\RR\times\Sigma'\to\Nb$ (with the canonical orientations and other properties discussed in section~\ref{sect:spacetimes}), thus giving a smooth map 
$\Psi=\pr_{\Sigma'}\circ(\rho')^{-1}\circ \psi\circ \rho_0:\Sigma\to\Sigma'$, where $\pr_{\Sigma'}$ is the projection onto $\Sigma'$ and $\rho_0(\cdot)=\rho(0,\cdot)$. 

Now $\Psi$ is an immersion (and hence also a submersion) because the kernel of $(\pr_{\Sigma'}\circ(\rho')^{-1})_*$ is timelike while the image of $(\psi\circ \rho_0)_*$ is spacelike;
it is also injective ($\pr_{\Sigma'}\circ(\rho')^{-1}$ identifies points only if they are connected by a timelike curve, while $\rho_0(\Sigma)$ is achronal and $\psi(\Mb)$ is causally convex) and surjective 
(by definition of $\Sigma'=\psi(\Sigma)$ and because $\pr_{\Sigma'}\circ(\rho')^{-1}\circ\rho'_0=\id_{\Sigma'}$).
Accordingly, $\Psi$ is a diffeomorphism (see, e.g., \cite[Thm 7.15]{Lee:intro_to_smooth_manifolds}) that preserves orientations because
$\psi$ preserves orientation and time-orientation. In particular, $\Psi$ is a homeomorphism
and so all Cauchy surfaces of $\Mb$ and $\Nb$ are homeomorphic. 
$\square$
 
\medskip
{\noindent\em Proof of~\ref{prop:Cauchy_chain} (converse):}  Our argument is a slight elaboration
and variant of that in \cite{FullingNarcowichWald} in order to incorporate detail on orientations. We
also take the opportunity to simplify the argument slightly, while also being more specific on some details. We suppose $\Mb$ and $\Nb$ have
oriented-diffeomorphic smooth spacelike Cauchy surfaces $\Sigma$ and $\Sigma'$ with canonical orientations $\wgth$ and $\wgth'$. Using any
oriented-diffeomorphism between $\Sigma$ and $\Sigma'$ we may put both $\Mb$ and $\Nb$
into normal form on $\RR\times\Sigma$ equipped with the orientation $dt\wedge\wgth$ by
means of oriented-diffeomorphisms $\rho_\Mb:\RR\times\Sigma\to\Mb$ and $\rho_\Nb:\RR\times\Sigma\to\Nb$. The two pulled back metrics on $\RR\times\Sigma$ may be written as
\[
\rho_\Mb^*\gb_\Mb = \beta_\Mb dt\otimes dt -\hb_t, \qquad
\rho_\Nb^*\gb_\Nb =\beta_\Nb dt\otimes dt - \kb_t
\]
where $\beta_\Mb,\beta_\Nb\in C^\infty(\RR\times\Sigma)$ are strictly positive and $\hb_t$ and $\kb_t$
are smooth Riemannian metrics on $\Sigma$ depending smoothly on $t$. One may find smooth positive functions $K,H\in C^\infty(\RR\times\Sigma)$ such that  $\kb_{t,\sigma}\ge K(t,\sigma) \hb_{t,\sigma}$
and $\hb_{t,\sigma}\ge H(t,\sigma) \kb_{t,\sigma}$ as 
quadratic forms.\footnote{E.g., use $K=[(\hb_t)^{i}_{\phantom{i}j}(\hb_t)^{j}_{\phantom{j}i}]^{-1/2}$ with $\kb_t$ used to raise indices, and the analogous expression for $H$.} Fixing $t_0>0$, let $F=(t_0,\infty)\times\Sigma$ and $P=(-\infty,-t_0)\times\Sigma$ and
choose any nonnegative $\chi\in C^\infty(\RR)$ such that $\chi$ equals unity on $F$ and vanishes on $P$. Construct a metric 
\[
\gb= \beta dt\otimes dt - (\chi \hb_t +(1-\chi) \kb_t)
\]
where $\beta$ is chosen to be a smooth positive function such that 
\[
\beta\le (\chi + (1-\chi)K)\beta_\Mb 
\]
on $t>-\frac{1}{2}t_0$, with equality for $t\ge t_0$, and
\[
\beta\le (1-\chi + \chi H)\beta_\Nb 
\]
on $t< \frac{1}{2}t_0$, with equality for $t\le -t_0$. Then it is easily seen that every $\gb$-causal curve 
is $\rho_\Mb^*\gb_\Mb$-causal in $(-\frac{1}{2}t_0,\infty)\times\Sigma$ and $\rho_\Nb^*\gb_\Nb$-causal in $(-\infty,\frac{1}{2} t_0)\times \Sigma$. But these metrics are globally hyperbolic, so every inextendible $\gb$-timelike curve intersects each $\{t\}\times\Sigma$ surface exactly once. Accordingly, 
$\RR\times\Sigma$, with the metric $\gb$, orientation $dt\wedge\wgth$ and time-orientation so that $\partial/\partial t$ is future-pointing, is a globally hyperbolic spacetime in $\Mand$ (or $\Man$
as appropriate), which we denote $\Ib$. The metric $\gb$ clearly coincides with $\rho_\Mb^*\gb_\Mb$
in $F$ and with $\rho_\Nb^*\gb_\Nb$ on $P$.

Finally, the regions $F$ and $P$ are open
globally hyperbolic subsets of $\Ib$ containing Cauchy surfaces of $\Ib$ and their images 
$\rho_\Mb(F)$ and $\rho_\Nb(P)$ evidently contain Cauchy surfaces for $\Mb$ and $\Nb$. Setting $\Fb=\Ib|_F$ and $\Pb=\Ib|_P$, we then have a diagram of the form \eqref{eq:Cauchy_chain}
\[
\Mb\leftarrow \Fb \rightarrow \Ib \leftarrow \Pb \rightarrow \Nb,
\]
with the canonical inclusions $\iota_{\Ib;F}$ and $\iota_{\Ib;P}$ providing the
inner Cauchy morphisms and the restrictions $\rho_\Mb|_F$ and $\rho_\Nb|_P$ as the outer two
Cauchy morphisms. $\square$. 

\subsection{Covariance of hyperbolic perturbations}\label{sect:cov_hyp}

Next, we turn to a number of results used in the discussion of relative Cauchy 
evolution in Sect.~\ref{sect:rce}. We recall that the chronological future($+$)/past($-$) $I^\pm_\Mb(p)$
of $p$ consists of all points (excluding $p$) that can be reached from $p$ along a future/past-directed piecewise smooth timelike curve in $\Mb$; by smoothing results such as \cite[Prop.~2.23]{Penrose1972} we obtain the same set if we only admit smooth timelike curves (which may even be chosen to be 
geodesic near their endpoints). Similarly, the causal future/past $J_\Mb^\pm(p)$ consists of all points (including $p$) that can be reached from $p$ by 
future/past directed piecewise smooth (or, equivalently, smooth) causal curves. Note that any causal curve is confined to a single connected component of the spacetime. For a subset $S\subset\Mb$ we define $J^\pm_\Mb(S) =  \bigcup_{p\in S} J^\pm_\Mb(p)$ etc. Extensive use will be made of the fact 
that globally hyperbolic spacetimes are causally simple: for every compact set
$K$, the sets $J_\Mb^\pm(K)$ are closed (see, e.g., Prop.~6.6.1 in \cite{HawkingEllis};
 Theorem 8.3.11 in~\cite{Wald_gr}). 

For any subset $S\subset\Mb$ we define the future($+$)/past($-$) Cauchy development $D^\pm(S)$ of $S$ to be the set of points $p$ such that every past/future-inextendible piecewise smooth causal curve through $p$ intersects $S$; $D_\Mb(S)=D_\Mb^+(S)\cup D_\Mb^-(S)$. If $S$ is either achronal or closed, we may replace `piecewise smooth' by `smooth' without loss, but more generally, this can result in a different set. 

{\noindent\em Proof of Lemma~\ref{lem:geom1}:} If $\Mb$ is connected, this is immediate from the special case $K=\supp \hb$ of the following result, Lemma~\ref{lem:pre_geom1}. If $\Mb$ has
more than one connected component, the result follows by applying Lemma~\ref{lem:pre_geom1} to each component. $\square$
\begin{Lem} \label{lem:pre_geom1}
Let $K$ be a compact subset of the underlying manifold $\Mc$ of $\Mb\in\Man$ and define
$\Mc^\pm = \Mc\setminus
J_\Mb^\mp(K)$. Then $\Mc^\pm$ are open, connected, globally hyperbolic subsets
of $\Mb[\hb]$ for any $\hb\in H(\Mb;K)$. Moreover, $\Mb|_{\Mc^\pm} =
\Mb[\hb]|_{\Mc^\pm}$ and the canonical inclusions
$\Mb[\hb]|_{\Mc^\pm}\to\Mb[\hb]$ are Cauchy morphisms. 
\end{Lem}
{\noindent\em Remark:} We do not assume that $K$ is connected.\\
{\noindent\em Proof:} As $K$ is compact, $J_\Mb^\pm(K)$ are
closed so $\Mc^\pm$ are open. We now claim that
$J_{\Mb[\hb]}^\pm(K)=J_{\Mb}^\pm(K)$ for any $h\in H(\Mb;K)$. To show
this (for the ($+$) case), take any $q\in\Mc\setminus K$ with $q\in J_{\Mb[\hb]}^+(K)$.
Then there is a future-directed $\Mb[\hb]$-causal curve $\gamma:[0,1]\to\Mc$ with
$\gamma(0)\in K$ and $\gamma(1)=q$. Defining $\tau_*=\sup
\gamma^{-1}(K)$ we have $\tau_*<1$ and $\gamma(\tau_*)\in K$. As $\hb$
is supported in $K$, the curve $\gamma|_{[\tau_*,1]}$ is also
$\Mb[\hb']$-causal for any $\hb'\in H(\Mb;K)$, so $q\in
J_{\Mb[\hb']}^+(K)$. Thus $J_{\Mb[\hb]}^+(K)\subset J_{\Mb[\hb']}^+(K)$;
reversing the roles of $\hb$ and $\hb'$ the two sets are therefore
equal for arbitrary $\hb,\hb'\in H(\Mb;K)$. Setting
$\hb'=\Ob$ the ($+$)-case of the claim is established; the ($-$)-case is analogous.  

To establish global hyperbolicity, take $p,q\in\Mc^-$ and $\gamma$ a future-directed
$\Mb[\hb]$-causal curve from $p$ to $q$. If $\gamma$ leaves $\Mc^-$ then
it contains a point of $J^+_{\Mb[\hb]}(K)$; hence $q\in
J^+_{\Mb[\hb]}(K)$, which is a contradiction. Thus $\gamma$ is contained
within $\Mc$, as required.

Connectedness is proved as follows. Take any $p,q\in\Mc^-$; then we
may find a $\Mb$-Cauchy surface $\Sigma$ that is contained in $\Mc^-$ and lies
to the past of both $p,q$.\footnote{\label{fn:ctf_arg} Let $\mathcal{T}:\Mc\to\RR$ be a Cauchy temporal function for $\Mb$, which exists by \cite[Thm 1.1]{Bernal:2004gm};
then $\mathcal{T}(K)$ is compact and hence $\mathcal{T}(J_\Mb^{+}(K))\subset [\tau,\infty)$
for some $\tau\in\RR$, which, without loss of generality may be chosen so that 
$\tau<\min\{\mathcal{T}(p),\mathcal{T}(q),0\}$. Then $\Sigma=\mathcal{T}^{-1}(\{2\tau\})$ 
meets the requirements; by choice of $\mathcal{T}$ we may
additionally arrange that $\Sigma$ be spacelike.} As there are past-directed causal curves joining
each of $p$ and $q$ to $\Sigma$, which is path-connected,\footnote{It is connected
\cite[Prop.~14.31]{ONeill} and therefore path-connected, because it is a topological manifold.}
we conclude that $p$ is path-connected to $q$. As $p,q$ are arbitrary, we
deduce that $\Mc^-$ is path-connected and hence connected. 

Finally, as $\Mc^\pm$ contain $\Mb[\hb]$-Cauchy surfaces (using a
similar argument to that in footnote~\ref{fn:ctf_arg}), the canonical
inclusions of $\Mb[\hb]|_{\Mc^\pm}\to\Mb[\hb]$ are Cauchy morphisms.
Moreover, $\Mb[\hb]|_{\Mc^\pm}=
(\Mc^\pm,\gb|_{\Mc^\pm},\ogth|_{\Mc^\pm},\tgth|_{\Mc^\pm}) 
=\Mb|_{\Mc^\pm}$ for all $\hb\in H(\Mb)$. 
$\square$

{\noindent\em Proof of Lemma~\ref{lem:geom2}:} 
We prove the ($+$) case, thus supposing that the range of $\psi$ is contained in $\Mc\setminus J^-_\Mb(K)$. Then we have $\psi^*\hb=\Ob$ for
$\hb\in H(\Mb;K)$ and it follows straightforwardly that the
underlying embedding of $\psi$ induces 
$\psi[\hb]:\Lb\to\Mb[\hb]$.
By Lem.~\ref{lem:geom1} the set $\Mc^+=\Mc\setminus J_\Mb^-(\supp\hb)$ is a 
globally hyperbolic subset of $\Mb$ and $\Mb[\hb]$; as $\psi(\Lb)\subset \Mc^+\subset
\Mc$, the morphisms $\psi$ and $\psi[\hb]$ factor via the inclusion
morphisms $\imath^+_\Mb[\hb]:\Mb^+[\hb]\to\Mb$ and $\jmath^+_\Mb[\hb]:\Mb^+[\hb]\to\Mb[\hb]$ respectively,
i.e., 
\[
\psi=\imath_\Mb^+[\hb]\circ\varphi^+, \qquad \psi[\hb]=\jmath_\Mb^+[\hb]\circ\varphi^+.
\]
for $\varphi^+:\Lb\to\Mb^+[\hb]$. 

If $\psi$ is Cauchy then $\psi(\Lb)$ contains a Cauchy surface for $\Mb$
and hence $\Mb^+[\hb]$ (as $\psi(\Lb)\subset\Mc^+$). Thus
$\varphi^+$ is Cauchy. As $\jmath_\Mb^+[\hb]$ is Cauchy and the
composite of Cauchy morphisms is Cauchy, it follows that $\psi[\hb]$ is also
Cauchy.  
$\square$

The next task is to prove that the push-forward
of a globally hyperbolic perturbation under a $\Mand$ (or $\Man$) morphism is again a
globally hyperbolic perturbation (Lemma~\ref{lem:H_functorial} below).
This is broken into steps as follows. 

\begin{Lem} \label{lem:gh1}
Suppose that $K$ is a compact subset of a globally
hyperbolic spacetime $\Mb=(\Mc,\gb,\ogth,\tgth)\in\Mand$ and that $\gamma:I\to\Mc$ is an
inextendible future-directed $\Mb$-timelike curve, where $I$ is an open interval of $\RR$. Then $\gamma^{-1}(K)$ is bounded.
\end{Lem} 
{\noindent\em Proof:} Choose a Cauchy temporal function $\mathcal{T}$ on $\Mb$; 
then $\mathcal{T}(K)$ is compact and contained in some interval $(\tau^-,\tau^+)$. Then
$\Sigma^\pm = \mathcal{T}^{-1}(\tau^\pm)$
are Cauchy surfaces of $\Mb$ to the past ($-$) and future ($+$) of $K$, i.e., $J_\Mb^\pm(\Sigma^\pm)\cap K=\emptyset$. As it is inextendible,
$\gamma$ intersects $\Sigma^\pm$ at unique $t^\pm\in I$ and it
is clear that $\gamma^{-1}(K)\subset (t^+,t^-)$ because 
$\gamma(t)$ lies in $J_\Mb^-(\Sigma^-)$ for $t<t^-$ (resp., $J_\Mb^+(\Sigma^+)$ for $t>t^+$) and does not intersect $K$ in this interval. $\square$

\begin{Lem} \label{lem:gh2}
Suppose $\Mb=(\Mc,\gb,\ogth,\tgth)\in\Mand$ and let $K$ be
a compact subset of $\Mc$ contained in an open $\Mb$-causally convex subset
$U$ that has at most finitely many connected components. 
Let $\Sigma$ be a Cauchy surface of $\Mb$ to the past of $K$, i.e., $K\subset I^+_\Mb(\Sigma)$. 
Suppose $\gb'$ is a time-orientable Lorentz metric on $\Mc$ with time-orientation
$\tgth'$ such that $\gb'=\gb$, $\tgth'=\tgth$ outside $K$, and so that
$\Ub=(U,\gb'|_U,\ogth|_U,\tgth'|_U)\in\Mand$. Then:
\begin{enumerate}\addtolength{\itemsep}{-0.5\baselineskip}
\item[(i)] if $\gamma$ is a $(\gb',\tgth')$-causal curve in $\Mc$ with endpoints in $U$ then
$\gamma$ is contained in $U$;
\item[(ii)] if $\gamma:\RR\to\Mc$ is an inextendible $(\gb',\tgth')$-timelike curve
intersecting $K$ then
$\gamma^{-1}(U)$ is an open interval and $\gamma^{-1}(K)$ is bounded;
\item[(iii)] any inextendible $\gb'$-timelike curve $\gamma:\RR\to\Mc$
intersects $\Sigma$ exactly once;
\item[(iv)] the spacetime $\Mb'=(\Mc,\gb',\ogth,\tgth')$ is globally
hyperbolic, i.e., $\Mb'\in\Mand$. 
\end{enumerate}
\end{Lem}
{\noindent\em Proof:} 
(i) Suppose $\gamma:[0,1]\to\Mc$ is $(\gb',\tgth')$-causal with
$\gamma(0),\gamma(1)\in U$, but $\gamma(t)\not\in U$ for some
$t\in(0,1)$. Then there are $t_0,t_1$ with $0<t_0<t<t_1<1$ such that
$\gamma(t_0),\gamma(t_1)\in U$ but $\gamma|_{[t_0,t_1]}$ does not
intersect $K$. Hence $\gamma|_{[t_0,t_1]}$ is $\Mb$-causal and therefore
contained in $U$ by causal convexity. This is a contradiction.\\
(ii) An immediate corollary of (i) is that $I=\gamma^{-1}(U)$ is an open
convex subset of $\RR$, i.e., an open interval. Now the restriction of $\gamma$ to $I$ is an
inextendible future-directed timelike curve in the globally hyperbolic spacetime $\Ub$.
Applying Lemma~\ref{lem:gh1}, we find that $(\gamma|_I)^{-1}(K)=\gamma^{-1}(K)$ is
bounded.\\
(iii) If $\gamma$ does not intersect the interior of $K$, it is also $\Mb$-timelike
and therefore intersects $\Sigma$ exactly once. If $\gamma$ does intersect ${\rm
int}(K)\subset U$, then $\gamma^{-1}(K)$ is bounded from below by (ii)
and $t_0=\inf \gamma^{-1}(K)$ is finite. 
As any portion of $\gamma$ outside $K$ is $\Mb$-timelike and
future-directed, we have
\[
\gamma(t)\in J_\Mb^+(\gamma(\sup\{t'\le t:~\gamma(t')\in K\})) \subset
J_\Mb^+(K)
\]
for any $t>t_0$. Thus $\gamma|_{(t_0,\infty)}$ does not intersect
$\Sigma$, while the past-inextendible portion $\gamma|_{(-\infty,t_0]}$
intersects $\Sigma$ exactly once because $\gamma(t_0)\in K$ lies to
the future of $\Sigma$.\\
(iv) It follows immediately from (iii) that $\Sigma$ is a Cauchy
surface for the spacetime $(\Mc,\gb',\ogth',\tgth')$, which is therefore globally hyperbolic.
$\square$

We can now prove the covariance property of globally hyperbolic perturbations. 

\begin{Lem} \label{lem:H_functorial}
Suppose $\psi:\Lb\to\Mb$ in $\Mand$. Then $\psi_*(H(\Lb))\subset H(\Mb)$.
(In particular, this applies to all morphisms in $\Man$.)
\end{Lem} 
{\noindent\em Proof:} Write $\Mb=(\Mc,\gb,\ogth,\tgth)$,
$\Lb=(\Lc,\psi^*\gb,\psi^*\ogth,\psi^*\tgth)$ and $U=\psi(\Lc)$,
$K=\psi(\supp\hb)$, where $\hb\in H(\Lb)$. 
Then $\gb'=\gb+\psi_*\hb$ is a Lorentz metric on $\Mc$. To
show that it is time-orientable, let $T_1$ (resp., $T_2$) be a
$\Lb[\hb]$-timelike (resp., $\Mb$-timelike) nowhere zero, future-pointing vector field on
$\Lc$ (resp., $\Mc$). Let $\chi\in\CoinX{\Mc}$ be nonnegative, with
$\chi=1$ on $K$ and $\chi=0$ outside $U$. Then $\chi\psi_*T_1+(1-\chi)T_2$ is nowhere zero
and $\gb'$-timelike, and therefore defines a time-orientation $\tgth'$ of
$\gb'$ that agrees with $\tgth$ outside $\psi(K)$. As there exist
$\Mb$-Cauchy surfaces to the past of $K$,
 Lem.~\ref{lem:gh2}(iv) entails
that $(\Mc,\gb',\ogth,\tgth')\in\Mand$, i.e., $\psi_*\hb\in H(\Nb)$. 
$\square$

\subsection{Causal complements and (multi-)diamonds}

Finally, we give a number of results relating to causal structure and multi-diamonds
in globally hyperbolic spacetimes. 
Similar results appear elsewhere (e.g., \cite[Appx B]{BrunettiRuzzi_topsect}, \cite[\S 2]{Br&Ru05},
\cite[\S 3]{Ruzzi:2005}) but we are not aware of a full presentation of all the results
needed in the body of this paper. Notation and terminology varies in the literature and the definition of causal complement is not always made clear (the cited references are exceptions to this). 
It is hoped that this appendix may be useful more widely. 

Recall that we have two notions of causal complement in a globally
hyperbolic spacetime $\Mb$:
$O^\perp=\Mc\setminus J_\Mb(O)$ and $O'=\Mc\setminus \cl(J_\Mb(O))$. 
Clearly $O'$ is always an open set. 

\begin{Lem} \label{lem:Jofopenisopen}
Let $O$ be an open subset of a globally hyperbolic spacetime
$\Mb$. Then $J_\Mb^\pm(O)$ are open, and 
$O\subset O''$.
\end{Lem}
{\noindent\em Proof:} If $q\in J_\Mb^\pm(O)$
then $O$ has nontrivial intersection with the (closed) set $J_\Mb^\mp(q)$, which is
the closure of $I_\Mb^\mp(q)$ as $\Mb$ is globally hyperbolic 
(\cite{ONeill}, Lem.~14.6). Thus $O$ intersects  $I_\Mb^\mp(q)$, so $q\in I_\Mb^\pm(p')$ for some $p'\in
O$, and we have shown that $J_\Mb^\pm(O)\subset
I_\Mb^\pm(O)={\rm int}(J_\Mb^\pm(O))$. As $O'\cap J_\Mb(O)$ is empty,
so is  $J_\Mb(O')\cap O$ and hence $\cl(J_\Mb(O'))\cap O$.
Thus $O\subset O''$. $\square$

\begin{Lem} \label{lem:causal_completeness} 
Let $\Sigma$ be an acausal Cauchy surface in globally hyperbolic spacetime $\Mb$ and let 
$S$ be an open subset of $\Sigma$ such that 
$\cl S$ has nontrivial complement in $\Sigma$.  Then 
$S''=D_\Mb(S)=D_\Mb(S)''$. In particular, 
every multi-diamond $O$ is causally complete in the sense that $O=O''$. 
\end{Lem}
{\noindent\em Proof:} First, using $D_\Mb(S)\subset J_\Mb(S)$, observe that
\[
D_\Mb(S)' = \Mc\setminus \cl J_\Mb(D_\Mb(S)) = \Mc\setminus \cl J_\Mb(S) = S'
\]
and hence $S''=D_\Mb(S)''$ (this holds for any subset $S$ of $\Mb$; 
similarly, we also have $D_\Mb(S)^\perp = S^\perp$ and thus $S^{\perp\perp} = 
D_\Mb(S)^{\perp\perp}$ for any subset $S$); it remains to show that $D_\Mb(S)$ is causally complete. 

As $S$ is open in $\Sigma$, it inherits the property of being an acausal topological
hypersurface~\cite[14.24]{ONeill} from $\Sigma$; accordingly $O=D_\Mb(S)$ is an open
subset of $\Mb$~\cite[14.42]{ONeill}. We observe that $\Sigma\setminus\cl(S)\subset
O'$; if not, then we may find $q\in\Sigma\setminus\cl(S)$
and $q_n\to q$ with $q_n\in J_\Mb(O)$. Choose an open neighbourhood $U$ of
$q$ in $\Sigma$ that does not intersect $\cl(S)$, then $D_\Mb(U)$ is
an open neighbourhood of $q$ that contains $q_n$ for sufficiently large
$n$. But $q_n\in J_\Mb(O)=J_\Mb(S)$ contradicts $q_n\in D_\Mb(U)$. 

To establish causal completeness it suffices to show $O''\subset
O$.  If $p\notin O$ there is an inextendible causal curve through $p$ intersecting
$\Sigma$ at $q\notin S$. Assume without loss that $p\in J_\Mb^+(q)$.
Then there are points $p_n\to p$ with
$p_n\in I_\Mb^+(q)$ and hence neighbourhoods $U_n$ of $q$ with
$U_n\subset I_\Mb^-(p_n)$. Each $U_n$ must intersect
$\Sigma\setminus\cl(S)$ nontrivially, so $p_n\in
J_\Mb^+(\Sigma\setminus\cl(S))\subset J_\Mb(O')$. 
Hence $p\in\cl(J_\Mb(O'))$ i.e., $p\notin O''$.
Thus $O''\subset O$, so $O''=O$. 

If $O$ is a multi-diamond then $O=D_\Mb(S)$ where $S$ meets the above
hypotheses; hence $O=O''$. 
$\square$

\begin{Lem} \label{lem:perp_of_compacts}
Suppose $K$ is a compact subset of globally hyperbolic spacetime $\Mb$. Then (i) $K^\perp$ is open,
$K^{\perp\perp}$ is closed, and $K\subset K^{\perp\perp}$; (ii)
$K^{\perp\perp\perp}=K^{\perp}$ and $K^{\perp\perp}$ is causally
complete with respect to $\perp$; (iii) $K^{\perp\perp}$ is causally
convex. If, in addition, $K$ has a multi-diamond neighbourhood $O$ then
$K^{\perp\perp}$ is compact and contained in $\cl(O)$. 
\end{Lem}
{\noindent\bf Remark:} In general $K^{\perp\perp}$ need not be compact, e.g., if $K$
contains a Cauchy surface for $\Mb$. As another example,
let $K$ be a closed ball of radius $1$ in the $t=0$ plane of the
$|t|<1/2$ portion of Minkowski space in standard coordinates; then
$K^{\perp\perp}$ is the $|t|<1/2$ portion of the diamond based on the
interior of $K$, and is noncompact.

{\noindent\em Proof:} (i) As $K$ is compact, $J_\Mb(K)$ is closed and $K^\perp
=\Mc\setminus J_\Mb(K)$ is therefore open. Hence, by Lem.~\ref{lem:Jofopenisopen}, 
$J_\Mb(K^\perp)$ is open and
$K^{\perp\perp}$ is closed. Moreover, as $K^\perp\cap J_\Mb(K)$ is
empty, so is $J_\Mb(K^\perp)\cap K$; hence we see that $K\subset
K^{\perp\perp}$.\\
(ii) If $p\in K^{\perp\perp}$ then $J_\Mb(p)\subset J_\Mb(K)$; otherwise, $J_\Mb(p)$
would intersect $K^\perp$, giving  $p\in J_\Mb(K^\perp)$ and a contradiction. Thus
$J_\Mb(K^{\perp\perp})=J_\Mb(K)$ and so $K^{\perp\perp\perp}=K^\perp$. In
particular, $K^{\perp\perp}$ is casually complete with respect to
$\perp$. \\
(iii) Take any $p,q\in K^{\perp\perp}$.  If a future-directed causal curve $\gamma$ joins $p$
and $q$ but leaves $K^{\perp\perp}$ there must be $r\in J_\Mb(K^\perp)$
such that $q\in J_\Mb^+(r)$, $p\in J_\Mb^-(r)$. Thus one or both of $p,q$ belong
to $J_\Mb(K^\perp)$, which is a contradiction. Hence $K^{\perp\perp}$ is
causally convex and therefore a closed globally hyperbolic subset of
$\Mb$. \\
Finally, if $K$ has a multi-diamond neighbourhood $O$, then
$O'\subset K^\perp$, and hence $J_\Mb(O')\subset J_\Mb(K^\perp)$. Hence 
$K^{\perp\perp}\subset \Mc\setminus J_\Mb(O')=\cl(O'')=\cl(O)$, which is
compact. Accordingly, $K^{\perp\perp}$ is a closed subset contained in a compact
set, and hence compact. $\square$

\begin{Lem}\label{lem:perp_of_reg_int}
 Let $\Mb$ be a globally hyperbolic spacetime. (i) Suppose $O_1$ and $O_2$ are open subsets of $\Mb$, 
with $O_1$ relatively compact and $\cl(O_1)\subset O_2$. Then $\cl(J_\Mb(O_1))\subset J_\Mb(O_2)$. 
(ii) Suppose $O_n$ ($n\in\NN)$ is a sequence of relatively compact subsets of $\Mb$ with $\cl(O_{n+1})\subset O_n$ for all $n\in\NN$ and
$\bigcap_{n\in\NN} O_n = K$ compact. Then
\[
J_\Mb(K) = \bigcap_{n\in\NN} J_\Mb(O_n) = \bigcap_{n\in\NN} \cl(J_\Mb(O_n)), \quad\textrm{and hence}\quad 
K^\perp = \bigcup_{n\in\NN} O_n'.
\]
\end{Lem}
{\noindent\em Proof:} (i) We calculate
\[
\cl(J_\Mb(O_1)) \subset \cl (J_\Mb(\cl(O_1))) = J_\Mb(\cl(O_1)) \subset J_\Mb(O_2)
\]
using the fact that $J_\Mb(\cl(O_1))$ is closed. \\
(ii) The inclusion $J_\Mb(K)\subset \bigcap_{n\in\NN} J_\Mb(O_n)$ is immediate from $K\subset O_n$ for all $n$.
On the other hand, if $p\in \bigcap_{n\in\NN} J_\Mb(O_n)$ then there exist $q_n\in J_\Mb(p)\cap O_n$ for all $n$. 
As all $q_n$ are contained in the relatively compact set $O_1$ we may pass to a convergent subsequence $q_{n_r}$ with
limit $q\in \cl(O_1)$; as all but finitely many of the $q_{n_r}$ are contained in each $O_{m+1}$ $(m=1,2,\ldots$), 
we also have $q\in\cl(O_{m+1})\subset O_m$ for each $m\in\NN$ and hence $q\in K$. As the $q_{n_r}$ lie in the closed
set $J_\Mb(p)$, we additionally have $q\in K\cap J_\Mb(p)$ and hence conclude that $p\in J_\Mb(K)$. Accordingly we have proved
the first of the required equalities. By part (i) we have $\cl (J_\Mb(O_{n+1}))\subset J_\Mb(O_n)$ for all $n$ from which the second equality
follows. Taking complements in $\Mb$ we obtain the required formula for $K^\perp$.
$\square$

\begin{Lem} \label{lem:perp+prime}
Let $S$ be a subset of a time-oriented Lorentzian spacetime $\Mb$ such that $J_\Mb^+(\cl(S))$ is closed (for example, 
if $S$ is a relatively compact subset of a globally hyperbolic spacetime). Then
\begin{equation}\label{eq:ccs}
J_\Mb^+(\cl(S)) = \cl (I_\Mb^+(S)) = \cl (J_\Mb^+ (S)).
\end{equation}
The analogous result holds for causal and chronological pasts. If both $J_\Mb^\pm(\cl (S))$ are closed then 
$J_\Mb(\cl (S)) = \cl (J_\Mb(S))$ and hence $(\cl (S))^\perp = S'$. 
\end{Lem}
{\noindent\em Proof:} Owing to the hypothesis, we have
\[
J_\Mb^+(\cl (S)) = \cl (I_\Mb^+(\cl (S))) = \cl (I_\Mb^+(S)) \subset \cl (J_\Mb^+(S)) \subset \cl (J_\Mb^+(\cl (S))) = J_\Mb^+(\cl (S))
\]
using the standard results Lemma 14.6(2) in~\cite{ONeill} and Prop.~2.11 in~\cite{Penrose1972} for the first two equalities. This establishes Eq.~\eqref{eq:ccs}; the remaining statements are trivial. 
$\square$

\begin{Lem} \label{lem:domain_of_dependence_and_perpperp}
Let $S$ be any subset in a globally hyperbolic spacetime $\Mb$. 
Then the Cauchy development obeys
$D_\Mb(S)\subset S^{\perp\perp}$, with equality if $S$ lies in an acausal
Cauchy surface of $\Mb$.
\end{Lem}
{\noindent\bf Remark:} The example of $S=\{p,q\}$ for $q\in J_\Mb^+(p)$, 
for which $D_\Mb(S)=S$, $S^{\perp\perp}=J_\Mb^+(p)\cap J_\Mb^-(q)$, shows that equality cannot be expected in general.\\
{\noindent\em Proof:} If $p\in D_\Mb(S)$ then every inextendible causal curve through $p$ intersects $S$. Thus any point
causally connected to $p$ is causally connected to $S$, i.e., $p\notin J_\Mb(S^\perp)$ and hence $p\in S^{\perp\perp}$. 
If $S\subset \Sigma$, an acausal Cauchy surface of $\Mb$, then $\Sigma\setminus S\subset S^\perp$. Accordingly, any inextendible causal
curve through $p\in S^{\perp\perp}$ must cut $\Sigma$ in $S$, so $S^{\perp\perp}=D_\Mb(S)$ in this case. 
$\square$ 

\begin{Lem} \label{lem:exhaustion}
Suppose $D$ is a multi-diamond, with base $B$ in spacetime $\Mb\in\Mand$. 
If $K$ is any compact subset of $D$ then $K\subset \tilde{K}^{\perp\perp}$ for
a compact subset $\tilde{K}$ of $B$ (hence $\tilde{K}\in\KK(\Mb;D)$).
\end{Lem} 
{\noindent\em Proof:} Suppose $\Sigma$ is a spacelike Cauchy surface for $\Mb$ with $B\subset \Sigma$. Then $J_\Mb(K)\cap \Sigma$ is compact and contained in $B$, which has a finite number $R$ of connected components $B_r$. Each $B_r$ is contained in a chart $(U_r,\phi_r)$ of $\Sigma$ in which $\phi_r(B_r)$ is an open ball; we may choose a compact set $K_r$ so that $\phi_r(K_r)$ is the closure of a slightly smaller ball with the same centre 
and so that $K_r$ contains $J_\Mb(K)\cap \Sigma\cap B_r$. Then $\tilde{K}=\bigcup_{r=1}^R K_r$
is compact and contains $J_\Mb(K)\cap \Sigma$. 
Moreover, $K\subset D_\Mb(\tilde{K})=\tilde{K}^{\perp\perp}$ by  Lemma~\ref{lem:domain_of_dependence_and_perpperp} and the
fact that spacelike Cauchy surfaces are acausal \cite[Lem. 14.42]{ONeill}. 
Finally, $\tilde{K}$ has a multi-diamond neighbourhood $D$,
with base $B\subset D$, so $\tilde{K}\in \KK(\Mb;D)$. $\square$

\begin{Lem} \label{lem:psi_perpperp}
If $\psi:\Mb\to\Nb$ in $\Mand$  then $\psi(K^{\perp\perp}) = \psi(K)^{\perp\perp}$ for all $K\in\KK(\Mb)$.
\end{Lem}
{\noindent\em Proof:}
Observe first that for any subset $S\subset \Mb$, we have $J_\Nb(\psi(S))\cap \psi(\Mb) = 
\psi(J_\Mb(S))$ by causal convexity of $\psi(\Mb)$ and hence 
$\psi(S)^\perp \cap \psi(\Mb) = \psi(S^\perp)$, using also the injectivity of $\psi$. 
It follows that 
\[
\psi(K^{\perp\perp}) = \psi(K^\perp)^\perp\cap\psi(\Mb) = 
(\psi(K)^\perp\cap\psi(\Mb))^\perp\cap\psi(\Mb).
\]
But as $K$ has a multi-diamond neighbourhood
$D$ in $\Mb$, $K^{\perp\perp}\subset \cl(D)$ (Lem.~\ref{lem:perp_of_compacts});
similarly, as $\psi(D)$ is a multi-diamond in $\Nb$ we have $\psi(K)^{\perp\perp}
\subset \cl(\psi(D))\subset \psi(\Mb)$ and hence 
\[
\psi(K^{\perp\perp}) =
(\psi(K)^\perp\cap\psi(\Mb))^\perp \supset \psi(K)^{\perp\perp}.
\]
Now take any point $p\in K^{\perp\perp}$ and suppose for a contradiction
that $\psi(p) \in \Nb\setminus \psi(K)^{\perp\perp} = J_\Nb(\psi(K)^\perp)$. 
By causal convexity of $\psi(\Mb)$, $\psi(p)$ would lie in $J_\Nb(\psi(K)^\perp)$ only
if $p\in J_\Mb(K^\perp)$, which would contradict the assumption that $p\in K^{\perp\perp}$. 
Accordingly, we have $\psi(p)\subset J_\Nb(q)$ for some
$q\in \psi(K)^\perp\setminus\psi(K^\perp)$, which must therefore
lie outside $\psi(\Mb)$ because $\psi(K^\perp) = \psi(K)^\perp\cap \psi(\Mb)$ as shown above. 
Without loss
of generality, we may suppose that $\psi(p)$ lies to the future of $q$
along smooth causal curve $\gamma$.
The pre-image of $\gamma$ under $\psi$ is a connected future-directed smooth causal curve,
which is past-inextendible in $\Mb$ and therefore contains
points outside $J_\Mb^+(K)$. Take any such
point $r$; $r$ cannot lie in $J_\Mb^-(K)$ (otherwise $q\in J_\Nb^-(\psi(K))$)
and hence $r\in K^\perp$. But this entails that $p\in J_\Mb(K^\perp)$, 
contradicting the initial assumption $p\in K^{\perp\perp}$. $\square$ 

\section{Subobjects, intersections and unions}\label{appx:subob}

We summarise the basic properties of subobjects that are used in the body of the text.
For completeness, we also include some standard definitions of category theory (although
we take the basic definition of a category for granted). To a large extent we 
follow~\cite{DikranjanTholen}. 

In a general category $\Ct$, then, a morphism $f$ is described as {\em monic} (or as a {\em monomorphism}) iff it
is left-cancellable, so $f\circ g=f\circ h$ implies $g=h$, 
and as {\em epic} (or as an {\em epimorphism}) iff it is right-cancellable, so 
$g\circ f=h\circ f$ implies $g=h$.  
An object $\mho$ of $\Ct$ is {\em initial} if
there is a unique morphism $\mho_A:\mho\to A$ for each object $A$ of $\Ct$. 
A monic will be equivalently described as defining a {\em subobject} of its codomain,
so that $m:M\to A$ is a subobject of $A$. 
In cases where the morphism $\mho_A$ is monic, we will describe this as the trivial
subobject of $A$. Subobjects $M\stackrel{m}{\to}A$ and
$M'\stackrel{m'}{\to}A$ are isomorphic iff there exists an isomorphism $f:M\to
M'$ such that $m=m'\circ f$, in which case we write $m\cong m'$; in the case
where $m=m'\circ f$ for some $f$ that is not necessarily an isomorphism,
we write $m\le m'$ ($f$ is uniquely specified because $m'$ is monic). 

A category $\Ct$ has {\em equalizers} if it satisfies the following condition: for every pair of morphisms $f,g:A\to B$ there is a morphism $h$  such that $f\circ h= g \circ h$ and such that
if $k$ is any morphism such that $f\circ k=g\circ k$ then $k$ factorizes
uniquely via $h$, i.e., $k=h\circ m$ for a unique morphism $m$; $h$ is said to be an
equalizer of $f$ and $g$ in this situation.

Given a collection $(m_i)_{i\in I}$ [in which $I$ is a class] of subobjects of $A$ their
intersection and union may be defined as follows: An {\em
intersection} is a subobject $m:M\to A$ with the following properties:
\begin{enumerate}\addtolength{\itemsep}{-0.5\baselineskip}
\item $m$ factorises via each $m_i$ as $m=m_i\circ j_i$; 
\item given any $f:B\to A$ factorising  via each $m_i$ as $f=m_i\circ
k_i$, there exists a unique $g:B\to M$ such that 
$j_i\circ g = k_i$ for all $i\in I$, and hence $f=m\circ g$.
\end{enumerate}
These properties define $m$ up to isomorphism and we write
\[
m\cong \bigwedge_{i\in I} m_i: \bigwedge_{i\in I} M_i \to A.
\]
The category $\Ct$ is said to have intersections (with respect to monics) if every such
collection of subjobjects has an intersection. More generally, one can define intersections
with respect to a subclass $\cal M$ of monics~\cite{DikranjanTholen}.
\begin{Lem} \label{lem:intersections}
(a) With the above notation, if $(v_i)_{i\in I}$ are isomorphisms pre-composable with
the $(m_i)$ then $(m_i)_{i\in I}$ has an intersection if and only if $(m_i\circ v_i)_{i\in I}$ does,
and 
\[
\bigwedge_{i\in I} m_i\circ v_i \cong \bigwedge_{i\in I} m_i
\]
(b) If $k:A\to A'$ is monic then $(k\circ m_i)_{i\in I}$ has an intersection if and only if $(m_i)_{i\in I}$ does; provided that $I$ is nonempty\footnote{The intersection of an empty class of subobjects of $A$ is $\id_A$.} we have
\[
k\circ \bigwedge_{i\in I} m_i \cong \bigwedge_{i\in I} k\circ m_i
\]
\end{Lem}
{\noindent\em Proof:} (a) Suppose $(m_i)$ has an intersection $m$ with factorizations $m=m_i\circ j_i$. 
Then $m$ also factorizes as $m=m_i\circ v_i \circ j'_i$ for $j'_i=v_i^{-1}\circ j_i$ and we will show
that this defines an intersection of $(m_i\circ v_i)_{i\in I}$. Suppose 
$f$ factorizes as $f=m_i\circ v_i \circ k_i$, then the intersection property of the $(m_i)$ implies
that there is a unique $g$ such that $v_i\circ k_i=j_i\circ g$ and hence $k_i = j'_i\circ g$ for all $i\in I$. 
Thus $(m_i\circ v_i)_{i\in I}$ has $m$ as an intersection. The reverse implication also follows from this argument.\\
(b) Suppose $(m_i)$ has an intersection $m$ with factorizations $m=m_i\circ j_i$; we must
show that $k\circ m$ is an intersection of the $k\circ m_i$, with factorizations $k\circ m = (k\circ m_i)\circ j_i$. To this end, suppose there are factorizations $f=k\circ m_i\circ l_i$ for all $i$. As
$k$ is monic, this implies the existence of $h$ such that $m_i\circ l_i = h$ for all $i$ and (because $\bigwedge_i m_i$ exists), 
the existence of a unique $g$ with $l_i =  j_i\circ g$ for all $i$, which was to be shown. 
On the other hand, suppose that $(k\circ m_i)_{i\in I}$ have an intersection $h = k\circ m_i\circ j_i$. 
Again, as $k$ is monic, we may write $h=k\circ m$ with $m=m_i\circ j_i$ for all $i$. To see that
this defines an intersection of $(m_i)_{i\in I}$, suppose $f=m_i\circ l_i$ for all $i$. Then $k\circ f=
k\circ m_i\circ l_i$ and (because $\bigwedge_i  k\circ m_i$ exists) there is a unique $g$ such that $l_i = j_i\circ g$, which was to be shown.  $\square$

On the other hand, the {\em union} is
a subobject $m:M\to A$ with the following properties
\begin{enumerate}\addtolength{\itemsep}{-0.5\baselineskip}
\item every $m_i$ factorises as $m_i=m\circ \tilde{m}_i$ (in which
$\tilde{m}_i:M_i\to M$)
\item given any $f:A\to B$, if there exists a subobject $n:N\to B$ such that every $f\circ m_i$ factorises as
$n\circ\tilde{n}_i$, then there is a unique morphism $\tilde{f}:M\to N$
such that $n\circ \tilde{f}=f\circ m$ and $\tilde{f}\circ \tilde{m}_i=\tilde{n}_i$ for all $i\in I$.
\end{enumerate}
Property (2) can be displayed diagrammatically as the commuting diagram
\begin{equation}\label{eq:property(2)}
\begin{tikzpicture}[description/.style={fill=white,inner sep=2pt}]
\matrix (m) [ampersand replacement=\&,matrix of math nodes, row sep=4em,
column sep=6em, text height=1.5ex, text depth=0.25ex]
{ \&  M_i \&    \\
M \&  \&  N  \&    \\
A  \&  \&  B
\\};
\path[->]
(m-1-2)  edge node[above,sloped] {$ \tilde{m}_i  $} (m-2-1)
             edge node[above,sloped] {$ \tilde{n}_i $} (m-2-3)
(m-2-1)  edge node[above] {$ \tilde{f} $} (m-2-3)
             edge node[above,left] {$ m $} (m-3-1)
(m-3-1)  edge node[above] {$ f $} (m-3-3)
(m-2-3)  edge node[left] {$ n $} (m-3-3);
\end{tikzpicture}
\end{equation}
%
%
(in which it is tacit that $m_i=m\circ \tilde{m}_i$).

It is easy to see that this defines the union subobject up to
isomorphism; we therefore write [following \cite{DikranjanTholen} \S1.9]
\[
m\cong \bigvee_{i\in I} m_i : \bigvee_{i\in I} M_i\to A
\]
The union always exists if $\Ct$ has intersections and also has pull-backs with respect
to monics in the following sense: whenever $f:X\to Y$ and $n:N\to Y$ is a subobject, there
is a subobject $m:M\to X$ and a morphism $f':M\to N$ such that $n\circ f'=f\circ m$, and
if there are morphisms $g$ and $h$ such that $n\circ h=f\circ g$ then there is a unique $t$ 
such that $m\circ t = g$, whereupon also $h=f'\circ t$.

\begin{Lem}\label{lem:union_refine}
Let $(m_i)_{i\in I}$ (resp., $(n_j)_{j\in J}$) be a class-indexed family of subobjects of $A\in \Ct$ with union
$m:M\to A$ (resp., $n:N\to A$). If, to each $i\in I$ there is $j(i)\in J$ such that
$m_i=n_{j(i)}\circ\mu_i$ for some $\mu_i$, then there is a unique $\xi:M\to N$
such that $n\circ\xi=m$. If, additionally, $J\subset I$ and $n_j\cong m_j$ for each $j\in J$ then $\xi$ is an isomorphism.
\end{Lem}
{\noindent\em Proof:} Let $n_j=n\circ \hat{n}_j$ be the factorizations associated
with $\bigvee_{j\in J} n_j$, and consider diagram~\eqref{eq:property(2)}, with $B=A$, $f=\id_A$ and $\tilde{n}_i = \hat{n}_{j(i)}\circ\mu_i$. As the outer portion commutes
we deduce the existence of a unique $\xi$ (replacing $\tilde{f}$) with the property stated. In the special case, we may apply this result again with the roles of $m_i$ and $n_j$ reversed, giving a unique $\eta$ such that $m\circ\eta=n$. As $m$ and $n$ are monic, it follows that $\eta$ and $\xi$ are mutual inverses, hence isomorphisms.
$\square$

A useful consequence is that if $I$ is a class and for each $i\in I$ there is
a nonempty class $J_i$ labelling subobjects $m_{ij}$, then we have the `Fubini property'
\begin{equation}\label{eq:Fubini}
\bigvee_{i\in I} \bigvee_{j\in J_i} m_{ij}\cong
\bigvee_{(i,j)\in K} m_{ij} \cong \bigvee_{j\in J}\bigvee_{i\in I_j} m_{ij}
\end{equation}
where $J=\bigcup_{i\in I}J_i$, $K=\{(i,j)\in I\times J:j\in J_i\}$ and
$I_j=\{i\in I:j\in J_i\}$ ($j\in J$).

\begin{Lem}\label{lem:union_invariance}
Suppose a category $\Ct$ has equalizers,
and intersections and pullbacks with respect to monics. Let
 $(m_i)_{i\in I}$ be a class-indexed family of subobjects of $A\in \Ct$ with union
$m:M\to A$. If $h:A\to A$ obeys $h\circ m_i=m_i$ for all $i\in I$ then $h\circ m=m$. 
\end{Lem}
{\noindent\em Proof:} We have $h\circ m_i=\id_A\circ m_i$ and hence a factorisation 
$m_i=g\circ \tilde{g}_i$ for each $i\in I$ where $g$ is an equalizer of $h$ and $\id_A$ (and is
necessarily monic). In conjunction with the factorisation $m_i=m\circ\tilde{m}_i$ this induces
a factorisation $m_i=n\circ\tilde{n}_i$ via the intersection (=pullback) $n:N\to A$ of $g$ and $m$,
corresponding to $n=g\circ k = m\circ \ell$. The outer portion of the 
diagram~\eqref{eq:property(2)} commutes for all $i\in I$, 
with $B=A$, $f=\id_A$, and there is therefore a morphism $\tilde{f}$ to make the diagram commute in full. 
Consequently, $h\circ m = h\circ n\circ \tilde{f} = h\circ g\circ k\circ \tilde{f} = g\circ k\circ \tilde{f} =
n\circ \tilde{f} = m$ as required. 
$\square$

As we study categories in which all morphisms are monic, the existence
of pull-backs with respect to monics follows from existence of intersections. 
%
%
%
%


\begin{thebibliography}{10}\setlength{\itemsep}{-1.5mm}
\providecommand{\url}[1]{{#1}}
\providecommand{\urlprefix}{URL }
\expandafter\ifx\csname urlstyle\endcsname\relax
  \providecommand{\doi}[1]{DOI~\discretionary{}{}{}#1}\else
  \providecommand{\doi}{DOI~\discretionary{}{}{}\begingroup
  \urlstyle{rm}\Url}\fi

\bibitem{AdamekHerrlichStrecker}
Ad{\'a}mek, J., Herrlich, H., Strecker, G.E.: Abstract and concrete categories:
  the joy of cats.
\newblock Repr. Theory Appl. Categ. pp. 1--507 (2006), reprint of the 1990
  original [Wiley, New York]

\bibitem{BeemEhrlichEasley}
Beem, J.K., Ehrlich, P.E., Easley, K.L.: Global {L}orentzian geometry,
  \emph{Monographs and Textbooks in Pure and Applied Mathematics}, vol. 202,
  second edn.
\newblock Marcel Dekker Inc., New York (1996)

\bibitem{Bernal:2003jb}
Bernal, A.N., S{\'{a}}nchez, M.: {On smooth Cauchy hypersurfaces and Geroch's
  splitting theorem}.
\newblock Commun. Math. Phys. \textbf{243}, 461--470 (2003)

\bibitem{Bernal:2004gm}
Bernal, A.N., S{\'{a}}nchez, M.: {Smoothness of time functions and the metric
  splitting of globally hyperbolic spacetimes}.
\newblock Commun. Math. Phys. \textbf{257}, 43--50 (2005), gr-qc/0401112

\bibitem{Bernal:2005qf}
Bernal, A.N., S{\'{a}}nchez, M.: {Further results on the smoothability of
  Cauchy hypersurfaces and Cauchy time functions}.
\newblock Lett. Math. Phys. \textbf{77}, 183--197 (2006), gr-qc/0512095

\bibitem{Bernal:2006xf}
Bernal, A.N., S{\'{a}}nchez, M.: {Globally hyperbolic spacetimes can be defined
  as causal instead of strongly causal}.
\newblock Class. Quantum Grav. \textbf{24}, 745--750 (2007), gr-qc/0611138

\bibitem{BrFr2000}
Brunetti, R., Fredenhagen, K.: Microlocal analysis and interacting quantum
  field theories: {R}enormalization on physical backgrounds.
\newblock Commun. Math. Phys. \textbf{208}, 623--661 (2000)

\bibitem{BruFre_LNP:2009}
Brunetti, R., Fredenhagen, K.: Quantum field theory on curved backgrounds.
\newblock In: Quantum field theory on curved spacetimes, \emph{Lecture Notes in
  Phys.}, vol. 786, pp. 129--155. Springer, Berlin (2009)

\bibitem{BrFrVe03}
Brunetti, R., Fredenhagen, K., Verch, R.: The generally covariant locality
  principle: A new paradigm for local quantum physics.
\newblock Commun. Math. Phys. \textbf{237}, 31--68 (2003)

\bibitem{BruGuiLo:2002}
Brunetti, R., Guido, D., Longo, R.: Modular localization and {W}igner
  particles.
\newblock Rev. Math. Phys. \textbf{14}, 759--785 (2002)

\bibitem{Br&Ru05}
Brunetti, R., Ruzzi, G.: Superselection sectors and general covariance. {I}.
\newblock Commun. Math. Phys. \textbf{270}, 69--108 (2007)

\bibitem{BrunettiRuzzi_topsect}
Brunetti, R., Ruzzi, G.: Quantum charges and spacetime topology: {T}he
  emergence of new superselection sectors.
\newblock Commun. Math. Phys. \textbf{287}, 523--563 (2009)

\bibitem{BDFS:2000}
Buchholz, D., Dreyer, O., Florig, M., Summers, S.J.: Geometric modular action
  and spacetime symmetry groups.
\newblock Rev. Math. Phys. \textbf{12}, 475--560 (2000)

\bibitem{BuLe:2004}
Buchholz, D., Lechner, G.: Modular nuclearity and localization.
\newblock Ann. Henri Poincar\'e \textbf{5}, 1065--1080 (2004)

\bibitem{BuMuSu:2002}
Buchholz, D., Mund, J., Summers, S.J.: Covariant and quasi-covariant quantum
  dynamics in {R}obertson-{W}alker spacetimes.
\newblock Class. Quantum Grav. \textbf{19}, 6417--6434 (2002)

\bibitem{BILY}
Budic, R., Isenberg, J., Lindblom, L., Yasskin, P.B.: On the determination of
  {C}auchy surfaces from intrinsic properties.
\newblock Commun. Math. Phys. \textbf{61}, 87--95 (1978)

\bibitem{CoRo:1994}
Connes, A., Rovelli, C.: von {N}eumann algebra automorphisms and
  time-thermodynamics relation in generally covariant quantum theories.
\newblock Class. Quantum Grav. \textbf{11}, 2899--2917 (1994)

\bibitem{DapFrePin2008}
Dappiaggi, C., Fredenhagen, K., Pinamonti, N.: {Stable cosmological models
  driven by a free quantum scalar field}.
\newblock Phys. Rev. \textbf{D77}, {104}{015} (2008)

\bibitem{DHP_dirac:2009}
Dappiaggi, C., Hack, T.P., Pinamonti, N.: The extended algebra of observables
  for {D}irac fields and the trace anomaly of their stress-energy tensor.
\newblock Rev. Math. Phys. \textbf{21}, 1241--1312 (2009)

\bibitem{DegVer2010}
Degner, A., Verch, R.: Cosmological particle creation in states of low energy.
\newblock J. Math. Phys. \textbf{51}, {022}{302} (2010)

\bibitem{DikranjanTholen}
Dikranjan, D., Tholen, W.: Categorical structure of closure operators,
  \emph{Mathematics and its Applications}, vol. 346.
\newblock Kluwer Academic Publishers Group, Dordrecht (1995)

\bibitem{Dimock1980}
Dimock, J.: Algebras of local observables on a manifold.
\newblock Commun. Math. Phys. \textbf{77}, 219--228 (1980)

\bibitem{Ferg_in_prep}
Ferguson, M.: Dynamical locality of the nonminimally coupled scalar field and
  enlarged algebra of {W}ick polynomials.
\newblock ArXiv:1203.2151

\bibitem{Fewster:gauge}
Fewster, C.J.: Endomorphisms and automorphisms of locally covariant quantum
  field theories.
\newblock ArXiv:1201.3295

\bibitem{Fewster2007}
Fewster, C.J.: Quantum energy inequalities and local covariance. {II}.
  {C}ategorical formulation.
\newblock Gen. Relativity Gravitation \textbf{39}, 1855--1890 (2007)

\bibitem{FewsterRegensburg}
Fewster, C.J.: On the notion of `the same physics in all spacetimes'.
\newblock In: F.~Finster, O.~M{\"u}ller, M.~Nardmann, J.~Tolksdorf, E.~Zeidler
  (eds.) Quantum Field Theory and Gravity. Conceptual and mathematical advances
  in the search for a unified framework. Birkh{\"a}user (2012).
\newblock ArXiv:1105.6202

\bibitem{Few&Pfen06}
Fewster, C.J., Pfenning, M.J.: Quantum energy inequalities and local
  covariance. {I}: Globally hyperbolic spacetimes.
\newblock J. Math. Phys. \textbf{47}, {082}{303} (2006)

\bibitem{FewVer:dynloc2}
Fewster, C.J., Verch, R.: Dynamical locality of the free scalar field.
\newblock ArXiv:1109.6732, to appear in {A}nnales H.~{P}oincar{\'e}

\bibitem{FullingNarcowichWald}
Fulling, S.A., Narcowich, F.J., Wald, R.M.: Singularity structure of the
  two-point function in quantum field theory in curved spacetime. {II}.
\newblock Ann. Physics \textbf{136}, 243--272 (1981)

\bibitem{Haag}
Haag, R.: Local Quantum Physics: Fields, Particles, Algebras.
\newblock Springer-Verlag, Berlin (1992)

\bibitem{HawkingEllis}
Hawking, S.W., Ellis, G.F.R.: The Large Scale Structure of Space-Time.
\newblock Cambridge University Press, London (1973)

\bibitem{Ho&Wa01}
Hollands, S., Wald, R.M.: Local {W}ick polynomials and time ordered products of
  quantum fields in curved spacetime.
\newblock Commun. Math. Phys. \textbf{223}, 289--326 (2001)

\bibitem{Ho&Wa02}
Hollands, S., Wald, R.M.: Existence of local covariant time ordered products of
  quantum fields in curved spacetime.
\newblock Commun. Math. Phys. \textbf{231}, 309--345 (2002)

\bibitem{Kay1978}
Kay, B.S.: Linear spin-zero quantum fields in external gravitational and scalar
  fields. {I}. {A} one particle structure for the stationary case.
\newblock Commun. Math. Phys. \textbf{62}, 55--70 (1978)

\bibitem{Kay79}
Kay, B.S.: Casimir effect in quantum field theory.
\newblock Phys. Rev. \textbf{D20}, 3052--3062 (1979)

\bibitem{Kay_Flocality:1992}
Kay, B.S.: The principle of locality and quantum field theory on (non-globally
  hyperbolic) curved spacetimes.
\newblock Rev. Math. Phys. (Special Issue), 167--195 (1992)

\bibitem{Kuckert:2000}
Kuckert, B.: Localization regions of local observables.
\newblock Commun. Math. Phys. \textbf{215}, 197--216 (2000)

\bibitem{Landau1969}
Landau, L.J.: A note on extended locality.
\newblock Commun. Math. Phys. \textbf{13}, 246--253 (1969)

\bibitem{Landau:1974}
Landau, L.J.: On local functions of fields.
\newblock Commun. Math. Phys. \textbf{39}, 49--62 (1974)

\bibitem{Lee:intro_to_smooth_manifolds}
Lee, J.M.: Introduction to smooth manifolds, \emph{Graduate Texts in
  Mathematics}, vol. 218.
\newblock Springer-Verlag, New York (2003)

\bibitem{Maclane}
Mac~Lane, S.: Categories for the Working Mathematician, 2nd edn.
\newblock Springer-Verlag, New York (1998)

\bibitem{Muellner2009}
M{\"u}llner, D.: Orientation reversal of manifolds.
\newblock Algebr. Geom. Topol. \textbf{9}, 2361--2390 (2009)

\bibitem{NomizuHideki1961}
Nomizu, K., Ozeki, H.: The existence of complete {R}iemannian metrics.
\newblock Proc. Amer. Math. Soc. \textbf{12}, 889--891 (1961)

\bibitem{ONeill}
O'Neill, B.: Semi-Riemannian Geometry.
\newblock Academic Press, New York (1983)

\bibitem{Penrose1972}
Penrose, R.: Techniques of differential topology in relativity.
\newblock Society for Industrial and Applied Mathematics, Philadelphia, Pa.
  (1972).
\newblock Conference Board of the Mathematical Sciences Regional Conference
  Series in Applied Mathematics, No. 7

\bibitem{Ruzzi:2005}
Ruzzi, G.: Homotopy of posets, net-cohomology and superselection sectors in
  globally hyperbolic space-times.
\newblock Rev. Math. Phys. \textbf{17}, 1021--1070 (2005)

\bibitem{Sanders_ReehSchlieder}
Sanders, K.: On the {R}eeh-{S}chlieder property in curved spacetime.
\newblock Commun. Math. Phys. \textbf{288}, 271--285 (2009)

\bibitem{Sanders_dirac:2010}
Sanders, K.: The locally covariant {D}irac field.
\newblock Rev. Math. Phys. \textbf{22}, 381--430 (2010)

\bibitem{Schoch1968}
Schoch, A.: On the simplicity of {H}aag fields.
\newblock Int. J. Theor. Phys. \textbf{1}, 107--113 (1968)

\bibitem{Verch01}
Verch, R.: A spin-statistics theorem for quantum fields on curved spacetime
  manifolds in a generally covariant framework.
\newblock Commun. Math. Phys. \textbf{223}, 261--288 (2001)

\bibitem{VerchRegensburg}
Verch, R.: Local covariance, renormalization ambiguity, and local thermal
  equilibrium in cosmology.
\newblock In: F.~Finster, O.~M{\"u}ller, M.~Nardmann, J.~Tolksdorf, E.~Zeidler
  (eds.) Quantum Field Theory and Gravity. Conceptual and mathematical advances
  in the search for a unified framework. Birkh{\"a}user (2012).
\newblock ArXiv:1105.6249

\bibitem{Wald_gr}
Wald, R.M.: General Relativity.
\newblock University of Chicago Press, Chicago (1984)

\bibitem{Wiesbrock:1998}
Wiesbrock, H.W.: Modular intersections of von {N}eumann algebras in quantum
  field theory.
\newblock Commun. Math. Phys. \textbf{193}, 269--285 (1998)

\end{thebibliography}
\end{document}